\documentclass[12pt,english]{article}
\usepackage{times}
\usepackage[T1]{fontenc}
\usepackage{geometry}
\usepackage{array}
\usepackage{rotating}
\usepackage{graphicx}
\usepackage{setspace}
\usepackage{amssymb}

\makeatletter

\providecommand{\tabularnewline}{\\}

\usepackage{bm}

\usepackage{babel}
\makeatother
\begin{document}

\section*{Multi-Resolution Subspace-Based Optimization Method for the Retrieval
of \emph{2D} Perfect Electric Conductors}

\noindent \vfill

\noindent X. Ye,$^{(1)}$, F. Zardi,$^{(2)(3)}$, M. Salucci,$^{(2)(3)}$\emph{,}
and A. Massa,$^{(2)(3)(4)(5)}$

\noindent \vfill

\noindent {\footnotesize ~}{\footnotesize \par}

\noindent {\footnotesize $^{(1)}$} \emph{\footnotesize Beijing Institute
of Technology}{\footnotesize \par}

\noindent {\footnotesize School of Information and Electronics, Beijing
1008 - China}{\footnotesize \par}

\noindent \textit{\emph{\footnotesize E-mail:}} \emph{\footnotesize xiuzhuye}{\footnotesize @}\emph{\footnotesize outlook.com}{\footnotesize \par}

\noindent {\footnotesize ~}{\footnotesize \par}

\noindent {\footnotesize $^{(2)}$} \emph{\footnotesize ELEDIA Research
Center} {\footnotesize (}\emph{\footnotesize ELEDIA}{\footnotesize @}\emph{\footnotesize UniTN}
{\footnotesize - University of Trento)}{\footnotesize \par}

\noindent {\footnotesize DICAM - Department of Civil, Environmental,
and Mechanical Engineering}{\footnotesize \par}

\noindent {\footnotesize Via Mesiano 77, 38123 Trento - Italy}{\footnotesize \par}

\noindent \textit{\emph{\footnotesize E-mail:}} {\footnotesize \{}\emph{\footnotesize francesco.zardi,
marco.salucci, andrea.massa}{\footnotesize \}@}\emph{\footnotesize unitn.it}{\footnotesize \par}

\noindent {\footnotesize Website:} \emph{\footnotesize www.eledia.org/eledia-unitn}{\footnotesize \par}

\noindent {\footnotesize ~}{\footnotesize \par}

\noindent {\footnotesize $^{(3)}$} \emph{\footnotesize CNIT - \char`\"{}University
of Trento\char`\"{} ELEDIA Research Unit }{\footnotesize \par}

\noindent {\footnotesize Via Sommarive 9, 38123 Trento - Italy}{\footnotesize \par}

\noindent {\footnotesize Website:} \emph{\footnotesize www.eledia.org/eledia-unitn}{\footnotesize \par}

\noindent {\footnotesize ~}{\footnotesize \par}

\noindent {\footnotesize $^{(4)}$} \emph{\footnotesize ELEDIA Research
Center} {\footnotesize (}\emph{\footnotesize ELEDIA}{\footnotesize @}\emph{\footnotesize UESTC}
{\footnotesize - UESTC)}{\footnotesize \par}

\noindent {\footnotesize School of Electronic Engineering, Chengdu
611731 - China}{\footnotesize \par}

\noindent \textit{\emph{\footnotesize E-mail:}} \emph{\footnotesize andrea.massa@uestc.edu.cn}{\footnotesize \par}

\noindent {\footnotesize Website:} \emph{\footnotesize www.eledia.org/eledia}{\footnotesize -}\emph{\footnotesize uestc}{\footnotesize \par}

\noindent {\footnotesize ~}{\footnotesize \par}

\noindent {\footnotesize $^{(5)}$} \emph{\footnotesize ELEDIA Research
Center} {\footnotesize (}\emph{\footnotesize ELEDIA@TSINGHUA} {\footnotesize -
Tsinghua University)}{\footnotesize \par}

\noindent {\footnotesize 30 Shuangqing Rd, 100084 Haidian, Beijing
- China}{\footnotesize \par}

\noindent {\footnotesize E-mail:} \emph{\footnotesize andrea.massa@tsinghua.edu.cn}{\footnotesize \par}

\noindent {\footnotesize Website:} \emph{\footnotesize www.eledia.org/eledia-tsinghua}{\footnotesize \par}

\noindent \vfill

\noindent \textbf{\emph{This work has been submitted to the IEEE for
possible publication. Copyright may be transferred without notice,
after which this version may no longer be accessible.}}

\noindent \vfill

\newpage
\section*{Multi-Resolution Subspace-Based Optimization Method for the Retrieval
of \emph{2D} Perfect Electric Conductors}

~

\noindent ~

\noindent ~

\begin{flushleft}X. Ye, F. Zardi, M. Salucci, and A. Massa\end{flushleft}

\noindent \vfill

\begin{abstract}
\noindent Perfect Electric Conductors (\emph{PEC}s) are imaged integrating
the subspace-based optimization method (\emph{SOM}) within the iterative
multi-scaling scheme (\emph{IMSA}). Without \emph{a-priori} information
on the number or/and the locations of the scatterers and modelling
their \emph{EM} scattering interactions with a (known) probing source
in terms of surface electric field integral equations, a segment-based
representation of \emph{PEC}s is retrieved from the scattered field
samples. The proposed \emph{IMSA-SOM} inversion method is validated
against both synthetic and experimental data by assessing the reconstruction
accuracy, the robustness to the noise, and the computational efficiency
with some comparisons, as well.
\end{abstract}
\noindent \vfill

\noindent \textbf{Key words}: Inverse Problems; Inverse Scattering;
Microwave Imaging; Iterative Multi-Scaling Approach; Multi-Resolution;
Subspace Optimization Method; Perfect Electric Conductor; Metallic
Scatterers.

\newpage
\section{Introduction \label{sec:intro}}

\noindent Detecting metallic objects and retrieving the scatterer
shape with microwave imaging techniques is a common task in many practical
applications. For instance, the detection of concealed weapons is
of paramount importance \cite{Liu 2019}-\cite{Wang 2021} in critical
environments such as airports and living spaces. In the industrial
context, electromagnetic (\emph{EM}) imaging has been used to localize
metallic contaminants in food products \cite{Ok 2014} as well as
cracks in metallic pipelines \cite{Amineh 2008}-\cite{Chen 2021}.
As for the medical framework, online sensors have been designed to
guide surgeons in the removal of metallic shrapnels \cite{Sakthivel 2014}.
Moreover, subsoil imaging of metallic objects is of great importance
for the detection of landmines \cite{Bourgeois 1998}\cite{Bansal 2015}
and for the inspection of underground infrastructures \cite{Shaw 2005}-\cite{Sheiki 2015}.

\noindent Generally speaking, metallic objects are, in practice, often
modelled as perfect electric conductors (\emph{PEC}s) and their \emph{EM}
interactions with the probing source have been modeled with reliable
techniques \emph{}in the state-of-the-art literature. Depending on
the mathematical modeling, which is a consequence of the representation
chosen for the unknown \emph{PEC} objects, the \emph{PEC} imaging
methods can be roughly divided in volume-based and surface-based approaches.

\noindent In volume-based representations, the investigation domain
is discretized in sub-units, namely \emph{pixels}, of unknown dielectric
properties to be retrieved by often modeling the field scattered by
each \emph{PEC} pixel as a multipole expansion in cylindrical harmonics.
For instance, the \emph{Local Shape Function} (\emph{LSF}) is an example
of a volume-based formulation where a boolean variable is associated
to each pixel for denoting whether it is \emph{PEC} or empty \cite{Chew 1992}.
Accordingly, the \emph{PEC} inverse scattering problem (\emph{ISP})
is rephrased as a binary optimization one, then solved with Genetic
Algorithms (\emph{GA}s) \cite{Zhou 2002}\cite{Takenaka 1997}. Otherwise,
derivative-based optimization algorithms, such as \emph{Conjugate
Gradients} (\emph{CG}s) \cite{Chew 1992}\cite{Weedon 1993}\cite{Otto 1994},
are applied when the binary descriptors are relaxed to continuous
ones.

\noindent In \cite{Poli 2013}, a different volume-based representation
is proposed to exploit a fast \emph{Bayesian Compressive Sensing}
(\emph{BCS}) technique for retrieving the $0$-th order coefficients
of the multipole expansion at each pixel location, which are then
post-processed to classify each pixel as \emph{PEC} or empty. Such
an approach has been further developed in \cite{Stevanovic 2016}
by considering a multipole expansion of higher order. A higher-order
expansion has been also adopted in \cite{Ye 2013} to solve the \emph{ISP}
involving both \emph{PEC} and dielectric scatterers. Another volume-based
representation for mixed targets, which relies on integral scattering
equations instead of a multipole expansion, has been presented in
\cite{Yu 2005}-\cite{Sun 2018}. More specifically, a complex permittivity
value has been assigned to each pixel and metallic objects have been
modeled as dielectrics with a large, but finite, imaginary part.

\noindent Differently from volumetric models, surface-based representations
of the \emph{PEC} scatterers only consider the outer contour since
it is enough to model the scattering behaviour of objects impenetrable
to \emph{EM} waves. Subject to the (also approximate) knowledge of
the position and number of disconnected targets, arbitrary \emph{PEC}
shapes have been described with either the coefficients of Fourier
series \cite{Roger 1981}-\cite{Chien 2006 b} or the control points
of spline curves \cite{Chien 2006 a}-\cite{Chien 2005}. In both
cases, the surface descriptors have been retrieved by applying the
Newton-Kantorovich method \cite{Roger 1981}, the Differential Evolution
optimizer \cite{Qing 2003}\cite{Qing 2004}, and the \emph{GA}s \cite{Chiu 1996}-\cite{Chien 2006 b}\cite{Zhou 2003}\cite{Chien 2005}.

\noindent Without \emph{a-priori} information on the scattering scenario
(i.e., the number and the positions of the objects), the \emph{Level
Set} (\emph{LS}) method has been successfully used in \cite{Litman 1998}
to yield the \emph{PEC} contour. Under the same hypothesis (i.e.,
without \emph{a-priori} knowledge), another surface-based method,
which is based on the discretization of the investigation domain in
a grid of segments, has been introduced in \cite{Ye 2010} and further
refined in \cite{Ye 2011}. Each segment has been then classified
as \emph{PEC} or non-\emph{PEC} with an inversion procedure that relies
on the \emph{Subspace Optimization Method} (\emph{SOM}). Such an approach
has been successfully applied to both \emph{TM} and \emph{TE} illuminations
\cite{Ye 2011}\cite{Shen 2013}, as well.

\noindent Unlike \emph{PEC} inversion methods that require \emph{a-priori}
information, which might be inaccurate or even unavailable, the reconstruction
approaches that use pixel- or segment-based representations of the
investigation domain can effectively model arbitrary connected and
non-connected \emph{PEC} geometries, but they need a large number
of unknowns to achieve a high-resolution imaging. This causes a heavier
ill-posedness of the \emph{ISP} at hand, which implies the use of
effective regularization countermeasures, as well as an increase of
the computational cost that might result unacceptable (especially)
if computationally-intensive optimization methods are adopted for
the scattering-data inversion.

\noindent To address these issues, a novel strategy, which is based
on the Iterative Multi-Scaling Approach (\emph{IMSA}) as applied to
a segment-based representation of the investigation domain of the
\emph{ISP}, is proposed hereinafter for the retrieval of \emph{PEC}
objects. As a matter of fact, the \emph{IMSA} has been extensively
applied to mitigate the non-linearity and to alleviate the non-uniqueness
of the \emph{ISP} when dealing with dielectric scatterers \cite{Caorsi 2003}-\cite{Salucci 2021}.
The underlying idea is that the data inversion is iteratively performed
for the reconstruction of only a portion of the whole investigation
domain, called \emph{Region of Interest} (\emph{RoI}), which is iteratively
reduced to implement a synthetic zoom on the scatterers. In this way,
the number of unknowns is kept close to the limited amount of information
collectable from the measurements \cite{Bucci 1997} for both mitigating
the ill-posedness of the \emph{ISP} \cite{Caorsi 2003} and reducing
the computational burden, but the spatial resolution of the reconstruction
is increased only in the \emph{RoI}s where the scatterers have been
detected.

\noindent Thanks to its flexibility and being a meta-level strategy,
the \emph{IMSA} has been integrated with a variety of solution strategies
(e.g., the \emph{BCS} \cite{Anselmi 2018}, the Particle \emph{}Swarm
\emph{}Optimization \cite{Salucci 2017b}, and the \emph{SOM} \cite{Oliveri 2011}\cite{Ye 2015})
and different formulations (e.g., Lippmann-Schwinger Integral Equations
or Contraction Integral Equations \cite{Zhong 2020}). Moreover, it
has been used to deal with both two- \cite{Salucci 2021} and three-dimensional
\cite{Salucci 2017a} inversion problems.

\noindent In this paper, the \emph{IMSA} has been combined with the
\emph{SOM} to tackle the \emph{PEC} imaging problem since this latter
method independently reduces the ill-posedness of the \emph{ISP} \cite{Chen 2009}
by constraining the unknown currents, induced on the investigation
domain, to a low-dimensional subspace. Moreover, such an integration
has been already successfully exploited for the retrieval of dielectric
scatterers \cite{Oliveri 2011}\cite{Ye 2015} even though with a
pixel-based representation of the investigation domain. Compared to
other state-of-the-art techniques, the \emph{IMSA-SOM} tool for the
\emph{PEC} reconstruction does not require \emph{a-priori} information
on the location or the number of the scatterers and it has limited
computational needs thanks to the profitable integration of the \emph{IMSA}
with an efficient optimization strategy based on \emph{CG}.

\noindent The outline of the paper is as follows. The mathematical
formulation of the \emph{ISP} arising in \emph{PEC} imaging is described
in Sect. \ref{sec:mathematical-formulation}, while the proposed \emph{IMSA-SOM}
inversion method is detailed in Sect. \ref{sec:solution_approach}.
Section \ref{sec:results} is devoted to the numerical and experimental
assessment with some comparisons, as well. Finally, concluding remarks
are drawn (Sect. \ref{sec:conclusions}).

\section{\noindent Mathematical Formulation\label{sec:mathematical-formulation}}

\noindent Let $\mathcal{D}$ be a two-dimensional (\emph{2D}) square
investigation domain of side $\mathcal{L}_{\mathcal{D}}$ laying on
the $xy$-plane within an unbounded free-space background with permittivity
and permeability $\varepsilon_{0}$ and $\mu_{0}$, respectively (Fig.
1), where unknown \emph{PEC} scatterers have to be retrieved. Such
a scenario is probed by a set of $V$ Transverse Magnetic (\emph{TM})
$z$-polarized monochromatic ($f$ being the working frequency) plane
waves impinging from the angular directions \{$\phi_{v}$; $v=1,...,V$\}
and having (known) electric field equal to $\mathbf{E}_{inc}^{\left(v\right)}\left(\mathbf{r}\right)=E_{inc}^{\left(v\right)}\left(\mathbf{r}\right)\hat{\mathbf{z}}$
{[}$\mathbf{r}$ $=$ ($x$, $y$){]} ($v=1,...,V$).%
\footnote{\noindent Throughout this paper, a bold notation is used to indicate
a physical vector ($\mathbf{a}$), a single bar indicates an algebraic
vector or 1D tensor ($\overline{a}$), and double bars indicate a
matrix or 2D tensor ($\overline{\overline{a}}$).%
} For each $v$-th ($v=1,...,V$) incidence, the electric field%
\footnote{\noindent Hereinafter, the time-dependency factor $e^{-j2\pi ft}$
is assumed and omitted. %
} $\mathbf{E}_{tot}^{\left(v\right)}\left(\mathbf{r}\right)=E_{tot}^{\left(v\right)}\left(\mathbf{r}\right)\hat{\mathbf{z}}$
is measured at $M$ positions, \{$\mathbf{r}_{m}^{\left(v\right)}$;
$m=1,...,M$\}, of the observation domain $\mathcal{D}_{obs}$ external
to $\mathcal{D}$.

\noindent In order to numerically manage the problem at hand, the
investigation domain is partitioned into $N$ square pixels identified
by $Q$ line edges (Fig. 1). A binary variable $P_{q}$ is assigned
to each $q$-th ($q=1,...,Q$) segment to denote the membership to
the \emph{PEC} when $P_{q}=1$, while $P_{q}=0$ stands for the background,
so that the dielectric distribution in $\mathcal{D}$ is univocally
identified by the algebraic binary vector $\overline{P}$ ($\overline{P}$
$\triangleq$ \{$P_{q}$; $q=1,...,Q$\}). From an \emph{EM} viewpoint,
each $q$-th ($q=1,...,Q$) edge is modeled in terms of an associated
piece-wise constant equivalent current, $\mathbf{J}^{\left(v\right)}\left(\mathbf{r}_{q}\right)=J_{z}^{\left(v\right)}\left(\mathbf{r}_{q}\right)\hat{\mathbf{z}}$,
centered at the barycenter of the $q$-th segment, $\mathbf{r}_{q}$,
subject to the condition that for each $v$-th ($v=1,...,V$) view\begin{equation}
\left(\overline{\overline{I}}-\overline{\overline{P}}\right)\overline{J}^{\left(v\right)}=\overline{0},\label{eq:Currents.Equation}\end{equation}
which forces the equivalent current to be zero on non\emph{-PEC} segments.
In (\ref{eq:Currents.Equation}), $\overline{\overline{I}}$ is the
identity matrix, $\overline{\overline{P}}$ is the diagonal matrix
whose $q$-th ($q=1,...,Q$) diagonal element is equal to $\left.\overline{\overline{P}}\right\rfloor _{qq}=P_{q}$,
$\overline{J}^{\left(v\right)}$ is the $Q$-size complex-valued algebraic
vector whose $q$-th ($q=1,...,Q$) entry is equal to $\left.\overline{J}^{\left(v\right)}\right\rfloor _{q}=J_{z}^{\left(v\right)}(\mathbf{r}_{q})$,
and $\overline{0}$ is the $Q$-size null vector.

\noindent Moreover, the $v$-th ($v=1,...,V$) scattered field, which
is defined as $\mathbf{E}_{sca}^{\left(v\right)}\left(\mathbf{r}\right)$
$\triangleq$ $\mathbf{E}_{tot}^{\left(v\right)}\left(\mathbf{r}\right)$
$-$ $\mathbf{E}_{inc}^{\left(v\right)}\left(\mathbf{r}\right)$,
is related to the corresponding $v$-th equivalent current, $\mathbf{J}^{\left(v\right)}\left(\mathbf{r}\right)$,
through the following relationship

\noindent \begin{equation}
\overline{E}_{sca}^{\left(v\right)}=\overline{\overline{G}}^{\left(v\right)}\,\overline{J}^{\left(v\right)}\label{eq:Escatt}\end{equation}
where $\overline{\overline{G}}$ is the Green's matrix, mapping the
equivalent current in $\mathbf{r}_{q}$ ($q=1,...,Q$) ($\mathbf{r}_{q}\in\mathcal{D}$)
to the scattered field at $\mathbf{r}_{t}$, whose ($t$, $q$)-th
entry is equal to \cite{Peterson 1998}\begin{equation}
\left.\overline{\overline{G}}\right\rfloor _{tq}=\left\{ \begin{array}{ll}
-\frac{k\eta W}{4}H_{0}^{\left(1\right)}\left(k\left\Vert \mathbf{r}_{t}-\mathbf{r}_{q}\right\Vert \right) & \textrm{if }t\neq q\\
-\frac{k\eta W}{4}\left\{ 1+j\frac{2}{\pi}\left[\ln\left(\frac{\gamma kW}{4}\right)-1\right]\right\}  & \textrm{if }t=q\end{array}\right.,\end{equation}
$k$ ($k\triangleq\frac{2\pi}{\lambda}$) and $\eta$ ($\eta\triangleq\sqrt{\frac{\mu_{0}}{\varepsilon_{0}}}$)
being the wavenumber and the background impedance, respectively, while
$\lambda$ is the wavelength at $f$, $W$ is the length of one edge,
$H_{0}^{(1)}$ is the Hankel function of the first kind and $0$-th
order, $\gamma\approx1.781$ is a constant \cite{Peterson 1998},
and $\left\Vert \cdot\right\Vert $ stands for the $\ell_{2}$-norm.

\noindent Depending on the location of the probing point $\mathbf{r}_{t}$,
(\ref{eq:Escatt}) can be further customized. When $\mathbf{r}_{t}\in\mathcal{D}_{obs}$
(i.e., $\mathbf{r}_{t}=\mathbf{r}_{m}^{\left(v\right)}$), (\ref{eq:Escatt})
is rewritten as\begin{equation}
\overline{E}_{sca}^{\left(v\right)}=\overline{\overline{G}}_{ext}^{\left(v\right)}\,\overline{J}^{\left(v\right)}\label{eq:Data.Equation}\end{equation}
($v=1,...,V$) where $\overline{\overline{G}}_{ext}^{\left(v\right)}$
is now a $M\times Q$ matrix with the ($m$, $q$)-th ($m=1,...,M$;
$q=1,...,Q$) entry given by $\left.\overline{\overline{G}}_{ext}^{\left(v\right)}\right\rfloor _{mq}$
$=$ $\left.\overline{\overline{G}}\right\rfloor _{tq}$ ($t\ne q$,
$m=t$), while $\overline{E}_{sca}^{\left(v\right)}$ is the complex
algebraic vector of size $M$ of the scattered field samples (i.e.,
$\overline{E}_{sca}^{\left(v\right)}$ $=$ \{$E_{sca}^{\left(v\right)}\left(\mathbf{r}_{m}^{\left(v\right)}\right)$;
$m=1,...,M$\}). Otherwise {[}i.e., $\mathbf{r}_{t}\in\mathcal{D}$
$\to$ $\mathbf{r}_{t}=\mathbf{r}_{p}$ ($p=1,...,Q$){]}, it turns
out that\begin{equation}
\overline{E}_{sca}^{\left(v\right)}=\overline{\overline{G}}_{int}\,\overline{J}^{\left(v\right)}\label{eq:Escatt-int}\end{equation}
where $\overline{\overline{G}}_{int}$ is now a $Q\times Q$ matrix
with the ($p$, $q$)-th ($p$, $q=1,...,Q$) entry given by $\left.\overline{\overline{G}}_{int}\right\rfloor _{pq}$
$=$ $\left.\overline{\overline{G}}\right\rfloor _{tq}$ ($t=p$),
while $\overline{E}_{sca}^{\left(v\right)}$ $=$ \{$E_{sca}^{\left(v\right)}\left(\mathbf{r}_{q}\right)$;
$q=1,...,Q$\}.

\noindent Dealing \emph{}with \emph{PEC}s, \emph{}the boundary condition
(i.e., the tangential total field vanishes in correspondence with
the \emph{PEC} segments) can be exploited to derive the following\begin{equation}
\overline{\overline{P}}\left(\overline{E}_{tot}^{\left(v\right)}\right)=\overline{0}\label{eq:Boundary.Equation}\end{equation}
($v=1,...,V$), where $\overline{E}_{tot}^{\left(v\right)}$ $=$
\{$E_{tot}^{\left(v\right)}\left(\mathbf{r}_{q}\right)$; $q=1,...,Q$\}.
Since $\overline{E}_{tot}^{\left(v\right)}$ $=$ $\overline{E}_{inc}^{\left(v\right)}$
$+$ $\overline{E}_{sca}^{\left(v\right)}$, through (\ref{eq:Escatt-int}),
(\ref{eq:Boundary.Equation}) assumes the following form\begin{equation}
\overline{\overline{P}}\left(\overline{E}_{inc}^{\left(v\right)}+\overline{\overline{G}}_{int}\,\overline{J}^{\left(v\right)}\right)=\overline{0}\label{eq:Rossiglione}\end{equation}
($v=1,...,V$).

\noindent According to such a formulation and the segment-based representation
of the \emph{PEC} scatterers, the inverse problem at hand is then
rephrased as that of determining $\overline{P}$ and $\{\overline{J}^{\left(v\right)};\, v=1,...,V\}$
that fulfil (\ref{eq:Currents.Equation}), (\ref{eq:Data.Equation}),
and (\ref{eq:Rossiglione}).

\section{\noindent Inversion Method\label{sec:solution_approach}}

\noindent The solution of the inverse problem formulated in Sect.
\ref{sec:mathematical-formulation} is addressed with an approach
based on the integration of the \emph{SOM} within the \emph{IMSA}
scheme. More specifically, the scattering-data inversion is carried
out by means of an iterative strategy that performs $S$ successive
{}``zooming'' steps. At each $s$-th ($s=1,...,S$; $s$ being the
step index) step, the \emph{PEC} profile of the \emph{RoI}, which
is the portion of the investigation domain $\mathcal{D}$ where the
scatterer has been estimated to lie, is retrieved by means of a deterministic
algorithm based on the \emph{SOM} as applied to (\ref{eq:Currents.Equation}),
(\ref{eq:Data.Equation}), and (\ref{eq:Rossiglione}). Such a reconstruction
is then exploited to improve the \emph{RoI} estimate by also enhancing
the spatial resolution of the retrieval. The process is repeated until
a data-matching convergence criterion holds true.

\noindent The implementation of such a multi-level process needs:
(\emph{a}) to re-define the problem unknowns, \{$\overline{P}$ ;
($\overline{J}^{\left(v\right)}$; $v=1,...,V$)\}, so that the \emph{SOM},
which is a continuous-variable optimization method, can be fruitfully
applied (Sect. {}``\emph{Unknowns Coding}''); (\emph{b}) to customize
the \emph{SOM} \cite{Chen 2018} to the problem at hand (Sect. {}``\emph{SOM
Implementation}''); (\emph{c}) to define a suitable cost function
that faithfully links the \emph{ISP} at hand to its mathematical representation
so that the solution is the global minimum of the cost function itself
(Sect. {}``\emph{Cost Function Definition}''); (\emph{d}) to customize
the meta-level \emph{IMSA} strategy to both such a formulation (i.e.,
problem unknowns and cost function) and the integration with the optimization
level (i.e., the \emph{SOM}) (Sect. {}``\emph{IMSA Implementation}'').
These items will be detailed or briefly recalled in the following.

\subsubsection*{\noindent Unknowns Coding\label{sub:Unknowns-Coding}}

Since the unknown vector $\overline{P}$ is binary, it is rewritten
as a function of the set of continuous variables, \{$x_{q}$; $q=1,...,Q$\},
as follows\begin{equation}
P_{q}=\frac{1}{1+\exp\left(-bx_{q}\right)}\label{eq:indicator_vector}\end{equation}
($q=1,...,Q$), the coefficient $b$ being defined as in \cite{Ye 2011}.
Consequently, since $x_{q}>0$ ($x_{q}<0$) implies that $P_{q}=1$
($P_{q}=0$), that is, the $q$-th ($q=1,...,Q$) segment is \emph{PEC}
(empty), the same information coded in the $Q$-size binary vector
$\overline{P}$ can be now drawn from the same size continuous vector
$\overline{x}$ ($\overline{x}$ $=$ \{$x_{q}$; $q=1,...,Q$\});

\subsubsection*{\emph{SOM} Implementation\label{sub:SOM-Implementation}}

According to the \emph{SOM} guidelines \cite{Chen 2009}, the $v$-th
($v=1,...,V$) equivalent current $\overline{J}^{\left(v\right)}$
is decomposed in two terms \begin{equation}
\overline{J}^{\left(v\right)}=\overline{J}_{D}^{\left(v\right)}+\overline{J}_{A}^{\left(v\right)},\label{eq:current_decomposition}\end{equation}
$\overline{J}_{D}^{\left(v\right)}$ and $\overline{J}_{A}^{\left(v\right)}$
being the deterministic part and the ambiguous one, respectively.
The former, $\overline{J}_{D}^{\left(v\right)}$ ($v=1,...,V$), is
computed from (\ref{eq:Data.Equation}) by applying the singular value
decomposition (\emph{SVD}) to the external Green matrix\begin{equation}
\overline{\overline{G}}_{ext}^{\left(v\right)}=\overline{\overline{\mathcal{U}}}^{\left(v\right)}\,\overline{\overline{\sigma}}^{\left(v\right)}\left(\overline{\overline{\mathcal{V}}}^{\left(v\right)}\right)^{H}\label{eq:svd}\end{equation}
where $^{H}$ stands for conjugate transposition, $\overline{\overline{\mathcal{U}}}^{\left(v\right)}$
$=$ \{$\overline{\mathcal{U}}_{q}^{\left(v\right)}$; $q=1,...,Q$\}
is the $M\times Q$ matrix whose $q$-th ($q=1,...,Q$) column is
the $M$-size left-singular vector (i.e., $\overline{\mathcal{U}}_{q}^{\left(v\right)}$),
$\overline{\overline{\sigma}}^{\left(v\right)}$ is the $Q\times Q$
diagonal matrix of the $Q$ singular values of $\overline{\overline{G}}_{ext}^{\left(v\right)}$
(i.e., $\overline{\sigma}^{\left(v\right)}$ $=$ \{$\sigma_{q}^{\left(v\right)}$;
$q=1,...,Q$\}), and $\overline{\overline{\mathcal{V}}}^{\left(v\right)}$
$=$ \{$\overline{\mathcal{V}}_{q}^{\left(v\right)}$; $q=1,...,Q$\}
is the $Q\times Q$ matrix whose $q$-th ($q=1,...,Q$) column is
the $Q$-size right-singular vector (i.e., $\overline{\mathcal{V}}_{q}^{\left(v\right)}$).
More specifically, $\overline{J}_{D}^{\left(v\right)}$ ($v=1,...,V$)
is given by\begin{equation}
\overline{J}_{D}^{\left(v\right)}=\sum_{q=1}^{Q_{th}}\frac{\left(\overline{\mathcal{U}}_{q}^{\left(v\right)}\right)^{H}\cdot\overline{E}_{sca}^{\left(v\right)}}{\sigma_{q}^{\left(v\right)}}\overline{\mathcal{V}}_{q}^{\left(v\right)}\label{eq:deterministic_currents}\end{equation}

\noindent where $\cdot$ denotes the scalar product, while $Q_{th}$
is the \emph{SVD} truncation threshold, which is adaptively set as
follows \cite{Oliveri 2011}\begin{equation}
Q_{th}=\arg\min_{Q'}\left\{ \left|\frac{\sum_{q=1}^{Q'}\sigma_{q}^{\left(v\right)}}{\sum_{q=1}^{Q}\sigma_{q}^{\left(v\right)}}-\alpha\right|\right\} ,\label{eq:adaptive_svd_threshold}\end{equation}
$\alpha$ ($0<\alpha\le1$) being a real user-defined calibration
parameter (see Sect. \ref{sub:Control-Parameters-Calibration}).

\noindent The ambiguous part of the current, $\overline{J}_{A}^{\left(v\right)}$
($v=1,...,V$), is yielded as the linear combination of the remaining
($Q-Q_{th}$) right-singular vectors \begin{equation}
\overline{J}_{A}^{\left(v\right)}=\sum_{q=Q_{th}+1}^{Q}w_{q-Q_{th}}^{\left(v\right)}\overline{\mathcal{V}}_{q}^{\left(v\right)}\label{eq:ambiguous_current}\end{equation}
where $\overline{w}^{\left(v\right)}$ $=$ \{$w_{q}^{\left(v\right)}$;
$q=1,...,\left(Q-Q_{th}\right)$\} is the unknown complex algebraic
vector of the weights of the $v$-th ($v=1,...,V$) ambiguous current,
while $\overline{\overline{w}}$ is the corresponding $\left(Q-Q_{th}\right)\times V$
size matrix ($\overline{\overline{w}}$ $\triangleq$ \{$\overline{w}^{\left(v\right)}$;
$v=1,...,V$\}).

\subsubsection*{Cost Function Definition\label{sub:Cost-Function-Definition}}

\noindent The cost function $\mathcal{F}$ is defined to quantify
the mismatch between actual and estimated values in both the scattered
field and the induced equivalent currents\begin{equation}
\mathcal{F}\left(\overline{x},\,\overline{\overline{w}}\right)=\sum_{v=1}^{V}\left[\mathcal{F}_{Field}^{\left(v\right)}\left(\overline{x},\,\overline{w}^{\left(v\right)}\right)+\mathcal{F}_{Curr}^{\left(v\right)}\left(\overline{x},\,\overline{w}^{\left(v\right)}\right)\right]\label{eq:cost_function}\end{equation}
where the term $\mathcal{F}_{Field}^{\left(v\right)}$ ($v=1,...,V$)
gives the vectorial distance between the measured scattered field,
$\overline{E}_{sca}^{\left(v\right)}$, and the scattered field yielded
from the estimated $v$-th equivalent current $\overline{J}^{\left(v\right)}$\begin{equation}
\mathcal{F}_{Field}^{\left(v\right)}\left(\overline{x},\,\overline{w}^{\left(v\right)}\right)=\frac{\left\Vert \overline{E}_{sca}^{\left(v\right)}-\overline{\overline{G}}_{ext}^{\left(v\right)}\,\overline{J}^{\left(v\right)}\right\Vert ^{2}}{\left\Vert \overline{E}_{sca}^{\left(v\right)}\right\Vert ^{2}},\label{eq:data_term}\end{equation}
while $\mathcal{F}_{Curr}^{\left(v\right)}$ is given by

\noindent \begin{equation}
\mathcal{F}_{Curr}^{\left(v\right)}\left(\overline{x},\,\overline{w}^{\left(v\right)}\right)=\frac{\left\Vert \left(\overline{\overline{I}}-\overline{\overline{P}}\right)\overline{J}^{\left(v\right)}\right\Vert ^{2}}{\left\Vert \overline{J}_{D}^{\left(v\right)}\right\Vert ^{2}}+\frac{\left\Vert \overline{\overline{P}}\,\left(\overline{E}_{inc}^{\left(v\right)}+\overline{\overline{G}}_{int}\,\overline{J}^{\left(v\right)}\right)\right\Vert ^{2}}{\left\Vert \overline{\overline{G}}_{int}\,\overline{J}_{D}^{\left(v\right)}\right\Vert ^{2}}\label{eq:state_term}\end{equation}
being the normalized error in fulfilling the conditions (\ref{eq:Currents.Equation})
and (\ref{eq:Rossiglione}) on the $v$-th ($v=1,...,V$) induced
equivalent current $\overline{J}^{\left(v\right)}$, which is equal
to $\overline{J}^{\left(v\right)}$ $=$ $\sum_{q=1}^{Q_{th}}\frac{\left(\overline{\mathcal{U}}_{q}^{\left(v\right)}\right)^{H}\cdot\overline{E}_{sca}^{\left(v\right)}}{\sigma_{q}^{\left(v\right)}}\overline{\mathcal{V}}_{q}^{\left(v\right)}$
$+$ $\sum_{q=Q_{th}+1}^{Q}w_{q-Q_{th}}^{\left(v\right)}\overline{\mathcal{V}}_{q}^{\left(v\right)}$
according to (\ref{eq:deterministic_currents}) and (\ref{eq:ambiguous_current}).

\subsubsection*{\emph{IMSA} Implementation\label{sub:IMSA-Implementation}}

\noindent The algorithmic customization of the \emph{IMSA} meta-level
to integrate the \emph{SOM}-based optimization for the retrieval of
\emph{PEC}s can be described through the following multi-step iterative
($i$ being the iteration index) process:

\begin{itemize}
\item \noindent \emph{Initialization} - Set the step index to $s=1$ and
the \emph{RoI} to the whole investigation domain $\mathcal{D}$ ($\mathcal{D}^{\left(1\right)}=\mathcal{D}$);
\item \noindent \emph{IMSA Loop}

\begin{itemize}
\item \noindent \emph{Unknowns Setup} ($i=0$) - If $s=1$, then set $\overline{\overline{w}}_{i}^{\left(s\right)}=\overline{x}_{i}^{\left(s\right)}=\overline{0}$.
Otherwise (i.e., $s>1$), map the trial solution from the previous
zooming step into the current $s$-th discretization grid of the \emph{RoI}
$\mathcal{D}^{\left(s\right)}$ (i.e., $\overline{\overline{w}}_{i}^{\left(s\right)}=\Phi\left\{ \overline{\overline{w}}_{i}^{\left(s-1\right)};\,\mathcal{D}^{\left(s\right)}\right\} $
and $\overline{x}_{i}^{\left(s\right)}=\Phi\left\{ \overline{x}^{\left(s-1\right)};\,\mathcal{D}^{\left(s\right)}\right\} $,
$\Phi$ being the mapping operator from the grid of $\mathcal{D}^{\left(s-1\right)}$
to the finer one of $\mathcal{D}^{\left(s\right)}$;
\item \noindent \emph{Scattering-Data Inversion} - Compute the $s$-th step
trial solution ($\overline{x}^{\left(s\right)}$, $\overline{\overline{w}}^{\left(s\right)}$)
within the \emph{RoI} $\mathcal{D}^{\left(s\right)}$ by solving the
following optimization problem\begin{equation}
\left(\overline{x}^{\left(s\right)},\,\overline{\overline{w}}^{\left(s\right)}\right)=\arg\min_{\overline{x},\overline{\overline{w}}}\left\{ \mathcal{F}\left(\overline{x},\,\overline{\overline{w}}\right)\right\} \label{eq:min_problem}\end{equation}
 with $I$ iterations of the deterministic two-step \emph{CG} algorithm
in \cite{Ye 2011} (i.e., $\overline{x}^{\left(s\right)}=\left.\overline{x}_{i}^{\left(s\right)}\right\rfloor _{i=I}$,
$\overline{\overline{w}}^{\left(s\right)}=\left.\overline{\overline{w}}_{i}^{\left(s\right)}\right\rfloor _{i=I}$)
starting from $\left.\overline{x}_{i}^{\left(s\right)}\right\rfloor _{i=0}$
and $\left.\overline{\overline{w}}_{i}^{\left(s\right)}\right\rfloor _{i=0}$; 
\item \noindent \emph{Step Check} - Stop the \emph{IMSA} loop if the maximum
number of zooming steps is reached (i.e., $s=S$) and output the estimated
solution by setting $\overline{x}_{opt}=\overline{x}^{\left(S\right)}$
and $\overline{\overline{w}}_{opt}=\overline{\overline{w}}^{\left(S\right)}$;
\item \noindent \emph{RoI} \emph{Update} - Compute the $s$-th estimate
of the \emph{PEC} indicator vector $\overline{P}^{\left(s\right)}$
through (\ref{eq:indicator_vector}) with $x_{q}\leftarrow x_{q}^{\left(s\right)}$
and apply the {}``filtering and clustering'' operations \cite{Caorsi 2003}
to determine the new \emph{RoI}, $\mathcal{D}^{\left(s+1\right)}$,
by defining its center, $\mathbf{r}_{\mathcal{D}}^{\left(s+1\right)}$
{[}$\mathbf{r}_{\mathcal{D}}^{\left(s+1\right)}$ $=$ ($x_{\mathcal{D}}^{\left(s+1\right)}$,
$y_{\mathcal{D}}^{\left(s+1\right)}$){]}, and side, $\mathcal{L}_{\mathcal{D}}^{\left(s+1\right)}$,
as follows\begin{equation}
\xi_{\mathcal{D}}^{\left(s+1\right)}=\frac{\sum_{q=1}^{Q}\xi_{q}^{\left(s\right)}P_{q}^{\left(s\right)}}{\sum_{q=1}^{Q}P_{q}^{\left(s\right)}}\label{eq:doi_barycenter}\end{equation}
($\xi\in\left\{ x;\, y\right\} $) and\begin{equation}
\mathcal{L}_{\mathcal{D}}^{\left(s+1\right)}=2\times\frac{\sum_{q=1}^{Q}\left\Vert \mathbf{r}_{q}^{\left(s\right)}-\mathbf{r}_{\mathcal{D}}^{\left(s+1\right)}\right\Vert P_{q}^{\left(s\right)}}{\sum_{q=1}^{Q}P_{q}^{\left(s\right)}};\label{eq:doi_edge}\end{equation}

\item \noindent \emph{RoI Check} - Terminate the \emph{IMSA} loop if zooming
factor $\eta^{\left(s\right)}$, which is defined as\begin{equation}
\eta^{\left(s\right)}=\frac{\left|\mathcal{L}_{\mathcal{D}}^{\left(s+1\right)}-\mathcal{L}_{\mathcal{D}}^{\left(s\right)}\right|}{\mathcal{L}_{\mathcal{D}}^{\left(s+1\right)}},\label{eq:zooming}\end{equation}
is below a user-defined threshold $\eta_{min}$ (i.e., $\eta^{\left(s\right)}\le\eta_{min}$)
and set the problem solution to the current trial one (i.e., $\overline{x}_{opt}=\overline{x}^{\left(s\right)}$and
$\overline{\overline{w}}_{opt}=\overline{\overline{w}}^{\left(s\right)}$).
Otherwise, update the \emph{IMSA} loop index {[}i.e., $s\leftarrow\left(s+1\right)${]}
and restart the {}``\emph{IMSA Loop}''.
\end{itemize}
\end{itemize}

\section{\noindent Numerical and Experimental Validation\label{sec:results}}

\noindent This Section is devoted to illustrate the results of the
validation of the proposed inversion method and to give some indications
on its performance in different scenarios and under various conditions.
Towards this end, representative test cases, concerned with both synthetic
and experimental scattering data, will be discussed.

\noindent To quantitatively assess the effectiveness of the data inversion/reconstructions,
suitable error functions, customized to the segment-based representation
of \emph{PEC} scatterers, are defined and used. Namely, they are the
\emph{total reconstruction error}

\noindent \begin{equation}
\Xi_{tot}\triangleq\frac{1}{Q}\sum_{q=1}^{Q}\left[\left(1-P_{q}^{true}\right)P_{q}^{opt}+P_{q}^{true}\left(1-P_{q}^{opt}\right)\right]\label{eq:total_error}\end{equation}
($\Xi_{tot}=0$ if $\overline{P}^{opt}=\overline{P}^{true}$ and $\Xi_{tot}=1$
if $\overline{P}^{opt}=\overline{1}-\overline{P}^{true}$), the \emph{internal
reconstruction error}\begin{equation}
\Xi_{int}\triangleq\frac{1}{Q_{int}}\sum_{q=1}^{Q_{int}}P_{q}^{true}\left(1-P_{q}^{opt}\right)\label{eq:internal_error}\end{equation}
($\Xi_{int}=0$ if $P_{q}^{opt}=1$ and $\Xi_{int}=1$ if $P_{q}^{opt}=0$,
$0\le q\le Q_{int}$), and the \emph{external reconstruction error}\begin{equation}
\Xi_{ext}\triangleq\frac{1}{Q_{ext}}\sum_{q=1}^{Q_{ext}}\left(1-P_{q}^{true}\right)P_{q}^{opt}\label{eq:external_error}\end{equation}
($\Xi_{int}=0$ if $P_{q}^{opt}=0$ and $\Xi_{int}=1$ if $P_{q}^{opt}=1$,
$0\le q\le Q_{ext}$), $\overline{P}^{true}$ and $\overline{P}^{opt}$
being the true/actual and the reconstructed \emph{PEC} indicator vectors,
respectively, while $Q_{int}=Q-Q_{ext}$, $Q_{ext}$ being the segments
of the investigation domain external to the support of the \emph{PEC}
scatterer.

\noindent Unless stated otherwise, the following reference scenario
has been considered throughout the numerical assessment. A measurement
setup at a frequency of $f=300$ {[}MHz{]} where a plane wave probes
a square investigation domain, $\mathcal{D}$, of side $\mathcal{L}_{\mathcal{D}}=3\lambda$
by impinging from $V=27$ different angular directions \{$\phi_{v}=2\pi\frac{(v-1)}{V}$;
$v=1,...,V$\}. The electric field coming from the interaction between
the probing source and the scatterers laying in $\mathcal{D}$ has
been measured by $M=27$ ideal probes uniformly-spaced on a circular
observation domain, $\mathcal{D}_{obs}$, external to the investigation
domain, with radius $\rho_{obs}=2.2\,\lambda$. The $M\times V$ synthetic
scattering data have been numerically generated with a Method of Moments
(\emph{MoM}) solver by densely discretizing $\mathcal{D}$ with square
cells $\lambda/50$-sided. To emulate real-data, the scattered field
samples have been then blurred with an additive white Gaussian noise
characterized by a signal-to-noise ratio (\emph{SNR}). According to
the guidelines in \cite{Bucci 1997}, the inverse problem at hand
has been solved by uniformly partitioning $\mathcal{D}$ in $N=18\times18$
sub-domains. Moreover, the maximum number of zooming steps of the
\emph{IMSA-SOM} has been set to $S=6$, while the value of the zooming
threshold has been fixed to $\eta_{min}=0.2$ according to \cite{Salucci 2021}.

\subsection{Illustrative Example\label{sub:Illustrative-Example}}

In this section, a detailed step-by-step description of the \emph{IMSA-SOM}
method as applied to an illustrative example is provided. Towards
this end, a square \emph{PEC} object of side $0.6\lambda$ has been
chosen {[}see the red contour in Fig. 2(\emph{a}){]} and reconstructed
by processing noisy scattered data with $SNR=40$ {[}dB{]}. 

\noindent According to the multi-step iterative procedure described
in Sect. \ref{sec:solution_approach} ({}``\emph{IMSA Implementation}''),
the \emph{RoI} has been first initialized to the whole investigation
domain ($\mathcal{D}^{\left(1\right)}=\mathcal{D}$). At the first
\emph{IMSA} step ($s=1$), a \emph{SOM}-based reconstruction of $\mathcal{D}^{\left(1\right)}${[}green
pattern - Fig. 2(\emph{a}){]} has been performed. Figure 2(\emph{a})
shows the reconstructed \emph{PEC} profile with the cyan segments
of the \emph{PEC} indicator vector $\left.\overline{P}^{\left(s\right)}\right\rfloor _{s=1}$.
Starting from such an estimate of the target position and shape, the
\emph{RoI} is updated by defining the green patterned region $\mathcal{D}^{\left(2\right)}$
in Fig. 2(\emph{b}). The second ($s=2$) \emph{IMSA} step has been
then carried out by applying the \emph{SOM} inversion to image $\mathcal{D}^{\left(2\right)}$.
As expected, owing to the improved \emph{RoI} estimate, the reconstructed
\emph{PEC} map is significantly closer to the actual one {[}Fig. 2(\emph{b}){]}.
In turn, such an improvement enables a further shrinking of the \emph{RoI},
$\mathcal{D}^{\left(3\right)}$, towards the actual support of the
\emph{PEC} {[}Fig. 2(\emph{c}){]}. At the successive \emph{IMSA} step
($s=3$), the \emph{SOM} inversion is repeated by yielding the \emph{PEC}
indicator vector $\left.\overline{P}^{\left(s\right)}\right\rfloor _{s=3}$
as mapped in Fig. 2(\emph{c}). The \emph{IMSA} loop has been then
stopped since the zooming factor ($\left.\eta^{\left(s\right)}\right\rfloor _{s=4}=5.32\times10^{-2}$)
was below the threshold ($\eta_{min}=0.2$) as pointed out in Fig.
2(\emph{c}) where the \emph{RoI} matches very closely the shape of
the actual target, and no further reconstruction enhancements were
expected further zooming. 

\noindent For the sake of completeness, the behaviors of both the
cost function, $\mathcal{F}$, and the reconstruction error, $\Xi_{tot}$,
throughout the iterative ($i=1,...,I$) multi-step ($s=1,...,S$)
\emph{IMSA-SOM} process are shown in Fig. 3.

\subsection{\noindent Control-Parameters Calibration\label{sub:Control-Parameters-Calibration}}

\noindent The \emph{IMSA-SOM} inversion method depends on its control
parameters (Sect. \ref{sec:solution_approach}), namely the $\alpha$
threshold, which regulates the adaptive \emph{SVD} truncation process,
and the maximum number of iterations of the deterministic two-step
\emph{CG} algorithm \cite{Ye 2011}, $I$, performed at each \emph{}$s$-th
($s=1,...,S$) \emph{IMSA} step. 

\noindent To give the interested readers some insights on the sensitivity
of \emph{IMSA-SOM} to these control parameters by also motivating
the choice of the control setup used throughout the whole validation,
the results of a study on a circular \emph{PEC} object $\lambda/2$
in radius are reported in the following. More in detail, the inversion
process has been repeated for each choice of the values of $\alpha$
and $I$ by processing different noisy data so that the inferred indications
are independent on the noise level.

\noindent The outcomes of such a calibration phase are summarized
in Fig. 4 in terms of the total reconstruction error. In particular,
Figure 4(\emph{a}) shows the behavior of $\Xi_{tot}$ versus $\alpha$
when fixing $I=1000$, while the dependence on $I$ ($\alpha=0.6$)
is analyzed in Fig. 4(\emph{b}). It turns out {[}Fig. 4(\emph{a}){]}
that the $\alpha$ value impacts significantly on the inversion accuracy,
the fluctuations of $\Xi_{tot}$ being quite large since, for instance,
$\Xi_{tot}=0.1$($\Xi_{tot}=0.01$) corresponds to $10$($1)$ \%
of wrongly reconstructed segments on the total number. To better illustrate
such an outcome, let us observe the reconstructions when $\alpha=0.0$
{[}Fig. 5(\emph{b}){]} and $\alpha=1.0$ {[}Fig. 5(\emph{c}){]}. In
the first case, the \emph{SVD} truncation is too selective and only
a limited portion of the \emph{PEC} is correctly retrieved {[}Fig.
5(\emph{b}){]}, while there is no truncation when $\alpha=1.0$ and
the inversion generates many artifacts outside the \emph{PEC} domain
{[}Fig. 5(\emph{c}){]}.

\noindent As for the sensitivity of the \emph{IMSA-SOM} performance
on $I$, Figure 4(\emph{b}) indicates that, as expected, the higher
the $I$ value, the smaller the reconstruction error is, but only
until a threshold around $\approx1000$ iterations. Indeed, if the
deterministic minimization of $\mathcal{F}$ is stopped too early
(e.g., $I=30$), the optimum of (\ref{eq:cost_function}) has not
yet been reached and \emph{}the reconstruction is sub-optimal {[}e.g.,
Fig. 5(\emph{d}){]}. On the other hand, increasing $I$ beyond a threshold
value yields negligible improvements as confirmed by the comparison
between the inversion results when setting $I=1000$ {[}Fig. 5(\emph{a}){]}
or $I=1500$ {[}Fig. 5(\emph{e}){]}.

\noindent The optimal setup for $\alpha$ and $I$ has been then chosen
according to the following rule\begin{equation}
\varsigma^{(opt)}=\frac{{\displaystyle \int_{SNR}\arg\min_{\varsigma}\left\{ \left.\Xi_{tot}\right\rfloor _{SNR}^{\varsigma}\right\} dSNR}}{{\displaystyle \int_{SNR}dSNR}}\end{equation}
($\varsigma=\left\{ \alpha;I\right\} $) and the result has been $\alpha_{opt}=0.6$
and $I_{opt}=1000$.

\noindent A proof of the effectiveness of such a choice is shown in
Fig. 5(\emph{a}) where the \emph{IMSA-SOM} reconstruction when $SNR=10$
{[}dB{]} is reported, the total error being $\left.\Xi_{tot}\right\rfloor _{SNR=10[dB]}=1.77\times10^{-2}$.

\subsection{\noindent Numerical Assessment\label{sub:Numerical-Assessment}}

\noindent Once calibrated, the performance of the \emph{IMSA-SOM}
have been assessed in comparison with a competitive state-of-the-art
inversion approach. For a fair comparison, the single-resolution \emph{SOM}
in \cite{Ye 2011} has been considered as reference.%
\footnote{\noindent According to the guidelines in \cite{Ye 2011}, a $\lambda/10$-sided
uniform grid has been chosen to discretize the investigation domain
$\mathcal{D}$ for the single-resolution inversion so that ($N^{SOM}=30\times30$).%
}

\noindent The first test case is concerned with a {}``T''-shaped
\emph{PEC} object is considered whose larger edge is $0.6\,\lambda$.
Such a shape presents corners and cavities making the reconstruction
process even harder. Figure 6 shows the behavior of the error indexes
versus the $SNR$ for both the \emph{IMSA-SOM} and the \emph{SOM}.
As it can be noticed, the reconstruction accuracy of the \emph{IMSA}-based
method is significantly better than that of the single-resolution
\emph{SOM}, the improvement of the total reconstruction error ranging
from $\Delta\Xi_{tot}\approx35$ \% ($SNR=40$ {[}dB{]}) up to $\Delta\Xi_{tot}\approx55$
\% ($SNR=5$ {[}dB{]}) being $\Delta\Xi_{tot}\triangleq\frac{\Xi_{tot}^{SOM}-\Xi_{tot}^{IMSA-SOM}}{\Xi_{tot}^{SOM}}$.
Moreover, it turns out that the multi-zooming approach reduces mainly
the external error, while the internal one is almost equivalent for
both methods. Such an outcome is pictorially pointed out by the map
of the retrieved \emph{PEC} profiles in Fig. 7 where the representative
examples of inversion when processing data at $SNR=20$ {[}dB{]} {[}Figs.
7(\emph{a})-7(\emph{b}){]}, $SNR=10$ {[}dB{]} {[}Figs. 7(\emph{c})-7(\emph{d}){]},
and $SNR=5$ {[}dB{]} {[}Figs. 7(\emph{e})-7(\emph{f}){]} are reported.
More in detail, the pictures in Figs. 7(\emph{a})-7(\emph{b}) show
that both techniques are able to correctly localize the \emph{PEC}
object within the investigation domain, but the \emph{IMSA-SOM} better
shapes it. The differences between the two inversion approaches become
more and more evident increasing the noise level to $SNR=10$ {[}dB{]}
{[}Fig. 7(\emph{c}) vs. Fig. 7(\emph{d}){]} and up to $SNR=5$ {[}dB{]}
{[}Fig. 7(\emph{e}) vs. 7(\emph{f}){]} when the \emph{SOM} image reveals
some disconnected artifacts in the surrounding of the bottom-left
corner of the {}``T'' boundary, as well, so that $\Delta\Xi_{tot}\approx55$
\%.

\noindent The second experiment is aimed at assessing the reliability
of the \emph{IMSA-SOM} in reconstructing \emph{PEC} shapes that are
not exactly mapped into the gridding of $\mathcal{D}$. Towards this
purpose, a {}``Diamond'' object with diagonal equal to $d=0.5\,\lambda$
and edges tilted with respect to the discretization grid has been
considered as benchmark. The results in Fig. 8 indicate that the \emph{IMSA-SOM}
reduces the value of the total error of the \emph{SOM} up to $\Delta\Xi_{tot}\approx61$
\% ($SNR=5$ {[}dB{]}), the minimum improvement being equal to $\Delta\Xi_{tot}\approx47$
\% ($SNR=40$ {[}dB{]}). Moreover, the (expected) worsening of the
reconstruction accuracy for higher and higher level of noise is more
contained (i.e., $\Delta\delta_{tot}^{IMSA-SOM}\approx62$ \% vs.
$\Delta\delta_{tot}^{IMSA-SOM}\approx73$ \% being $\Delta\delta_{tot}\triangleq\frac{\left.\Xi_{tot}\right\rfloor _{SNR=5\,[dB]}-\left.\Xi_{tot}\right\rfloor _{SNR=40\,[dB]}}{\left.\Xi_{tot}\right\rfloor _{SNR=5\,[dB]}}$).

\noindent Such an enhanced robustness to the noise blurring the data
is also highlighted in Figs. 9(\emph{a})-9(\emph{f}) since the \emph{SOM}
profiles are increasingly over-estimated as the $SNR$ reduces {[}Fig.
9(\emph{b}), Fig. 9(\emph{d}), and Fig. 9(\emph{f}){]}, while there
are little differences in both size and shape when applying the \emph{IMSA-SOM}.
This latter mainly derives from the ability of the \emph{IMSA} to
model more accurately the tilted edges of the diamond contour thanks
to the use of a denser discretization in the \emph{RoI}.

\noindent Still referring to the {}``Diamond'' \emph{PEC}, the dependence
of the \emph{IMSA-SOM} inversion on the size of the scatterer has
been evaluated next by varying the diagonal, $d$, between $d=0.1\,\lambda$
and $d=1.5\,\lambda$. Figure 10 gives the values of the total reconstruction
error, $\Xi_{tot}$, versus $d$ for different $SNR$s. It turns out
that, whatever the combination of the target size and the noise level,
the \emph{IMSA-SOM} performs better than the \emph{SOM}. However,
it is worth noticing that, while the error values tend to be quite
close for larger dimensions of the \emph{PEC} ($\frac{d}{\lambda}\to1.5$)
and high $SNR$s, the advantage of using the \emph{IMSA} strategy
becomes greater as the size is lowered and the noise level is getting
heavier.

\noindent To further confirm these conclusions, Figures 11(\emph{a})-11(\emph{f})
show the \emph{PEC} profiles retrieved when processing data with $SNR=5$
{[}dB{]}. When $d=1.1\,\lambda$ {[}Figs. 11(\emph{a})-11(\emph{b}){]},
the \emph{SOM} image presents both spurious artifacts outside the
actual contour and a wrong empty internal region, which are properly
avoided by the \emph{IMSA-SOM}. Moving to the case $d=0.7\,\lambda$,
the \emph{IMSA-SOM} confirms to be more accurate in shaping the actual
scatterer as well as in estimating its support {[}Fig. 11(\emph{c})
vs. Fig. 11(\emph{d}){]}. This is even more evident when $d=0.3\,\lambda$
{[}Fig. 11(\emph{e}) vs. Fig. 11(\emph{f}){]}.

\noindent The successive experiment has been devoted to infer the
highest spatial resolution, $R$ (i.e., the minimum distance at which
two disconnected objects can be distinguished), achievable by the
\emph{IMSA-SOM}. More specifically, two \emph{PEC} circles of radius
$0.1\,\lambda$ have been considered and the minimum distance between
their boundaries, $D$, has been varied within the range $0.3\,\lambda$
$\le D\le$ $0.7\,\lambda$. 

\noindent Figure 12 shows the total reconstruction error, $\Xi_{tot}$,
as a function of the object distance, $D$, when $SNR=20$ {[}dB{]}.
Unlike the \emph{IMSA-SOM}, where the $\Xi_{tot}$ is almost flat
in all the range of variation of $D$ and equal to $\Xi_{tot}^{IMSA-SOM}\approx0.01$,
the plot of the \emph{SOM} error shows a hill-like behavior with $\Delta\Xi_{tot}\ge55$
\% when $\frac{D}{\lambda}\le0.57$. Figures 13(\emph{a})-13(\emph{f})
illustrate these deductions by showing the profiles reconstructed
in correspondence with three representative values of $D$, namely
$D=0.5\,\lambda$ {[}Figs. 13(\emph{a})-13(\emph{b}){]}, $D=0.35\,\lambda$
{[}Figs. 13(\emph{c})-13(\emph{d}){]}, and $D=0.3\,\lambda$ {[}Figs.
13(\emph{e})-13(\emph{f}){]}. When the inter-objects distance is $D=0.5\,\lambda$,
both single and multi-resolution \emph{SOM} techniques distinguish
two separated scatterers {[}Figs. 13(\emph{a})-13(\emph{b}){]}, even
though the {}``bare'' \emph{SOM} overestimates the size of the circles
and it also retrieves spurious artifacts in the proximity of the actual
\emph{PEC}s {[}Fig. 13(\emph{b}){]}. Decreasing the distance below
$D\approx0.5\,\lambda$ {[}e.g., $D=0.35\,\lambda$- Fig. 13(\emph{d}){]},
the single-resolution is no more able to recognize two objects as
correctly done by the \emph{IMSA}-based inversion {[}Fig. 13(\emph{c})
vs. Fig. 13(\emph{d}){]}. If $D$ is further shortened {[}e.g., $D=0.30\,\lambda$
- Figs. 13(\emph{e})-13(\emph{f}){]}, neither of the two methods can
resolve the two disconnected supports. However, while the \emph{IMSA-SOM}
map in Fig. 13(\emph{e}) shows some {}``ghost'' shadows in between
the \emph{PEC} circles, the \emph{SOM} also detects wrong artifacts
far from the actual scatterers {[}Fig. 13(\emph{f}){]}.

\noindent Similar results have been obtained in other test cases during
an exhaustive numerical assessment, thus we are quite confident to
state that the spatial resolution of the \emph{IMSA-SOM} is $R^{IMSA-SOM}=0.35\,\lambda$,
that is a $30$ \% improvement over the single-resolution \emph{SOM}
being $R^{SOM}=0.5\,\lambda$.

\subsection{\noindent Experimental Assessment\label{sub:Experimental-Assessment}}

\noindent After the numerical validation with synthetic scattering
data, this section is devoted to the assessment of the \emph{IMSA-SOM}
inversion strategy against experimental data measured in a real environment.
Towards this end, the {}``rectTM\_cent'' and {}``rectTM\_dece''
datasets, provided by the Institut Fresnel, have been considered \cite{Belkebir 2001}.
These two datasets have been generated by measuring the \emph{EM}
interactions between a metallic cylinder of section $1.27\textrm{ [cm]}\times2.45\textrm{ [cm]}$
and the \emph{EM} field radiated by a horn antenna located $72$ {[}cm{]}
away and working at $f=8$ {[}GHz{]}. More in detail, the scatterer
under test has been illuminated from $V=36$ different angular directions,
while the scattered electric field has been collected by $M=49$ measurement
probes uniformly located on a circular observation domain $\mathcal{D}_{obs}$
of radius $\rho_{obs}=76$ {[}cm{]}. The two datasets differ for the
position of the object with respect to the center of the measurements
systems (i.e., $(x_{0},\, y_{0})=(-0.5,-0.75)$ {[}cm{]} - Dataset
{}``rectTM\_cent''; $(x_{0},y_{0})=(0,4)$ {[}cm{]} - Dataset {}``rectTM\_dece'').

\noindent The inversions of the {}``rectTM\_cent'' dataset are shown
in Figs. 14(\emph{a})-14(\emph{b}). One can observe that both methods
correctly localize the unknown object, but the \emph{SOM} gets worse
since (once again) it overestimates the size of the cylinder as confirmed
by the values of the total reconstruction error, $\Xi_{tot}$, in
Tab. I (i.e., $\Delta\Xi_{tot}\approx55$ \%).

\noindent Concerning the computational issues, it turns out that the
\emph{IMSA-SOM} allows a computational saving%
\footnote{\noindent On a standard laptop computer with i5-8265U processor and
8 {[}GB{]} \emph{RAM}.%
} with respect to the bare \emph{SOM} of about $\Delta t\approx77$
\% ($\Delta t\triangleq\frac{\mathcal{T}^{SOM}-\mathcal{T}^{IMSA-SOM}}{\mathcal{T}^{SOM}}$).
Indeed, even though the \emph{IMSA-SOM} repeats up to $S=6$ times
the inversion process, while the \emph{SOM} does it only once, the
dimension of the inversion problem at hand is much smaller (i.e.\emph{,}
$N^{IMSA-SOM}\ll N^{SOM)}$) despite the higher spatial resolution
yielded at the convergence.

\noindent Similar outcomes can be drawn for the {}``rectTM\_dece''
dataset, the \emph{IMSA-SOM} improvements in both reconstruction accuracy
and computational costs being $\Delta\Xi_{tot}\approx50$ \% and $\Delta t\approx72$
\% (Tab. I). For completeness, the \emph{PEC} profiles retrieved by
the multi-resolution/steps and the single-step \emph{SOM} implementations
are shown in Fig. 15.

\section{\noindent Conclusions\label{sec:conclusions}}

\noindent A novel inversion strategy, named \emph{IMSA-SOM}, has been
developed to address the \emph{ISP} for \emph{2D} \emph{PEC}s. The
proposed strategy combines the \emph{IMSA} and the \emph{SOM} and
it has proved to be reliable and highly effective in a wide range
of scenarios and under different conditions. Indeed, the developed
inversion approach has been tested against both numerical and experimental
scattering data by considering complex shapes and closely-located
scatterers, as well. 

\noindent Compared to the state-of-the-art literature on the subject,
to the best of the authors' knowledge, the main outcome of this paper
and of the related research work is that the developed innovative
method for the \emph{PEC} reconstruction is a reliable, flexible,
and computationally efficient inversion tool, robust to the noise
on the scattering data, as well, that effectively mitigates the non-linearity
and the ill-posedness of the imaging problem at hand.

\noindent Future works, beyond the scope of the current manuscript,
will be aimed at extending the formulation to three-dimensional (\emph{3D})
geometries as well as at customizing the proposed implementation to
buried objects scenarios of great applicative interest.

\section*{Acknowledgements}

\noindent This work has been partially supported by the Italian Ministry
of Education, University, and Research within the PRIN 2017 Program,
for the Project {}``Cloaking Metasurfaces for a New Generation of
Intelligent Antenna Systems (MANTLES)'' (Grant No. 2017BHFZKH - CUP:
E64I19000560001) and the Project \char`\"{}CYBER-PHYSICAL ELECTROMAGNETIC
VISION: Context-Aware Electromagnetic Sensing and Smart Reaction (EMvisioning)\char`\"{}
(Grant no. 2017HZJXSZ - CUP: E64I19002530001), within the Program
\char`\"{}Progetti di Ricerca Industriale e Sviluppo Sperimentale
nelle 12 aree di specializzazione individuate dal PNR 2015-2020\char`\"{},
Specialization Area \char`\"{}Smart Secure \& Inclusive Communities\char`\"{}
for the Project \char`\"{}Mitigazione dei rischi naturali per la sicurezza
e la mobilita' nelle aree montane del Mezzogiorno (MITIGO)\char`\"{}
(Grant no. ARS01\_00964), and within the Program \char`\"{}Smart cities
and communities and Social Innovation\char`\"{} for the Project \char`\"{}Piattaforma
Intelligente per il Turismo (SMARTOUR)\char`\"{} (Grant no. SCN\_00166
- CUP: E44G14000040008). Moreover, it benefited from the networking
activities carried out within the Project {}``SPEED'' (Grant No.
61721001) funded by National Science Foundation of China under the
Chang-Jiang Visiting Professorship Program. A. Massa wishes to thank
E. Vico for her never-ending inspiration, support, guidance, and help.

\newpage
\section*{FIGURE CAPTIONS}

\begin{itemize}
\item \textbf{Figure 1.} \emph{Problem Scenario} - 2D \emph{TM} imaging
Setup.
\item \textbf{Figure 2.} \emph{Illustrative} \emph{Example} ({}``Square''
\emph{PEC} object, $M=V=27$) - \emph{RoI} and \emph{PEC} map retrieved
at different zooming steps of the \emph{IMSA-SOM} algorithm: (\emph{a})
$s=1$, (\emph{b}) $s=2$, and (\emph{c}) $s=3$.
\item \textbf{Figure 3.} \emph{Illustrative} \emph{Example} ({}``Square''
\emph{PEC} object, $M=V=27$; \emph{IMSA-SOM}) - Plots of the cost
function, $\mathcal{F}$, and the total reconstruction error, $\Xi_{tot}$,
versus the iteration index ($i=1,...,I$; $s=1,...,S$).
\item \textbf{Figure 4.} \emph{Control-Parameters Calibration} ({}``Circle''
\emph{PEC} object, $M=V=27$) - Plots of the total reconstruction
error, $\Xi_{tot}$, as a function of (\emph{a}) $\alpha$ ($I=1000$)
and (\emph{b}) $I$ ($\alpha=0.6$).
\item \textbf{Figure 5.} \emph{Control-Parameters Calibration} ({}``Circle''
\emph{PEC} object, $M=V=27$, $SNR=10$ {[}dB{]}; \emph{IMSA-SOM})
- Maps of the \emph{PEC} profile reconstructed when setting (\emph{a})
$\alpha=0.6$ and $I=1000$, (\emph{b}) $\alpha=0.0$ and $I=1000$,
(\emph{c}) $\alpha=1.0$ and $I=1000$, (\emph{d}) $\alpha=0.6$ and
$I=30$, and (\emph{e}) $\alpha=0.6$ and $I=1500$.
\item \textbf{Figure 6.} \emph{Numerical Assessment} ({}``T'' \emph{PEC}
object, $M=V=27$) - Plots of the reconstruction error indexes versus
the \emph{SNR} value.
\item \textbf{Figure 7.} \emph{Numerical Assessment} ({}``T'' \emph{PEC}
object, $M=V=27$) - Maps of the \emph{PEC} profile retrieved by (\emph{a})(\emph{c})(\emph{e})
the \emph{IMSA-SOM} and (\emph{b})(\emph{d})(\emph{f}) the \emph{SOM}
when processing noisy scattering data with (\emph{a})(\emph{b}) $SNR=20$
{[}dB{]}, (\emph{c})(\emph{d}) $SNR=10$ {[}dB{]}, and (\emph{e})(\emph{f})
$SNR=5$ {[}dB{]}.
\item \textbf{Figure 8.} \emph{Numerical Assessment} ({}``Diamond'' \emph{PEC}
object, $d=0.5\lambda$, $M=V=27$) - Plots of the reconstruction
error indexes versus the \emph{SNR} value.
\item \textbf{Figure 9.} \emph{Numerical Assessment} ({}``Diamond'' \emph{PEC}
object, $d=0.5\lambda$, $M=V=27$) - Maps of the \emph{PEC} profile
retrieved by (\emph{a})(\emph{c})(\emph{e}) the \emph{IMSA-SOM} and
(\emph{b})(\emph{d})(\emph{f}) the \emph{SOM} when processing noisy
scattering data with (\emph{a})(\emph{b}) $SNR=20$ {[}dB{]}, (\emph{c})(\emph{d})
$SNR=10$ {[}dB{]}, and (\emph{e})(\emph{f}) $SNR=5$ {[}dB{]}.
\item \textbf{Figure 10.} \emph{Numerical Assessment} ({}``Diamond'' \emph{PEC}
object, $M=V=27$) - Plots of the reconstruction error indexes as
a function of the diagonal of the scatterer, $d$.
\item \textbf{Figure 11.} \emph{Numerical Assessment} ({}``Diamond'' \emph{PEC}
object, $M=V=27$, $SNR=5$ {[}dB{]}) - Maps of the \emph{PEC} profile
retrieved by (\emph{a})(\emph{c})(\emph{e}) the \emph{IMSA-SOM} and
(\emph{b})(\emph{d})(\emph{f}) the \emph{SOM} when processing the
data scattered by the actual object with diagonal (\emph{a})(\emph{b})
$d=1.1\,\lambda$, (\emph{c})(\emph{d}) $d=0.7\,\lambda$, and (\emph{e})(\emph{f})
$d=0.3\,\lambda$.
\item \textbf{Figure 12.} \emph{Numerical Assessment} ({}``Two Circles''
\emph{PEC} objects, $M=V=27$, $SNR=20$ {[}dB{]}) - Plots of the
total reconstruction error, $\Xi_{tot}$, as a function of the inter-scatterers
distance, $D$.
\item \textbf{Figure 13.} \emph{Numerical Assessment} ({}``Two Circles''
\emph{PEC} objects, $M=V=27$, $SNR=20$ {[}dB{]}) - Maps of the \emph{PEC}
profile retrieved by (\emph{a})(\emph{c})(\emph{e}) the \emph{IMSA-SOM}
and (\emph{b})(\emph{d})(\emph{f}) the \emph{SOM} when processing
the data scattered by the actual \emph{PEC} objects spaced by (\emph{a})(\emph{b})
$D=0.50\,\lambda$, (\emph{c})(\emph{d}) $D=0.35\,\lambda$, and (\emph{e})(\emph{f})
$D=0.30\,\lambda$.
\item \textbf{Figure 14.} \emph{Experimental Assessment} (Dataset {}``rectTM\_cent'',
$V=36$, $M=49$) - Maps of the \emph{PEC} profile retrieved by (\emph{a})
the \emph{IMSA-SOM} and (\emph{b}) the \emph{SOM}.
\item \textbf{Figure 15.} \emph{Experimental Assessment} (Dataset {}``rectTM\_dece'',
$V=36$, $M=49$) - \emph{}Maps of the \emph{PEC} profile retrieved
by (\emph{a}) the \emph{IMSA-SOM} and (\emph{b}) the \emph{SOM}.
\end{itemize}

\section*{TABLE CAPTIONS}

\begin{itemize}
\item \textbf{Table I.} \emph{Experimental Assessment} ($V=36$, $M=49$)
- \emph{}Reconstruction error indexes and CPU time.
\end{itemize}
\newpage
\begin{center}~\vfill\end{center}

\begin{center}\begin{tabular}{c}
\includegraphics[%
  width=0.90\textwidth]{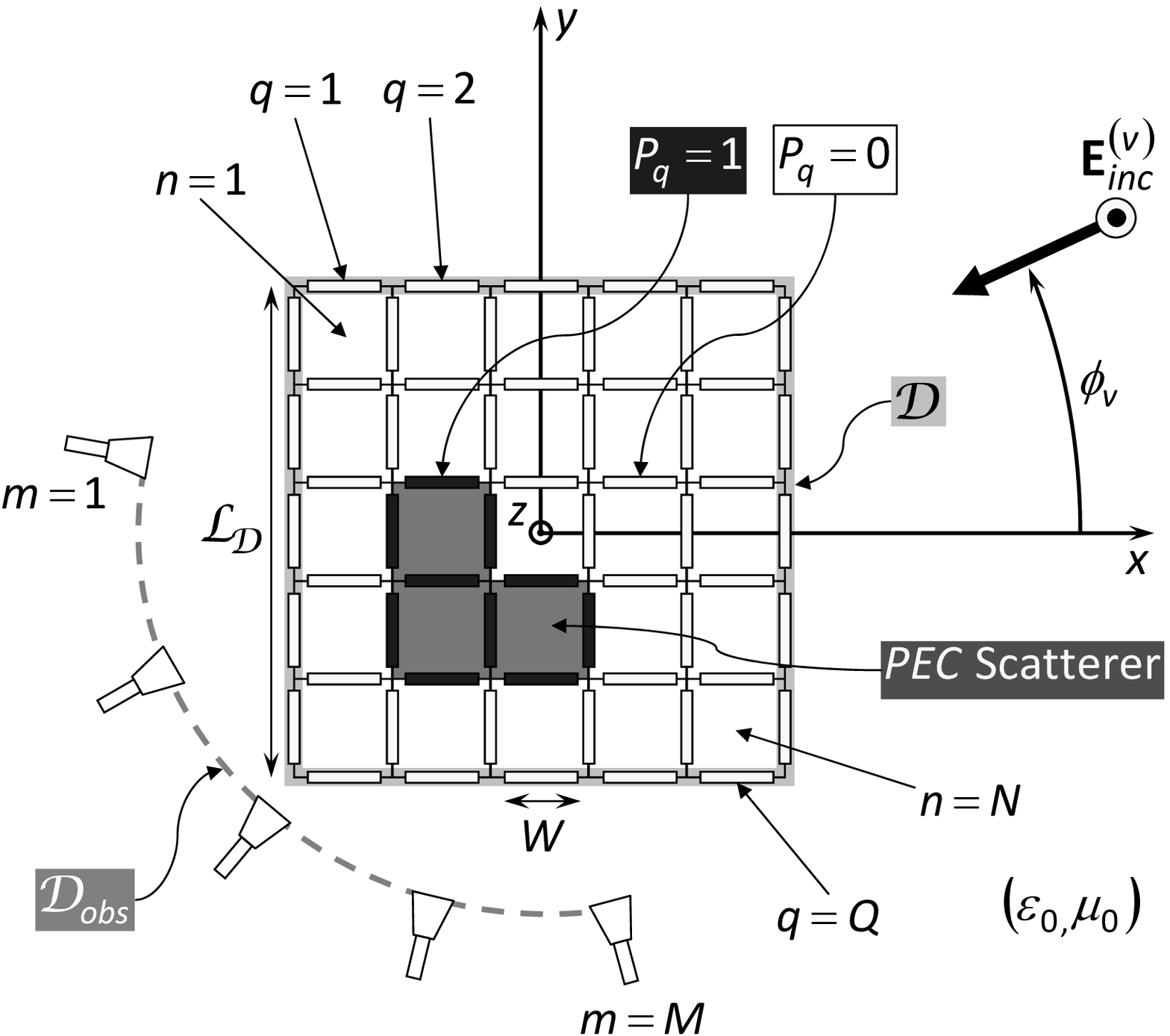}\tabularnewline
\end{tabular}\end{center}

\begin{center}~\vfill\end{center}

\begin{center}\textbf{Fig. 1 - Ye et} \textbf{\emph{al.}}\textbf{,}
\textbf{\emph{{}``}}Multi-Resolution Subspace-Based ...''\end{center}

\newpage
\begin{center}~\vfill\end{center}

\begin{center}\begin{tabular}{cc}
\includegraphics[%
  width=0.48\textwidth]{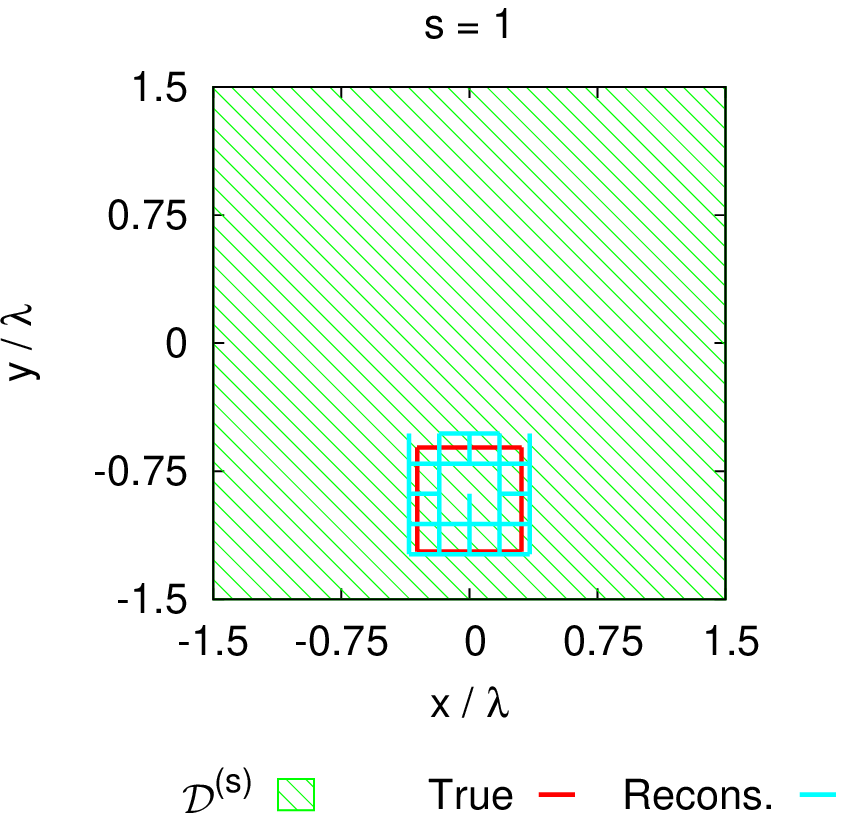}&
\includegraphics[%
  width=0.48\textwidth]{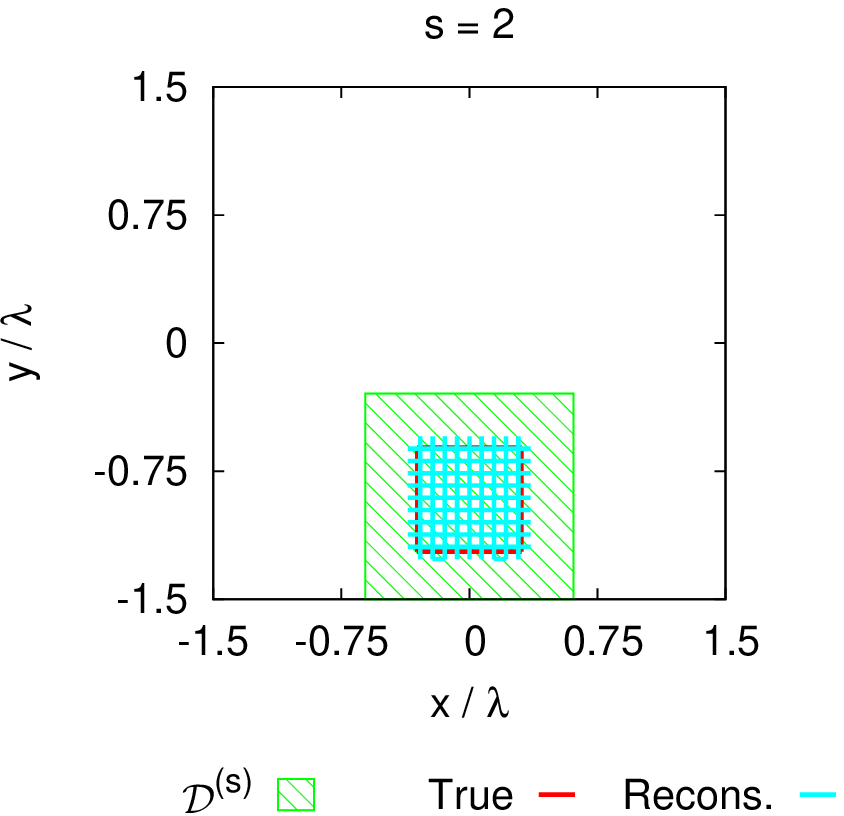}\tabularnewline
(\emph{a})&
(\emph{b})\tabularnewline
\multicolumn{2}{c}{\includegraphics[%
  width=0.48\textwidth]{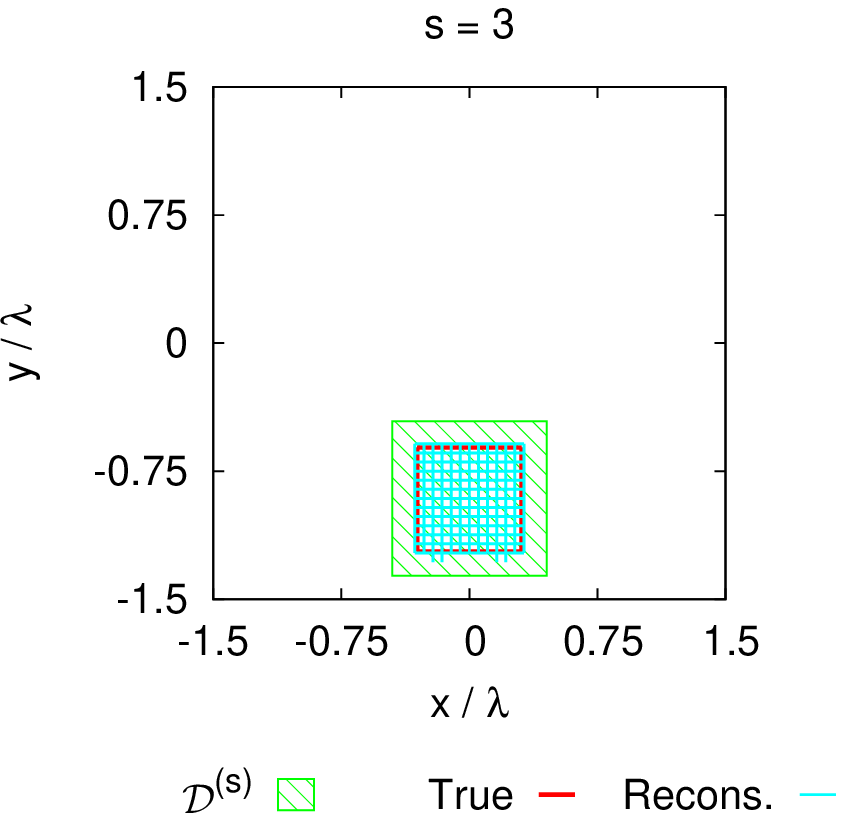}}\tabularnewline
\multicolumn{2}{c}{(\emph{c})}\tabularnewline
\end{tabular}\end{center}

\begin{center}~\vfill\end{center}

\begin{center}\textbf{Fig. 2 - Ye et} \textbf{\emph{al.}}\textbf{,}
\textbf{\emph{{}``}}Multi-Resolution Subspace-Based ...''\end{center}

\newpage
\begin{center}~\vfill\end{center}

\begin{center}\begin{tabular}{c}
\includegraphics[%
  width=0.80\textwidth]{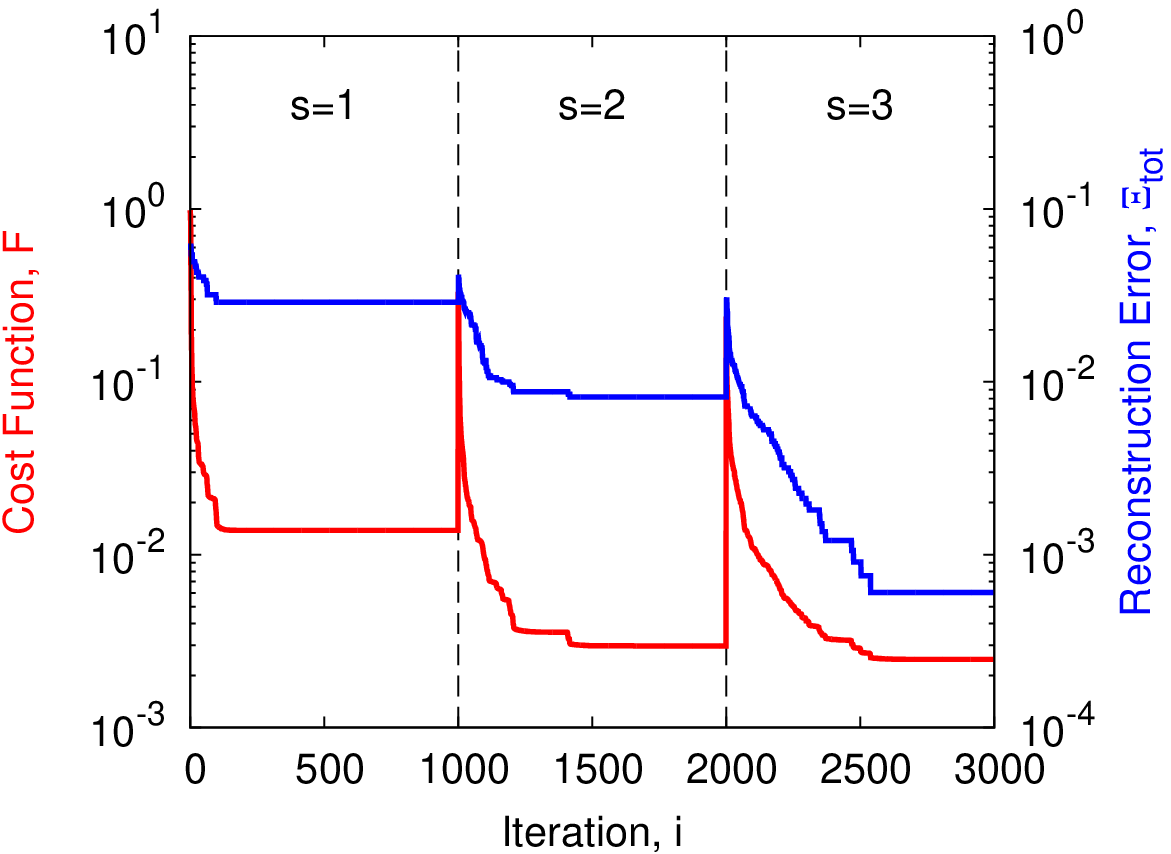}\tabularnewline
\end{tabular}\end{center}

\begin{center}~\vfill\end{center}

\begin{center}\textbf{Fig. 3 - Ye et} \textbf{\emph{al.}}\textbf{,}
\textbf{\emph{{}``}}Multi-Resolution Subspace-Based ...''\end{center}

\newpage
\begin{center}~\vfill\end{center}

\begin{center}\begin{tabular}{c}
\includegraphics[%
  width=0.80\textwidth,
  keepaspectratio]{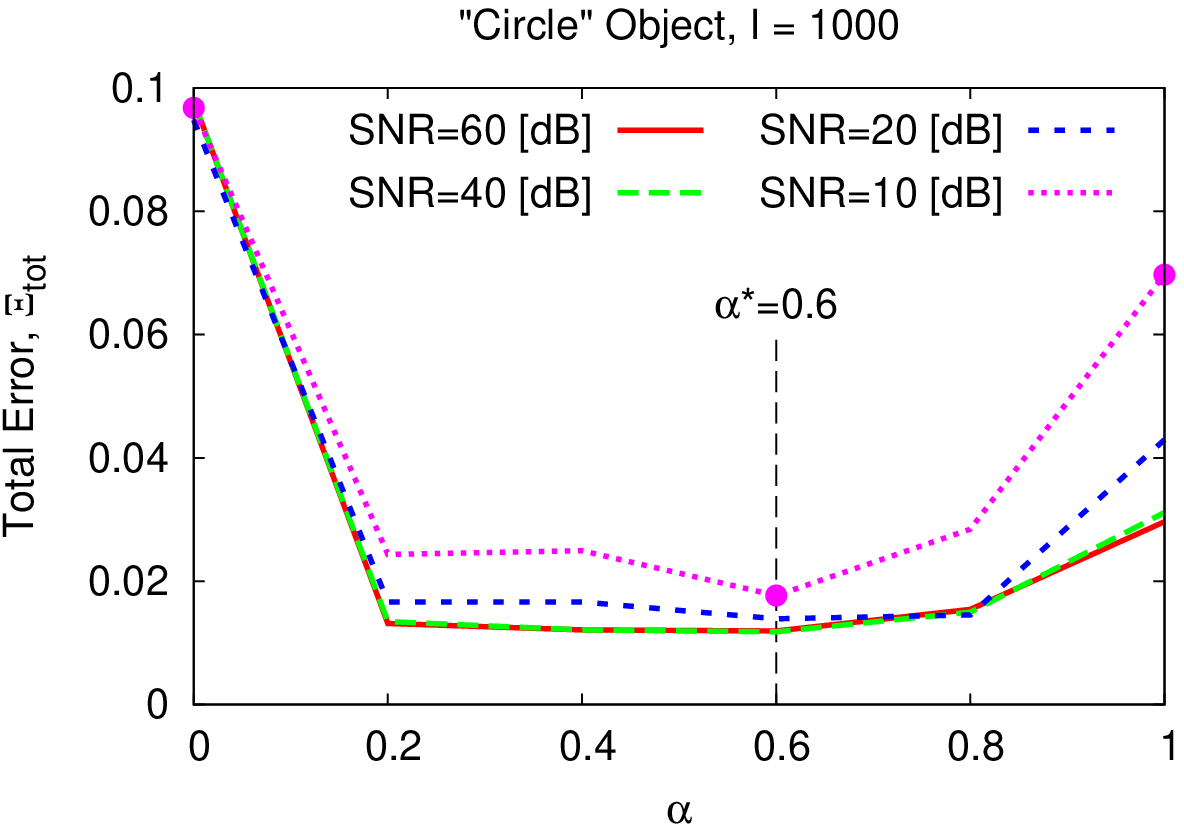}\tabularnewline
(\emph{a})\tabularnewline
\includegraphics[%
  width=0.80\textwidth,
  keepaspectratio]{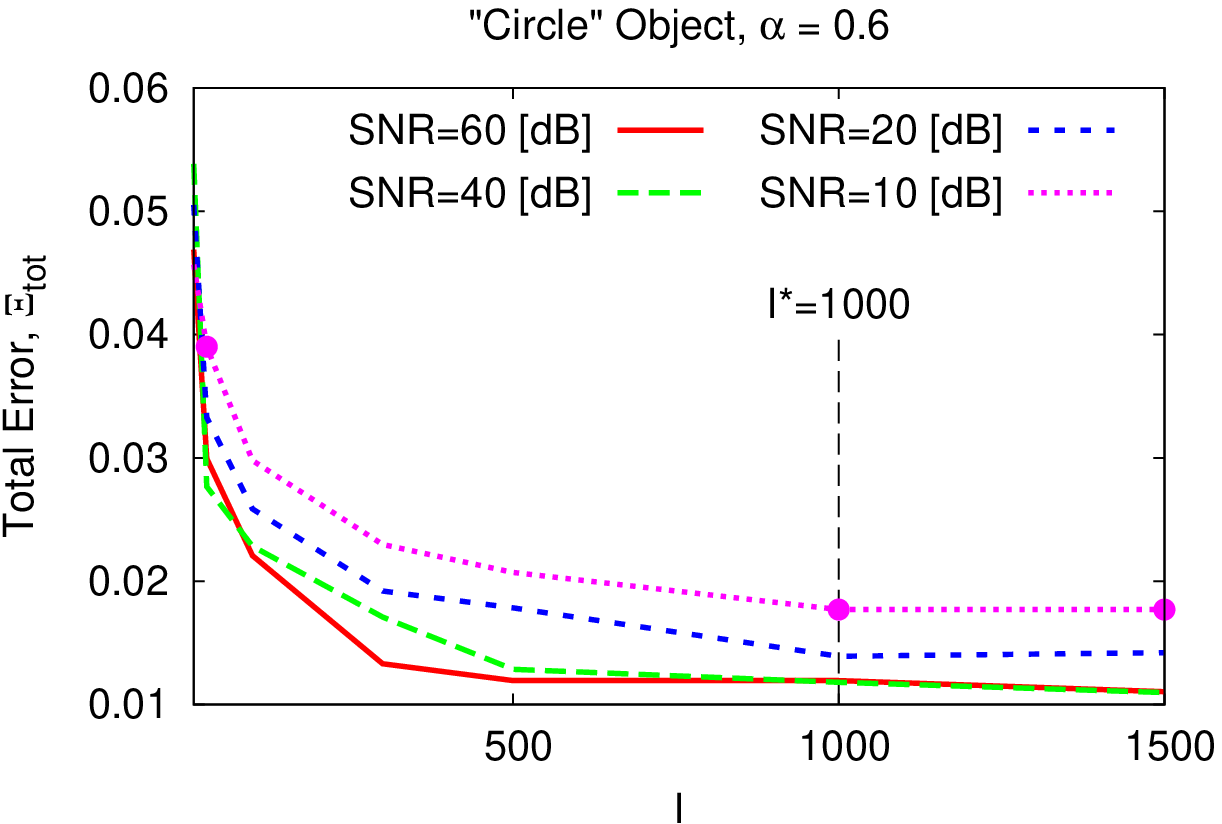}\tabularnewline
(\emph{b})\tabularnewline
\end{tabular}\end{center}

\begin{center}~\vfill\end{center}

\begin{center}\textbf{Fig. 4 - Ye et} \textbf{\emph{al.}}\textbf{,}
\textbf{\emph{{}``}}Multi-Resolution Subspace-Based ...''\end{center}

\newpage
\begin{center}~\vfill\end{center}

\begin{center}\begin{tabular}{cc}
\multicolumn{2}{c}{\includegraphics[%
  width=0.35\textwidth,
  keepaspectratio]{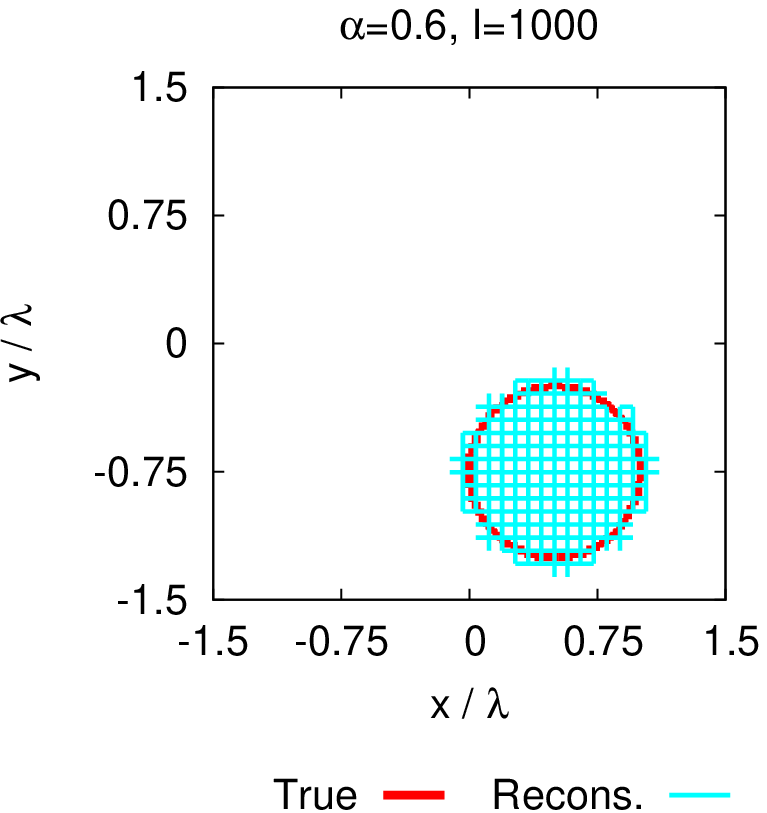}}\tabularnewline
\multicolumn{2}{c}{(\emph{a})}\tabularnewline
\includegraphics[%
  width=0.35\textwidth,
  keepaspectratio]{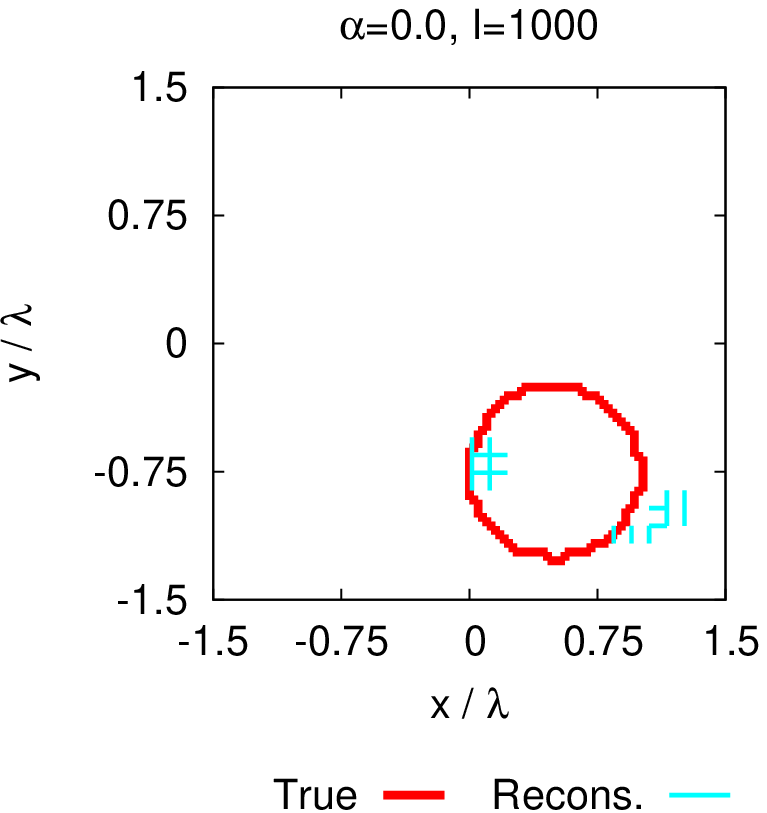}&
\includegraphics[%
  width=0.35\textwidth,
  keepaspectratio]{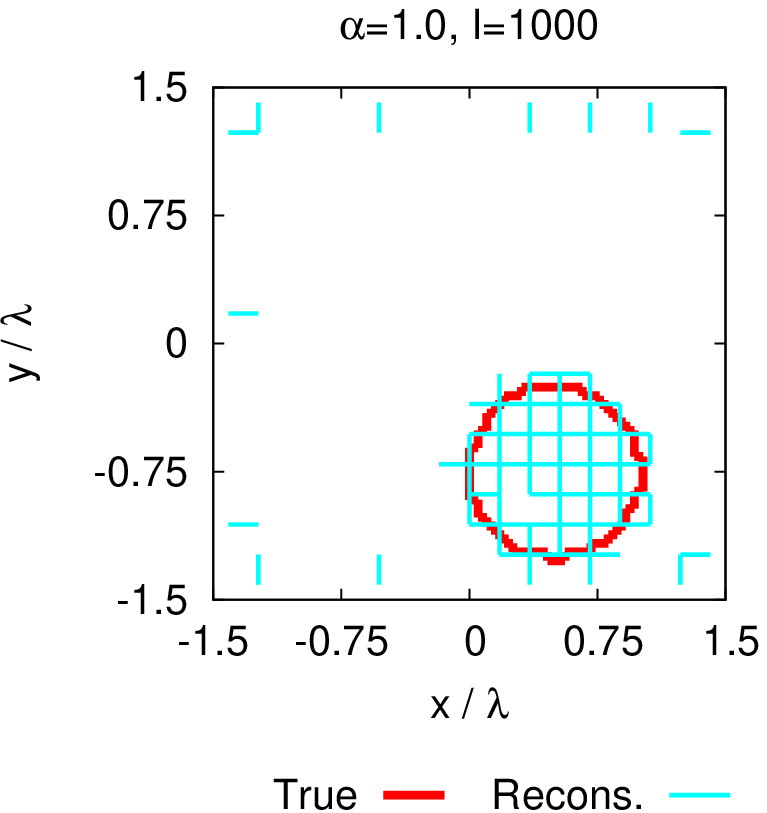}\tabularnewline
(\emph{b})&
(\emph{c})\tabularnewline
\includegraphics[%
  width=0.35\textwidth,
  keepaspectratio]{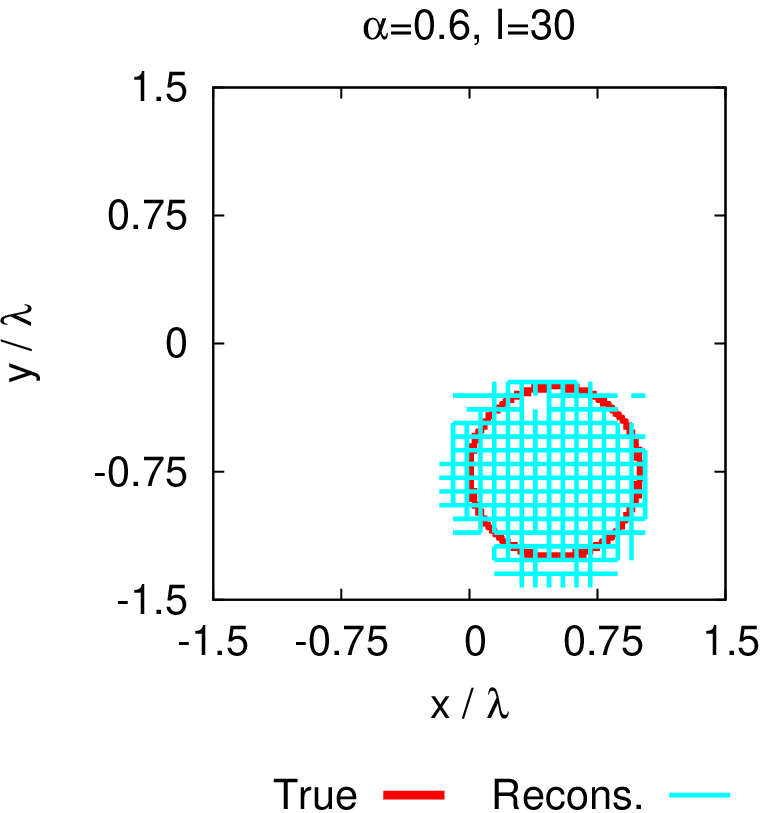}&
\includegraphics[%
  width=0.35\textwidth,
  keepaspectratio]{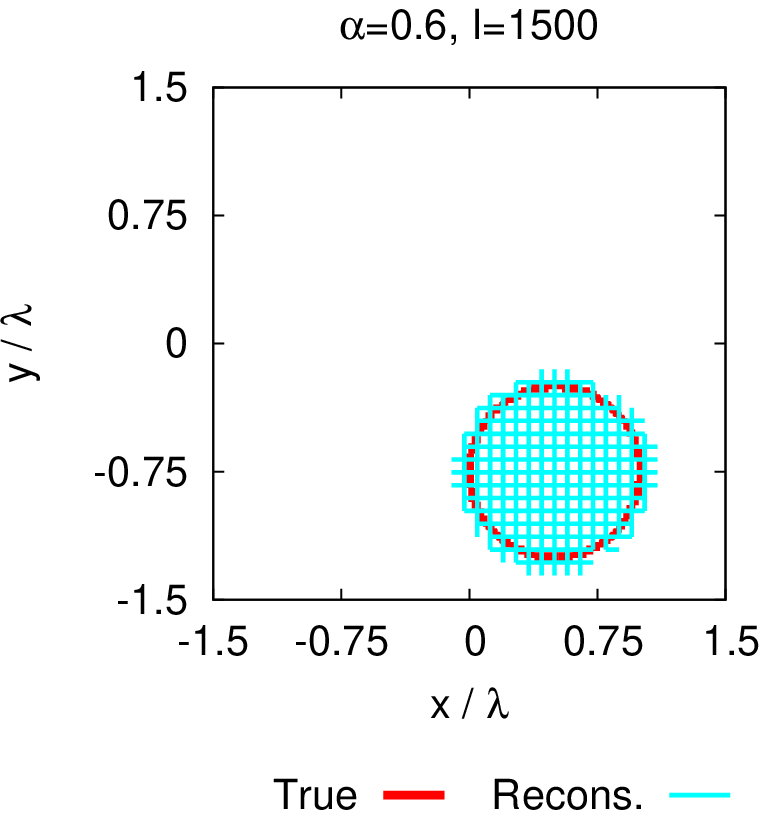}\tabularnewline
(\emph{d})&
(\emph{e})\tabularnewline
\end{tabular}\end{center}

\begin{center}~\vfill\end{center}

\begin{center}\textbf{Fig. 5 - Ye et} \textbf{\emph{al.}}\textbf{,}
\textbf{\emph{{}``}}Multi-Resolution Subspace-Based ...''\end{center}

\newpage
\begin{center}~\vfill\end{center}

\begin{center}\begin{tabular}{c}
\includegraphics[%
  width=1.0\textwidth,
  keepaspectratio]{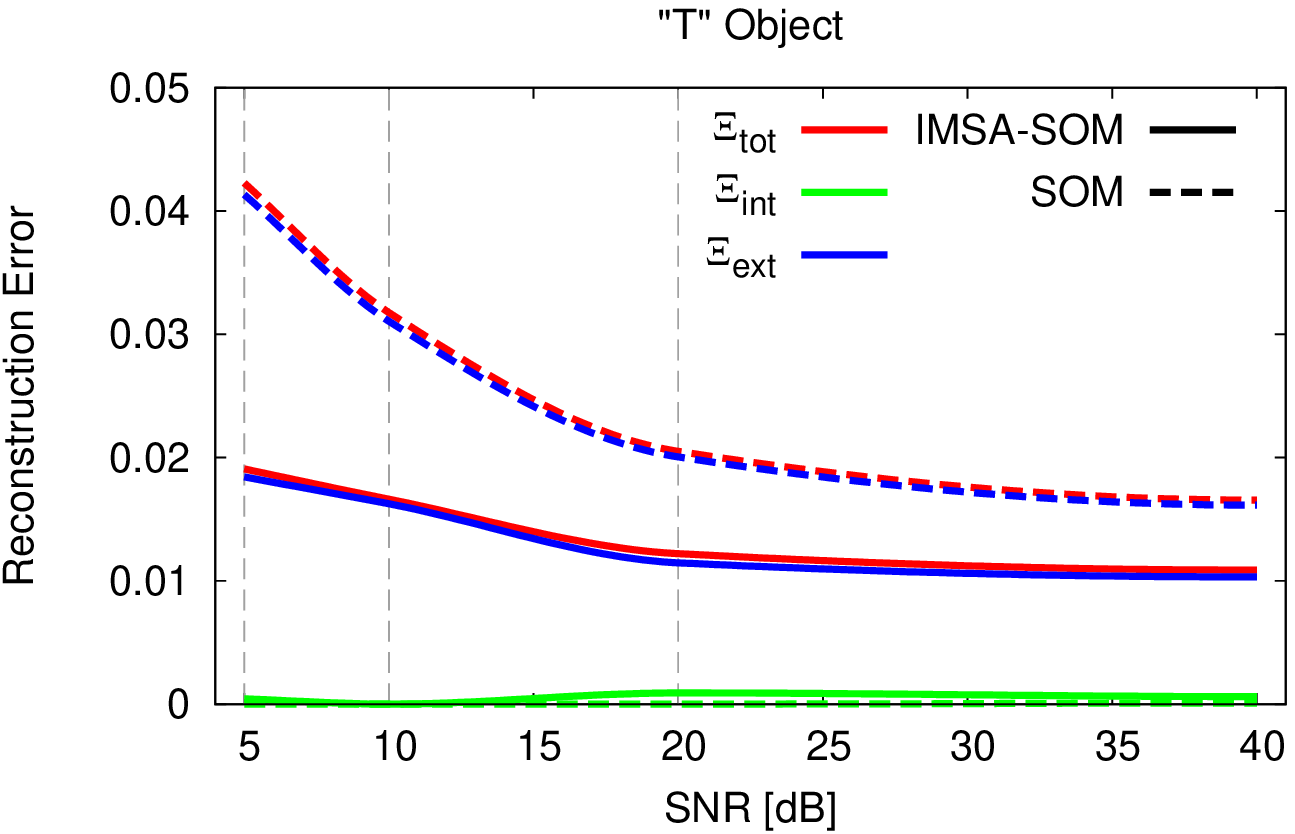}\tabularnewline
\end{tabular}\end{center}

\begin{center}~\vfill\end{center}

\begin{center}\textbf{Fig. 6 - Ye et} \textbf{\emph{al.}}\textbf{,}
\textbf{\emph{{}``}}Multi-Resolution Subspace-Based ...''\end{center}

\newpage
\begin{center}~\vfill\end{center}

\begin{center}\begin{tabular}{ccc}
&
\emph{~~~~~~~~~~IMSA-SOM}&
\emph{~~~~~~~~~~SOM}\tabularnewline
\begin{sideways}
\emph{~~~~~~~~~~~}~~~~~$SNR=20$ {[}dB{]}%
\end{sideways}&
\includegraphics[%
  width=0.40\columnwidth,
  keepaspectratio]{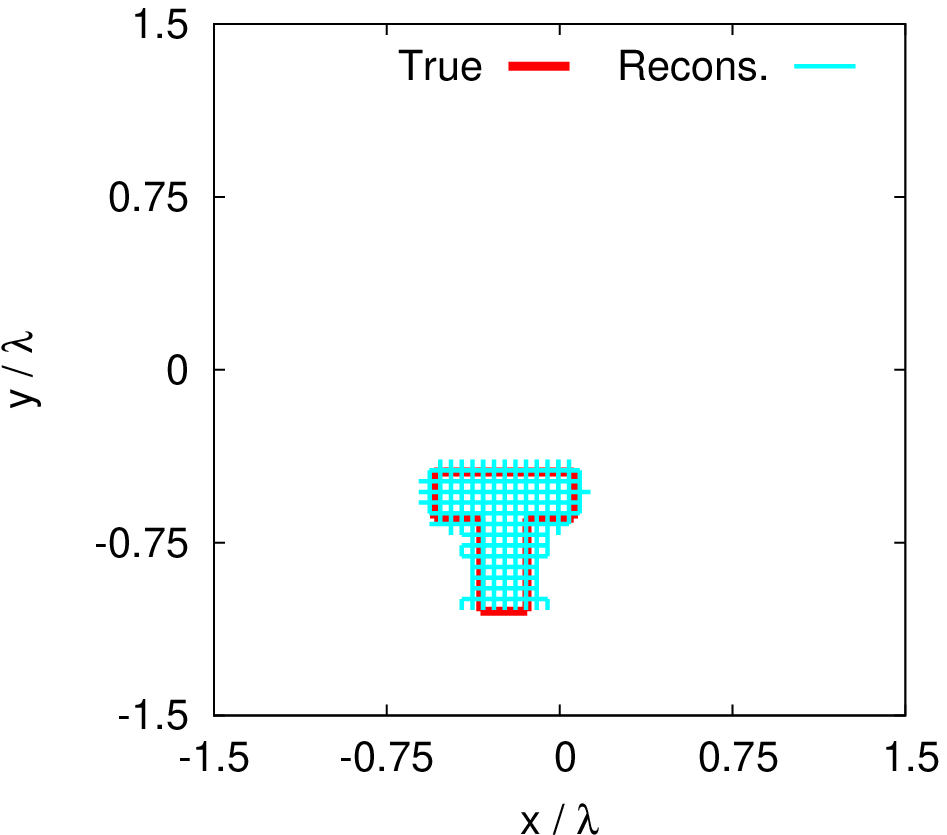}&
\includegraphics[%
  width=0.40\columnwidth,
  keepaspectratio]{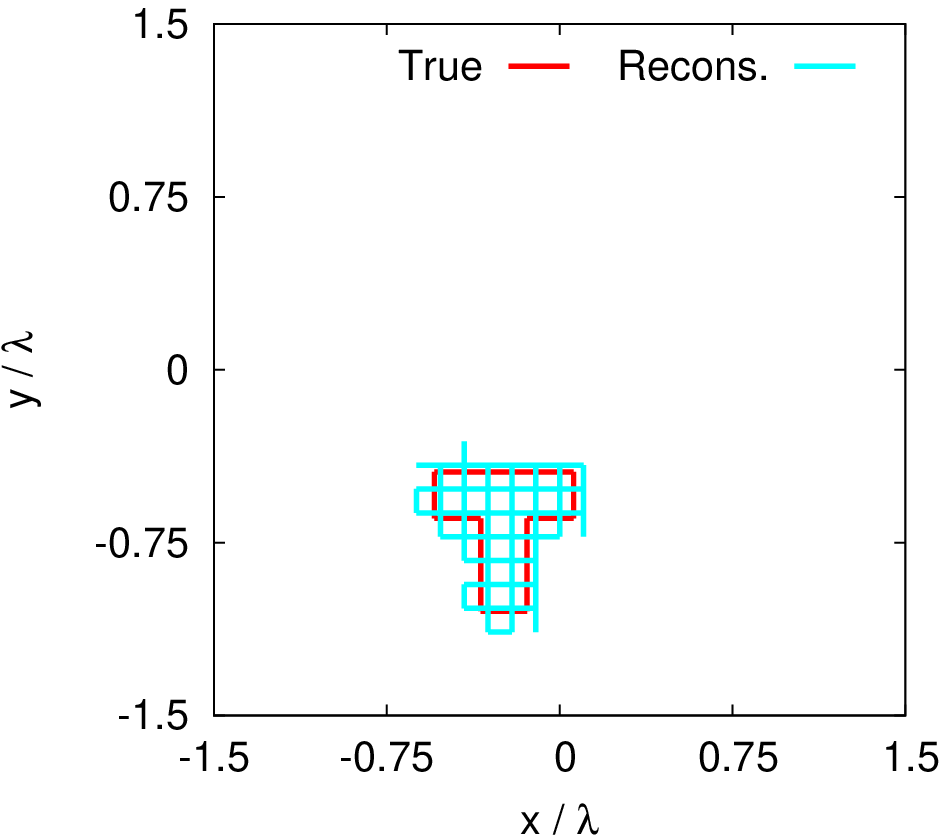}\tabularnewline
&
\emph{~~~~~~~~~~}(\emph{a})&
\emph{~~~~~~~~~~}(\emph{b})\tabularnewline
\begin{sideways}
\emph{~~~~~~~~~~~}~~~~~$SNR=10$ {[}dB{]}%
\end{sideways}&
\includegraphics[%
  width=0.40\columnwidth,
  keepaspectratio]{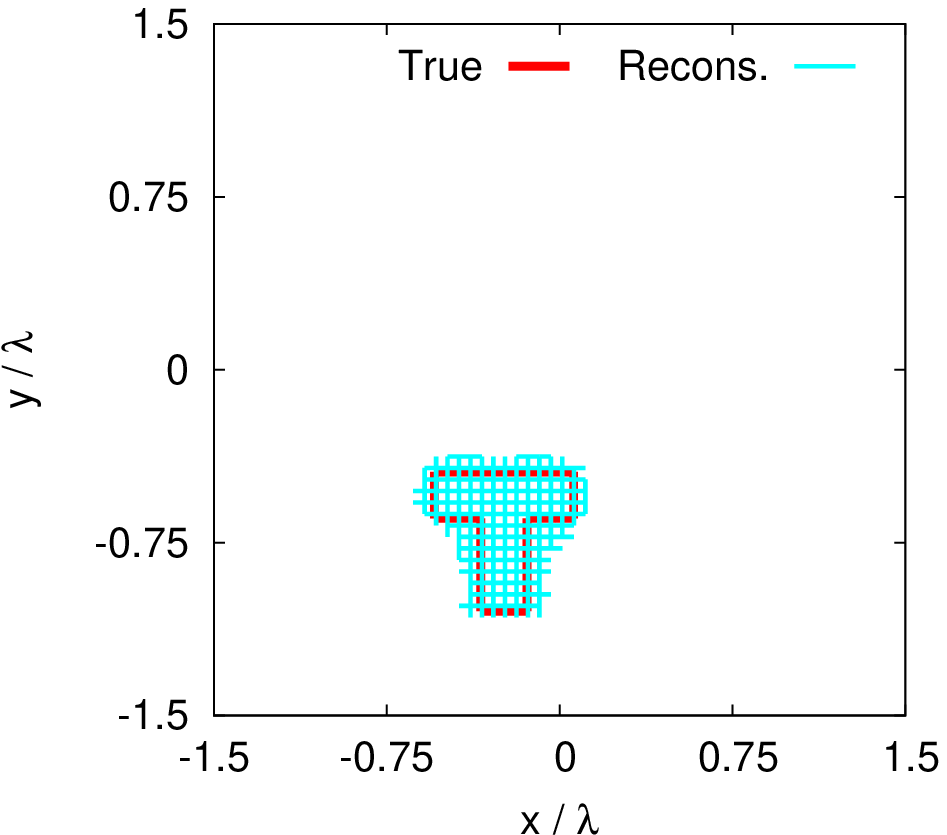}&
\includegraphics[%
  width=0.40\columnwidth,
  keepaspectratio]{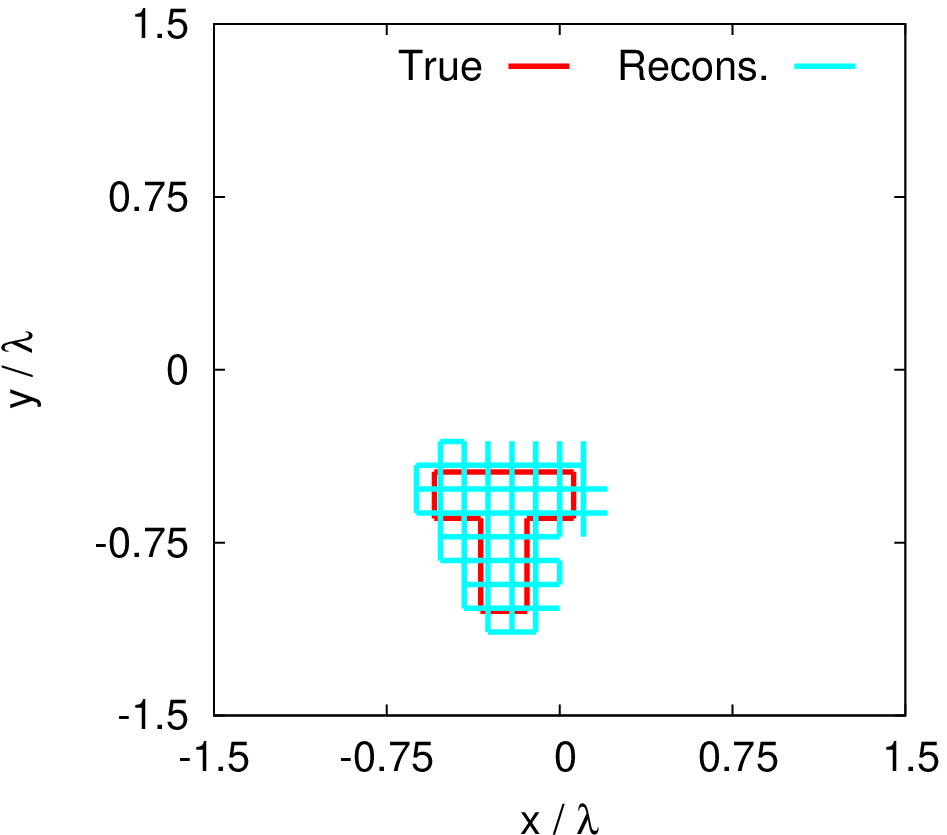}\tabularnewline
&
\emph{~~~~~~~~~~}(\emph{c})&
\emph{~~~~~~~~~~}(\emph{d})\tabularnewline
\begin{sideways}
\emph{~~~~~~~~~~~}~~~~~$SNR=5$ {[}dB{]}%
\end{sideways}&
\includegraphics[%
  width=0.40\columnwidth,
  keepaspectratio]{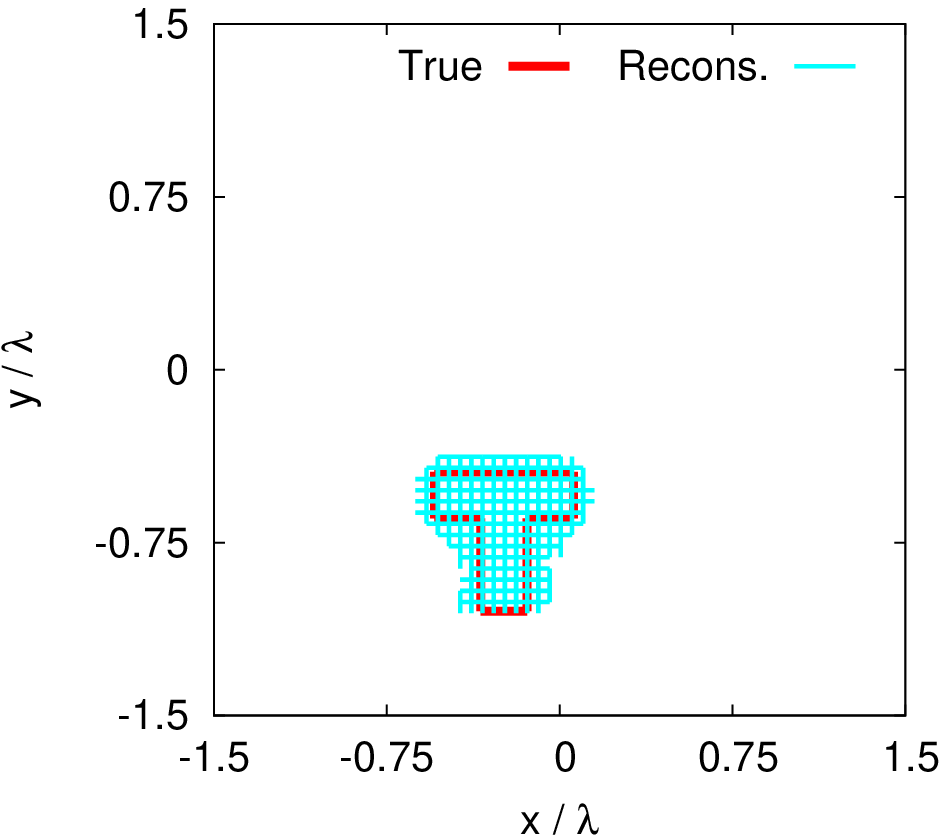}&
\includegraphics[%
  width=0.40\columnwidth,
  keepaspectratio]{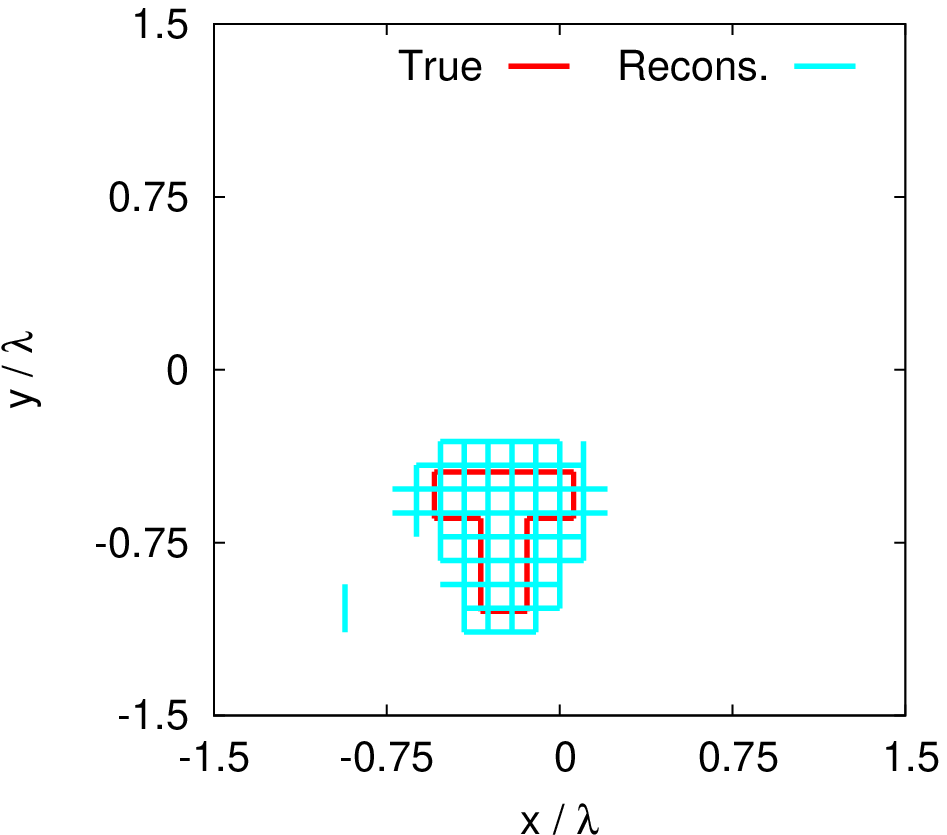}\tabularnewline
&
\emph{~~~~~~~~~~}(\emph{e})&
\emph{~~~~~~~~~~}(\emph{f})\tabularnewline
\end{tabular}\end{center}

\begin{center}~\vfill\end{center}

\begin{center}\textbf{Fig. 7 - Ye et} \textbf{\emph{al.}}\textbf{,}
\textbf{\emph{{}``}}Multi-Resolution Subspace-Based ...''\end{center}

\newpage
\begin{center}~\vfill\end{center}

\begin{center}\begin{tabular}{c}
\includegraphics[%
  width=1.0\textwidth,
  keepaspectratio]{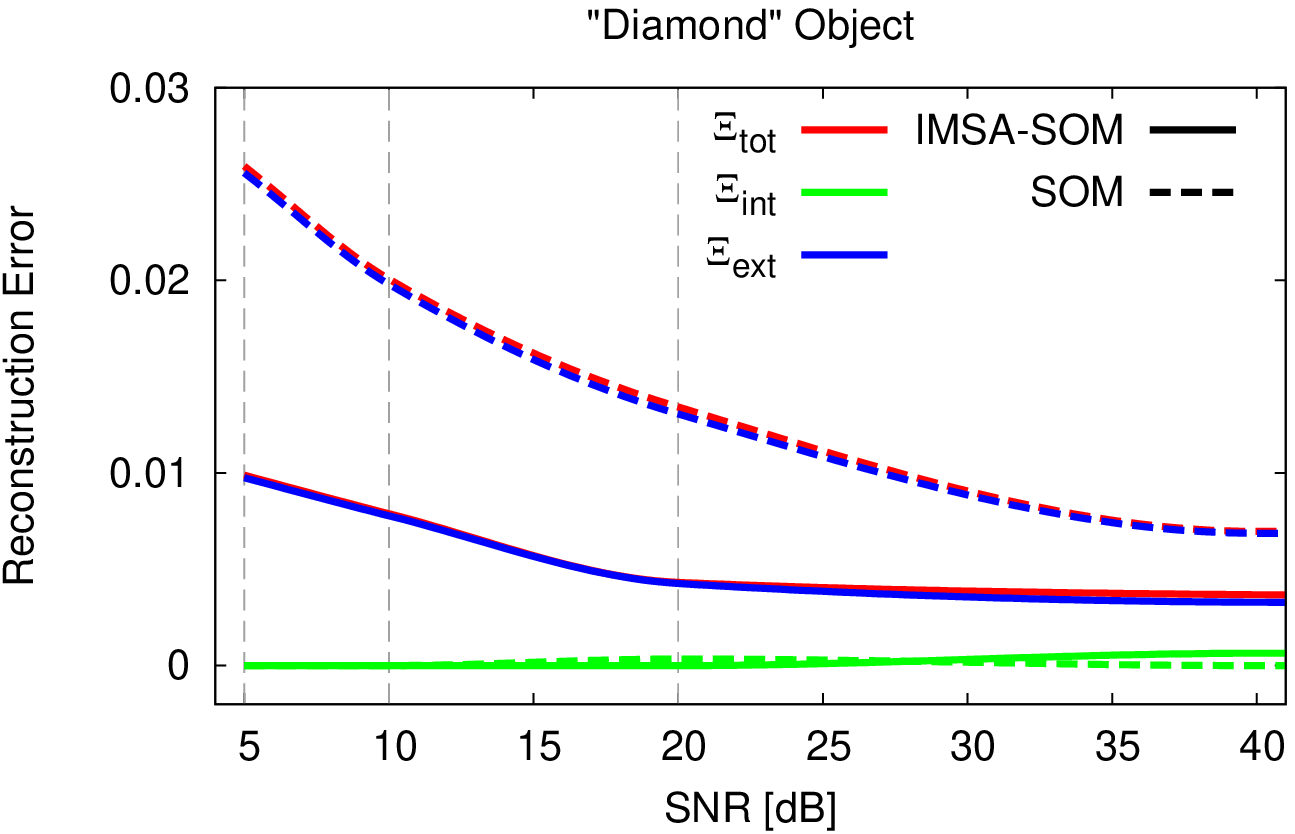}\tabularnewline
\end{tabular}\end{center}

\begin{center}~\vfill\end{center}

\begin{center}\textbf{Fig. 8 - Ye et} \textbf{\emph{al.}}\textbf{,}
\textbf{\emph{{}``}}Multi-Resolution Subspace-Based ...''\end{center}

\newpage
\begin{center}~\vfill\end{center}

\begin{center}\begin{tabular}{ccc}
&
\emph{~~~~~~~~~~IMSA-SOM}&
\emph{~~~~~~~~~~SOM}\tabularnewline
\begin{sideways}
\emph{~~~~~~~~~~~}~~~~~$SNR=20$ {[}dB{]}%
\end{sideways}&
\includegraphics[%
  width=0.40\columnwidth,
  keepaspectratio]{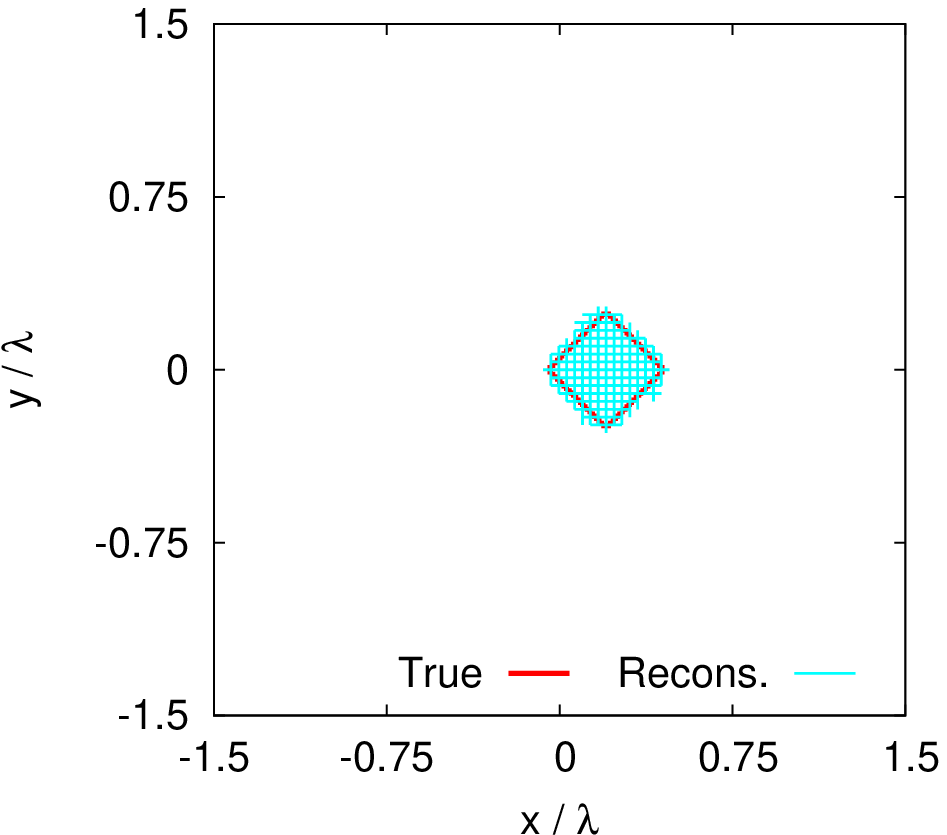}&
\includegraphics[%
  width=0.40\columnwidth,
  keepaspectratio]{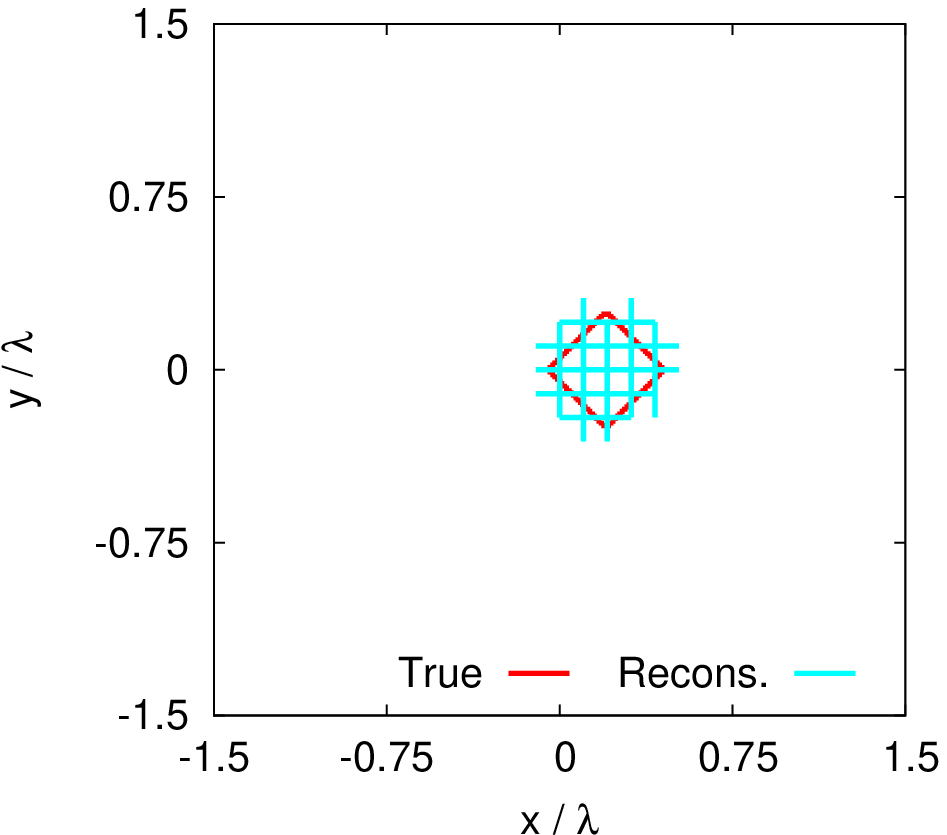}\tabularnewline
&
\emph{~~~~~~~~~~}(\emph{a})&
\emph{~~~~~~~~~~}(\emph{b})\tabularnewline
\begin{sideways}
\emph{~~~~~~~~~~~}~~~~~$SNR=10$ {[}dB{]}%
\end{sideways}&
\includegraphics[%
  width=0.40\columnwidth,
  keepaspectratio]{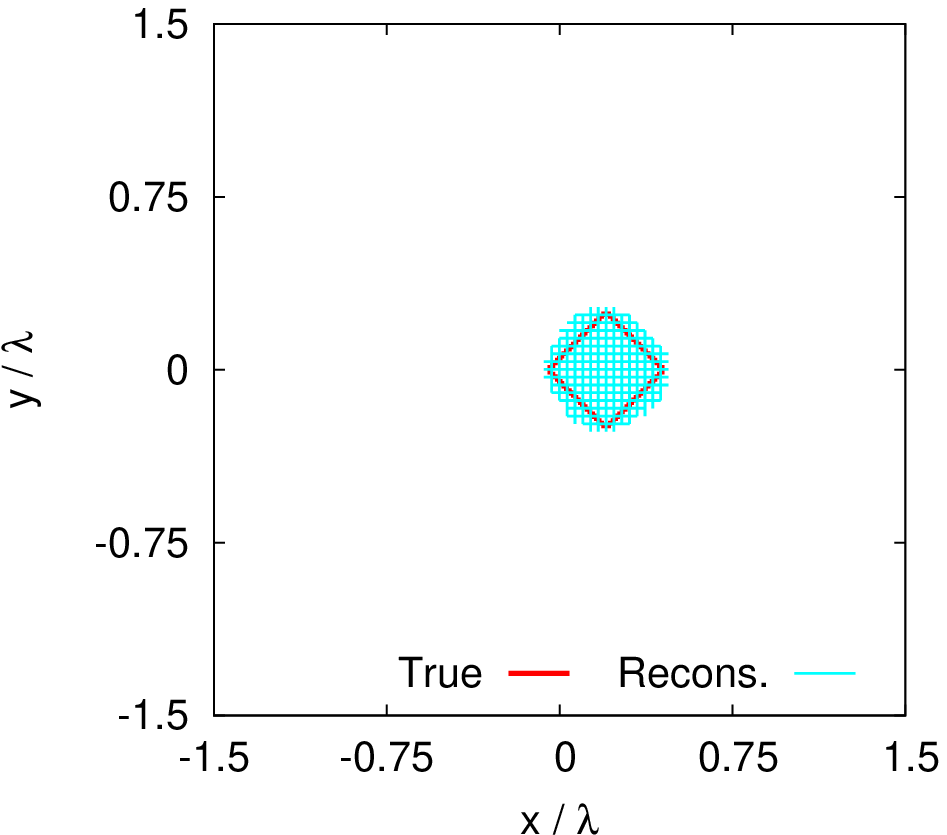}&
\includegraphics[%
  width=0.40\columnwidth,
  keepaspectratio]{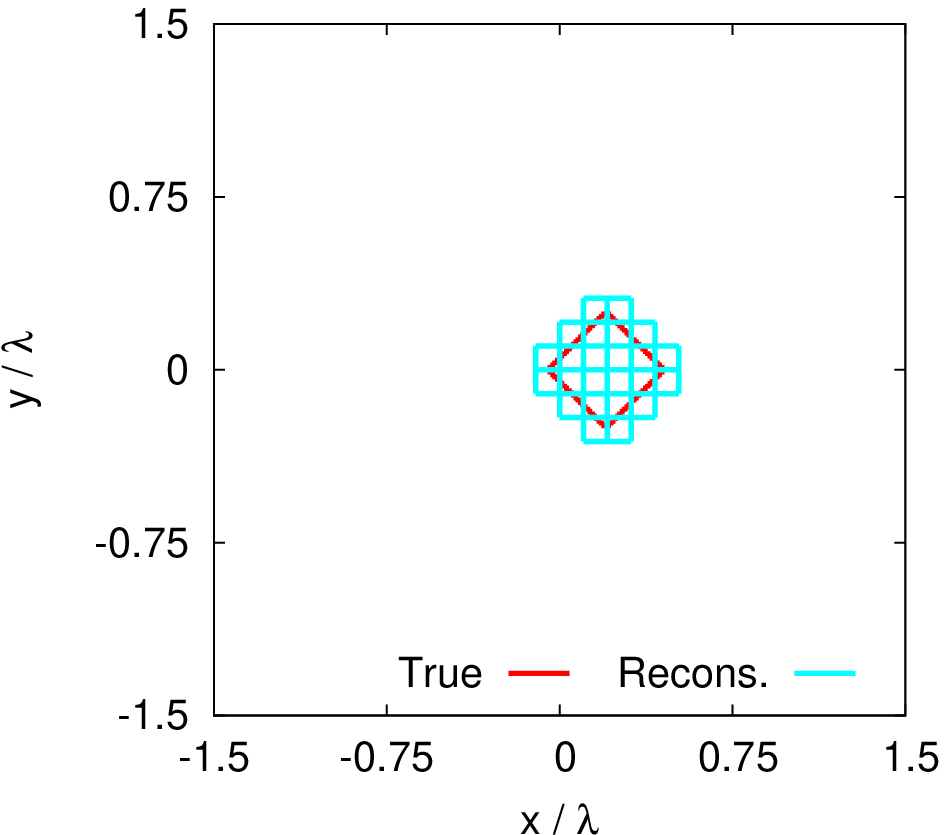}\tabularnewline
&
\emph{~~~~~~~~~~}(\emph{c})&
\emph{~~~~~~~~~~}(\emph{d})\tabularnewline
\begin{sideways}
\emph{~~~~~~~~~~~}~~~~~$SNR=5$ {[}dB{]}%
\end{sideways}&
\includegraphics[%
  width=0.40\columnwidth,
  keepaspectratio]{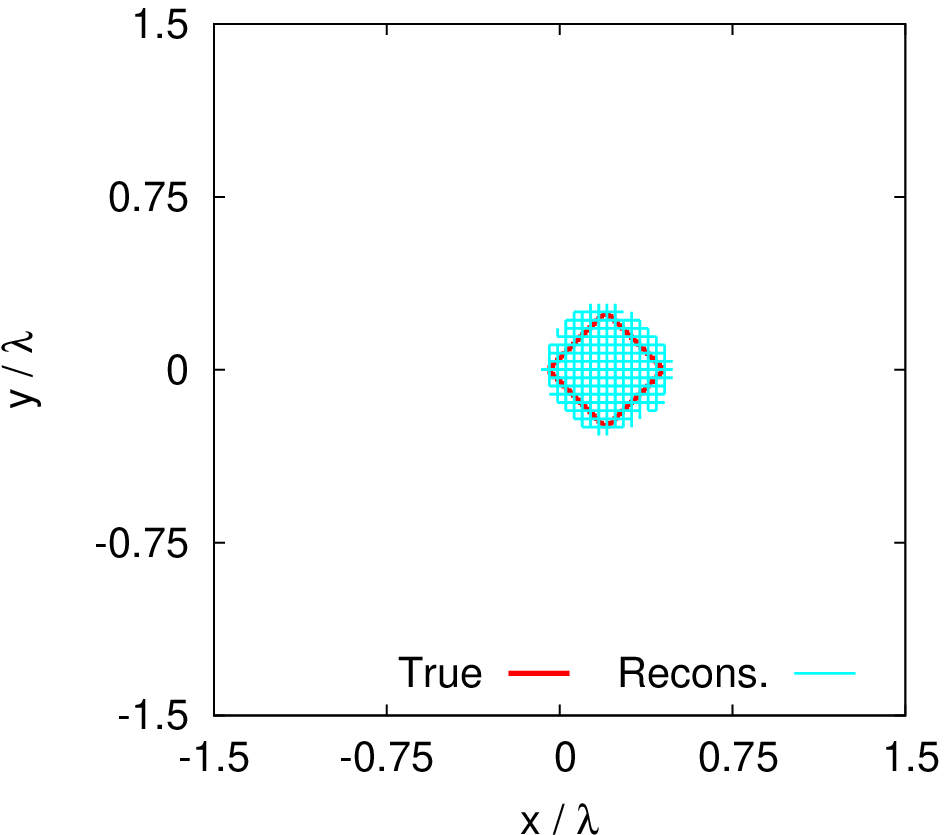}&
\includegraphics[%
  width=0.40\columnwidth,
  keepaspectratio]{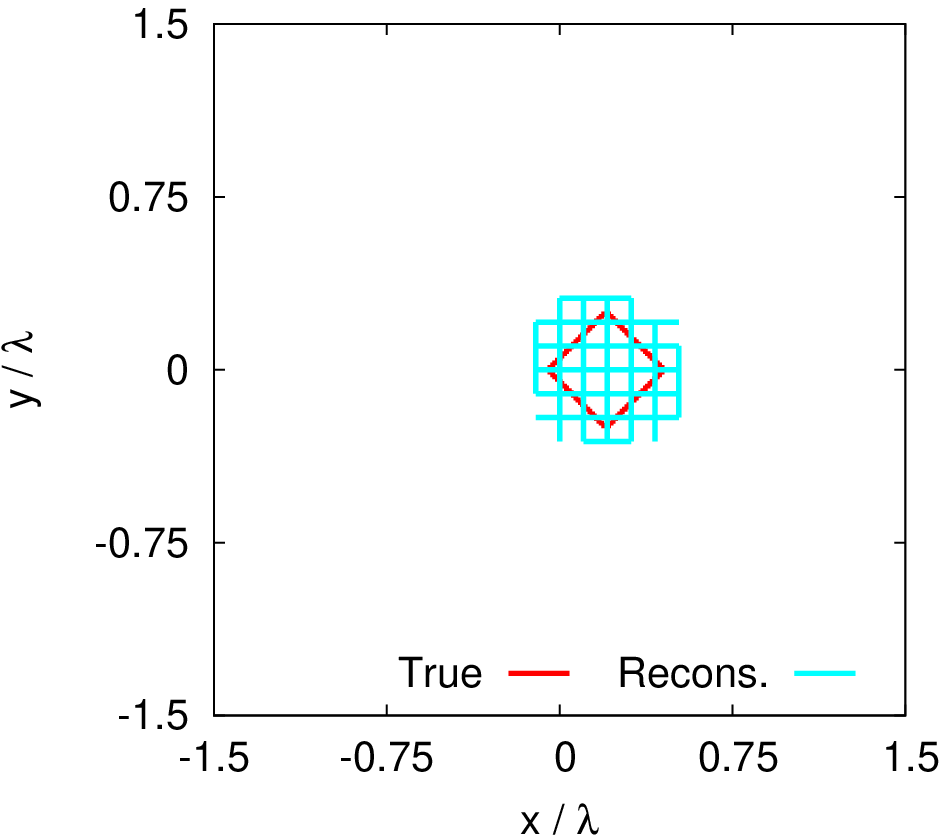}\tabularnewline
&
\emph{~~~~~~~~~~}(\emph{e})&
\emph{~~~~~~~~~~}(\emph{f})\tabularnewline
\end{tabular}\end{center}

\begin{center}~\vfill\end{center}

\begin{center}\textbf{Fig. 9 - Ye et} \textbf{\emph{al.}}\textbf{,}
\textbf{\emph{{}``}}Multi-Resolution Subspace-Based ...''\end{center}

\newpage
\begin{center}~\vfill\end{center}

\begin{center}\begin{tabular}{c}
\includegraphics[%
  width=0.80\textwidth]{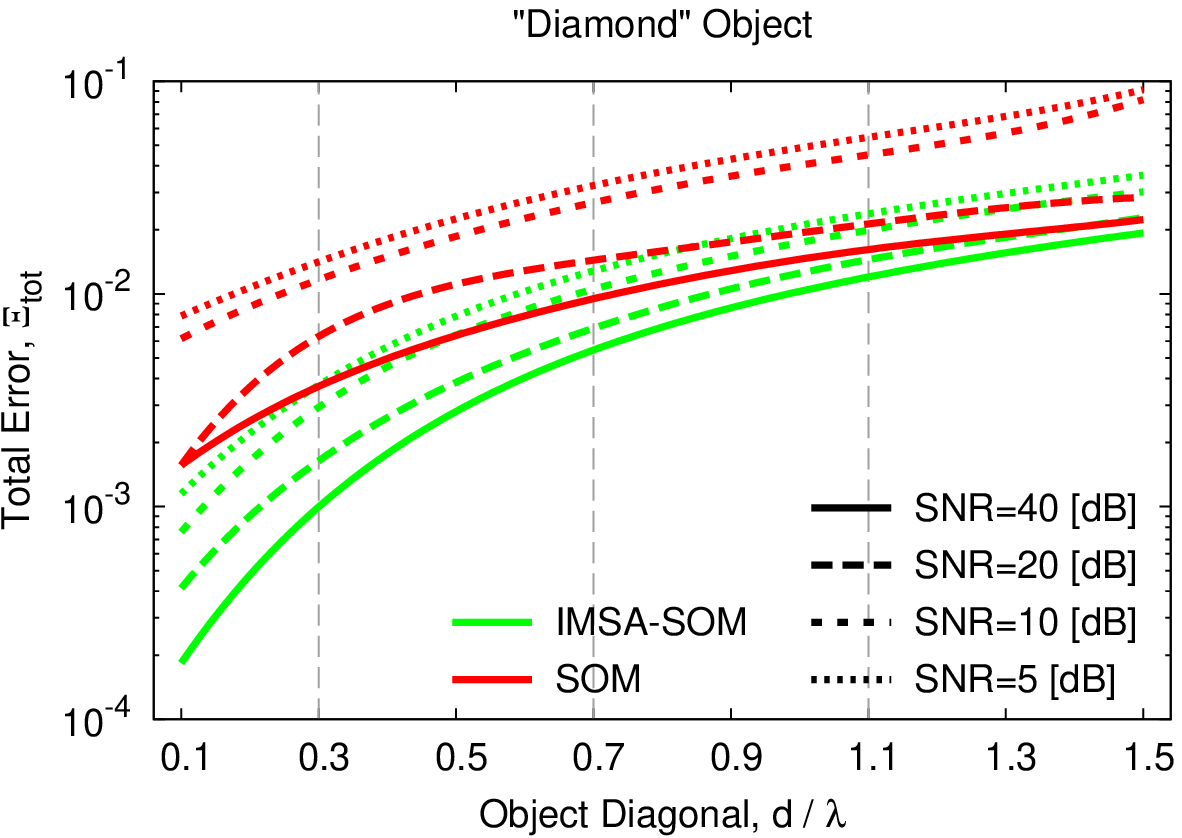}\tabularnewline
\end{tabular}\end{center}

\begin{center}~\vfill\end{center}

\begin{center}\textbf{Fig. 10 - Ye et} \textbf{\emph{al.}}\textbf{,}
\textbf{\emph{{}``}}Multi-Resolution Subspace-Based ...''\end{center}

\newpage
\begin{center}~\vfill\end{center}

\begin{center}\begin{tabular}{ccc}
&
\emph{~~~~~~~~~~IMSA-SOM}&
\emph{~~~~~~~~~~SOM}\tabularnewline
\begin{sideways}
\emph{~~~~~~~~~~~~~~~~~}~~~~~$d=1.1\,\,\lambda$%
\end{sideways}&
\includegraphics[%
  width=0.40\columnwidth]{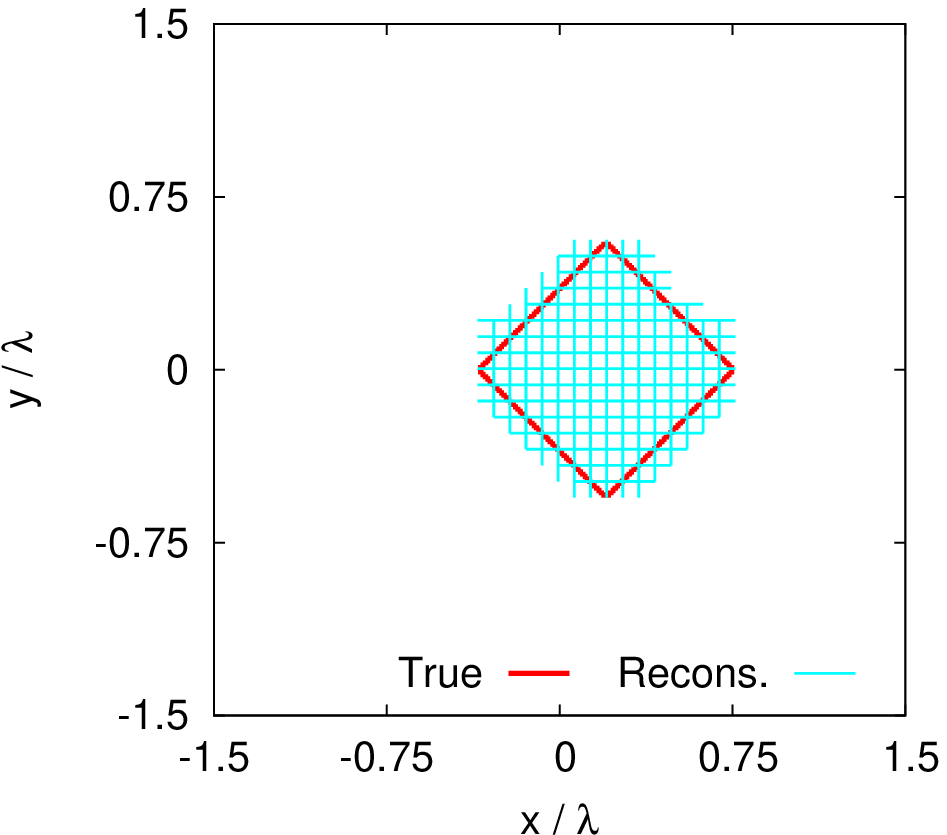}&
\includegraphics[%
  width=0.40\columnwidth]{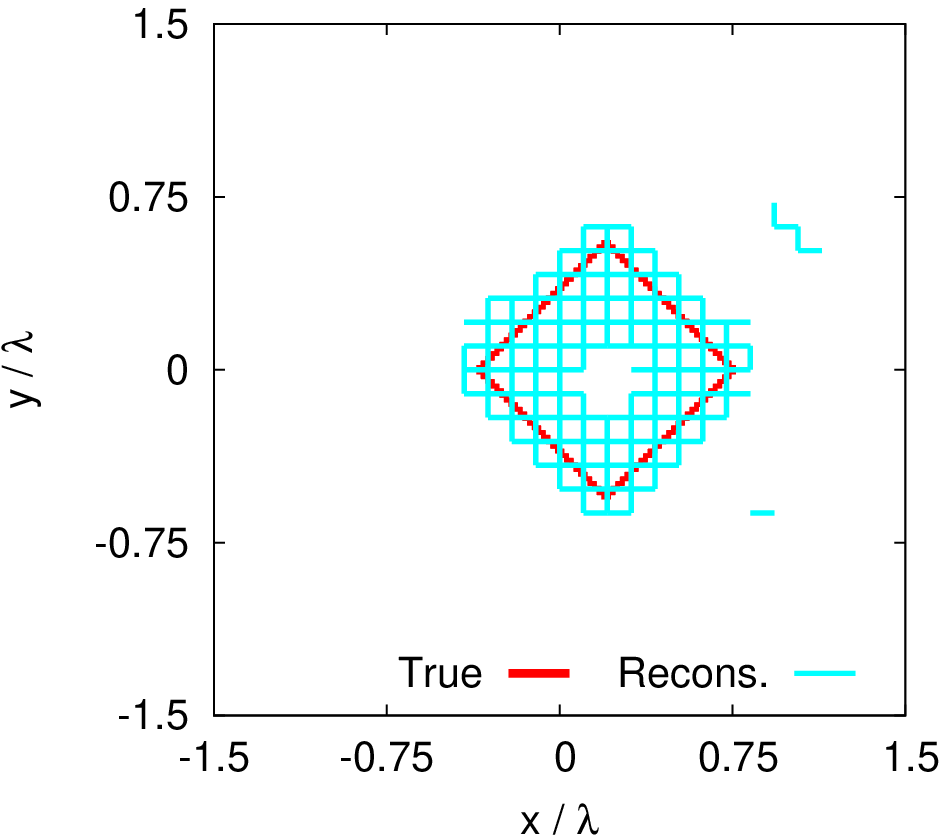}\tabularnewline
&
\emph{~~~~~~~~~~}(\emph{a})&
\emph{~~~~~~~~~~}(\emph{b})\tabularnewline
\begin{sideways}
\emph{~~~~~~~~~~~~~~~~~}~~~~~$d=0.7\,\,\lambda$%
\end{sideways}&
\includegraphics[%
  width=0.40\columnwidth]{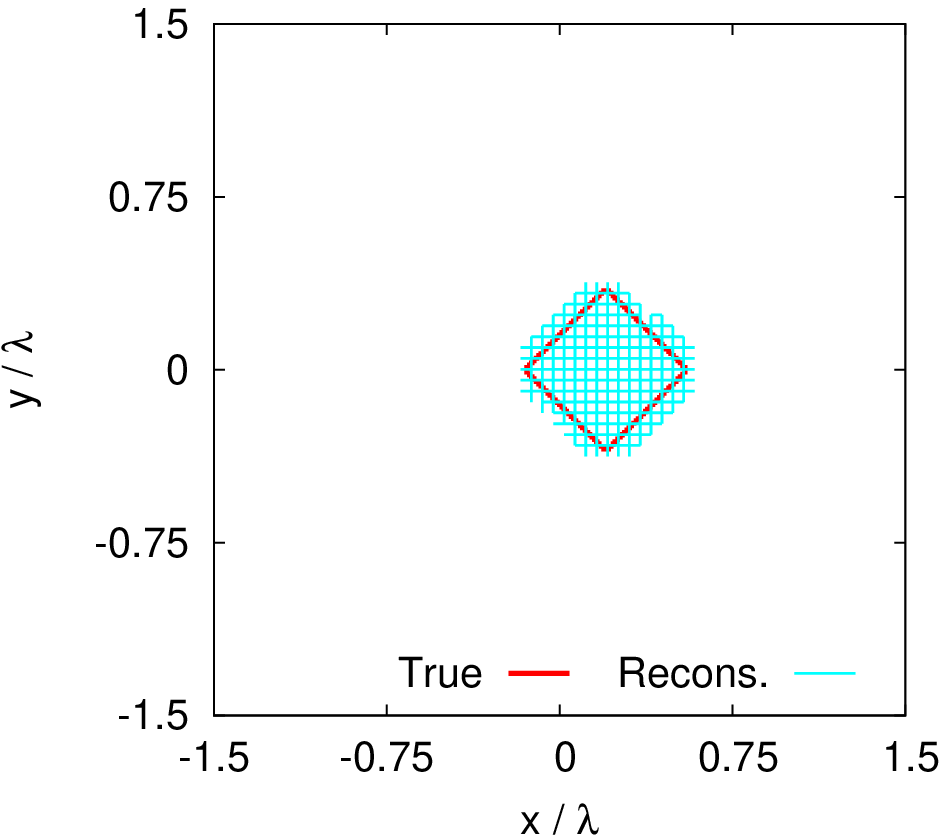}&
\includegraphics[%
  width=0.40\columnwidth]{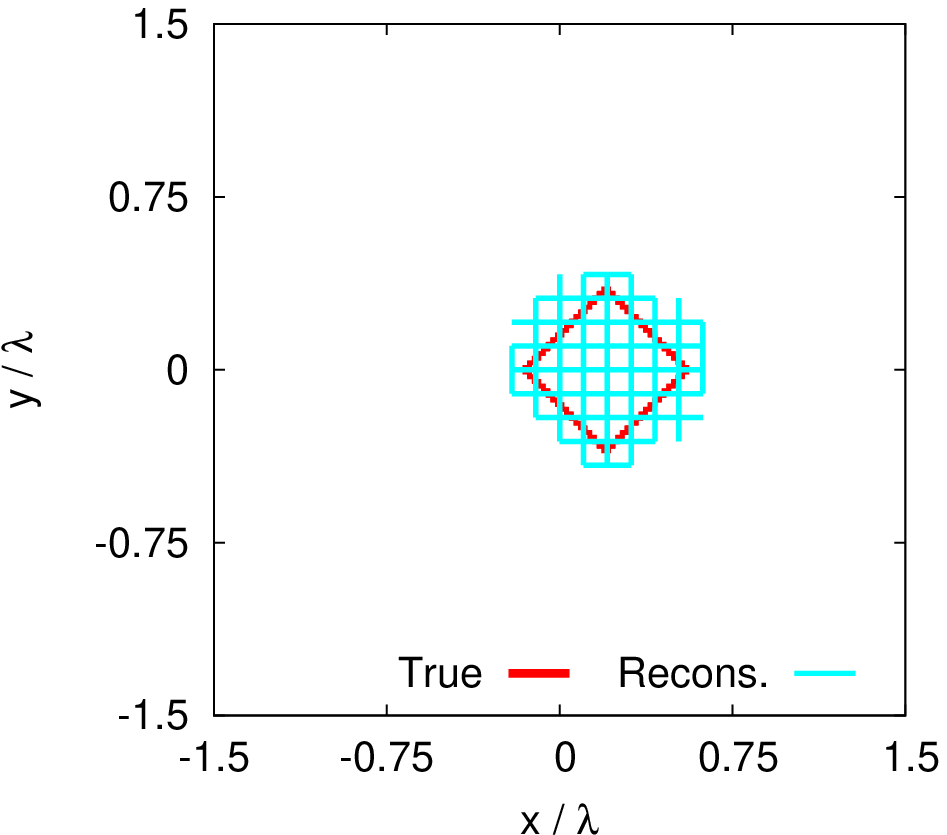}\tabularnewline
&
\emph{~~~~~~~~~~}(\emph{c})&
\emph{~~~~~~~~~~}(\emph{d})\tabularnewline
\begin{sideways}
\emph{~~~~~~~~~~~~~~~~~}~~~~~$d=0.3\,\,\lambda$%
\end{sideways}&
\includegraphics[%
  width=0.40\columnwidth]{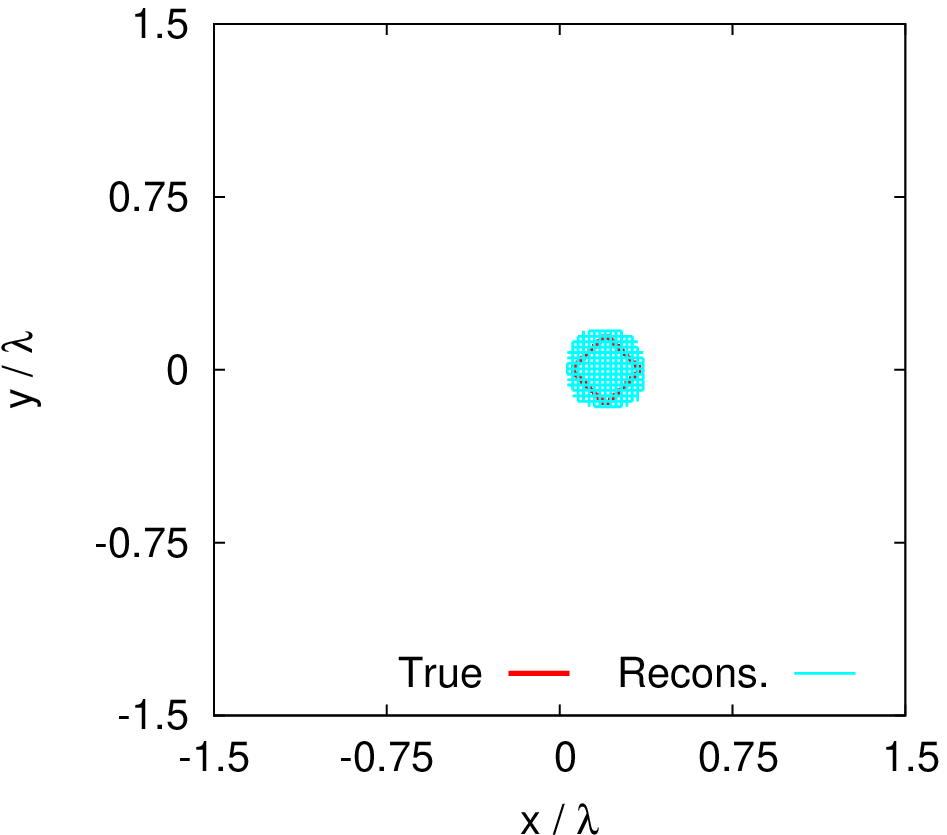}&
\includegraphics[%
  width=0.40\columnwidth]{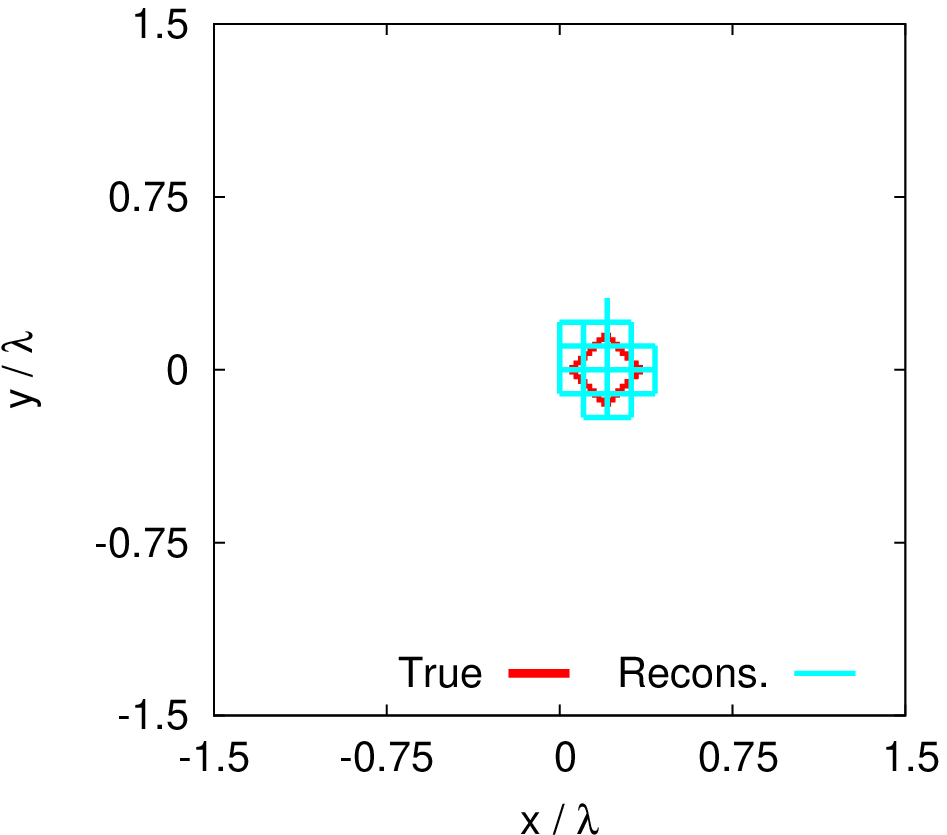}\tabularnewline
&
\emph{~~~~~~~~~~}(\emph{e})&
\emph{~~~~~~~~~~}(\emph{f})\tabularnewline
\end{tabular}\end{center}

\begin{center}~\vfill\end{center}

\begin{center}\textbf{Fig. 11 - Ye et} \textbf{\emph{al.}}\textbf{,}
\textbf{\emph{{}``}}Multi-Resolution Subspace-Based ...''\end{center}

\newpage
\begin{center}~\vfill\end{center}

\begin{center}\begin{tabular}{c}
\includegraphics[%
  width=1.0\textwidth,
  keepaspectratio]{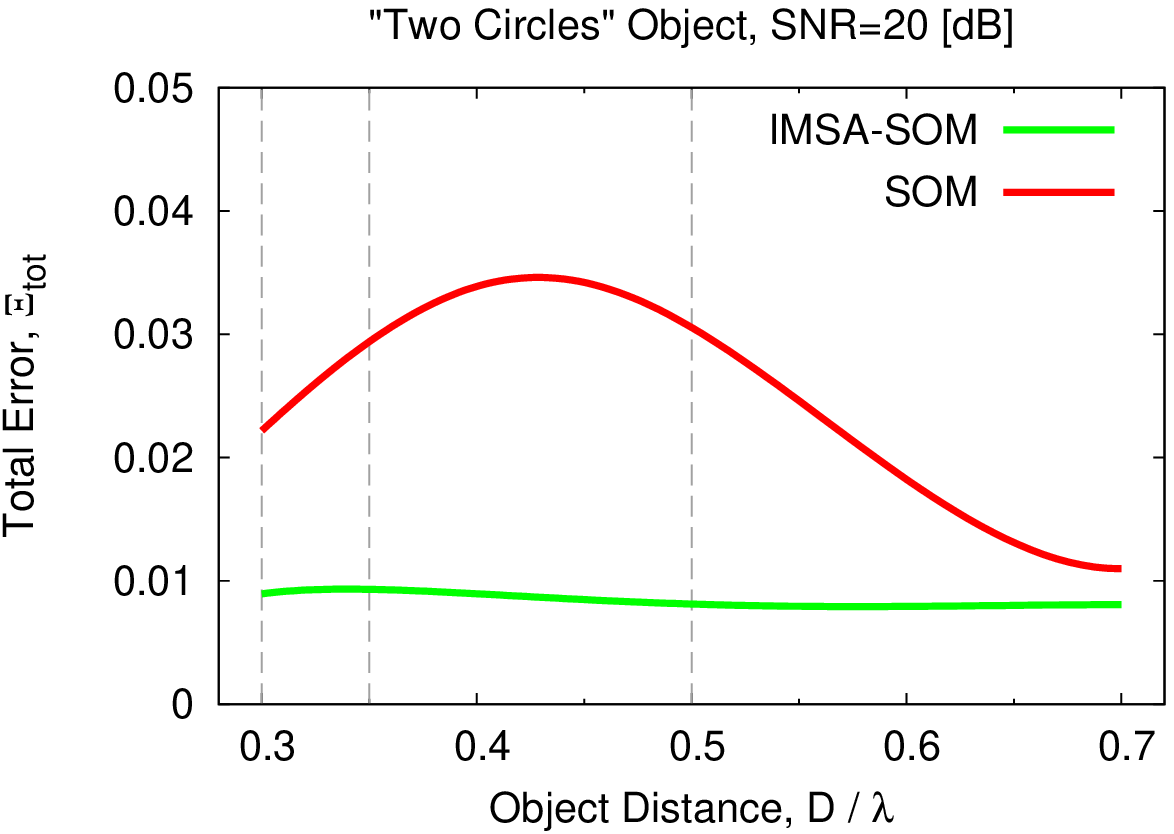}\tabularnewline
\end{tabular}\end{center}

\begin{center}~\vfill\end{center}

\begin{center}\textbf{Fig. 12 - Ye et} \textbf{\emph{al.}}\textbf{,}
\textbf{\emph{{}``}}Multi-Resolution Subspace-Based ...''\end{center}

\newpage
\begin{center}~\vfill\end{center}

\begin{center}\begin{tabular}{ccc}
&
\emph{~~~~~~~~~~IMSA-SOM}&
\emph{~~~~~~~~~~SOM}\tabularnewline
\begin{sideways}
\emph{~~~~~~~~~~~~~~~}~~~~~$D=0.50\,\,\lambda$%
\end{sideways}&
\includegraphics[%
  width=0.40\columnwidth]{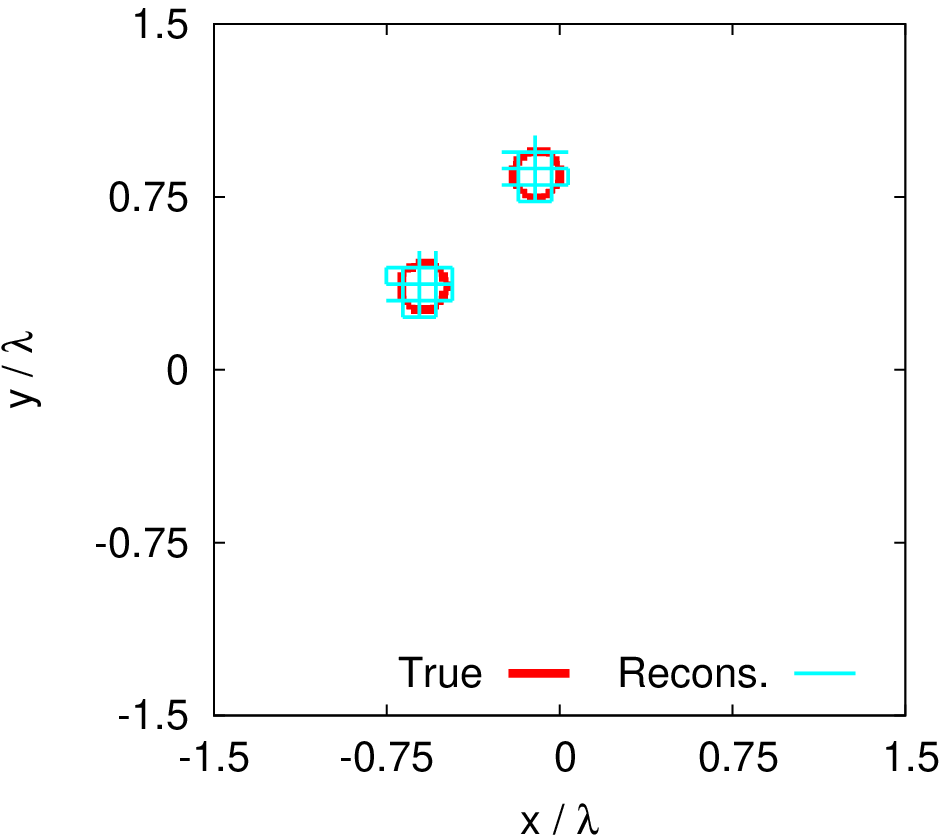}&
\includegraphics[%
  width=0.40\columnwidth]{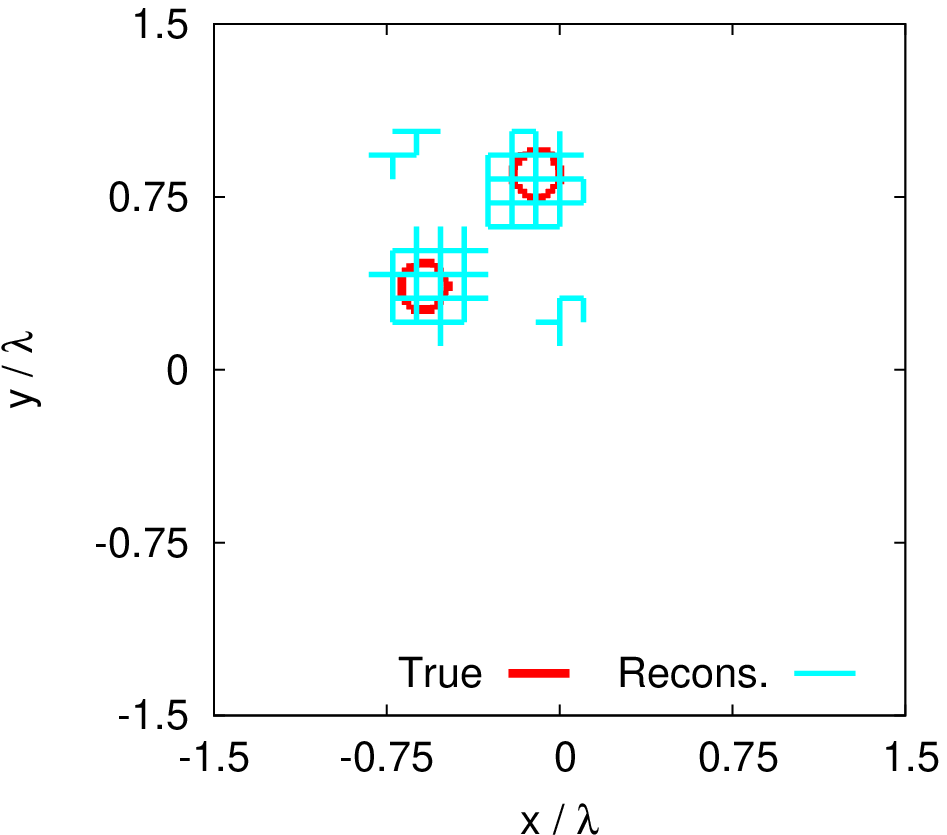}\tabularnewline
&
\emph{~~~~~~~~~~}(\emph{a})&
\emph{~~~~~~~~~~}(\emph{b})\tabularnewline
\begin{sideways}
\emph{~~~~~~~~~~~~~~~}~~~~~$D=0.35\,\,\lambda$%
\end{sideways}&
\includegraphics[%
  width=0.40\columnwidth]{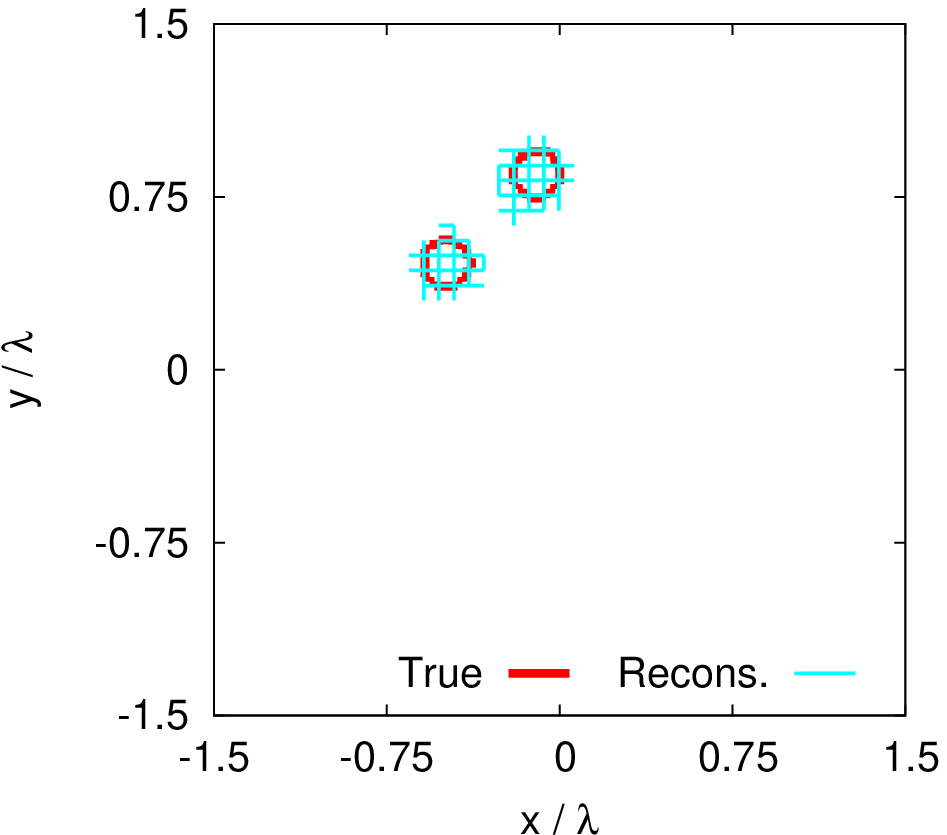}&
\includegraphics[%
  width=0.40\columnwidth]{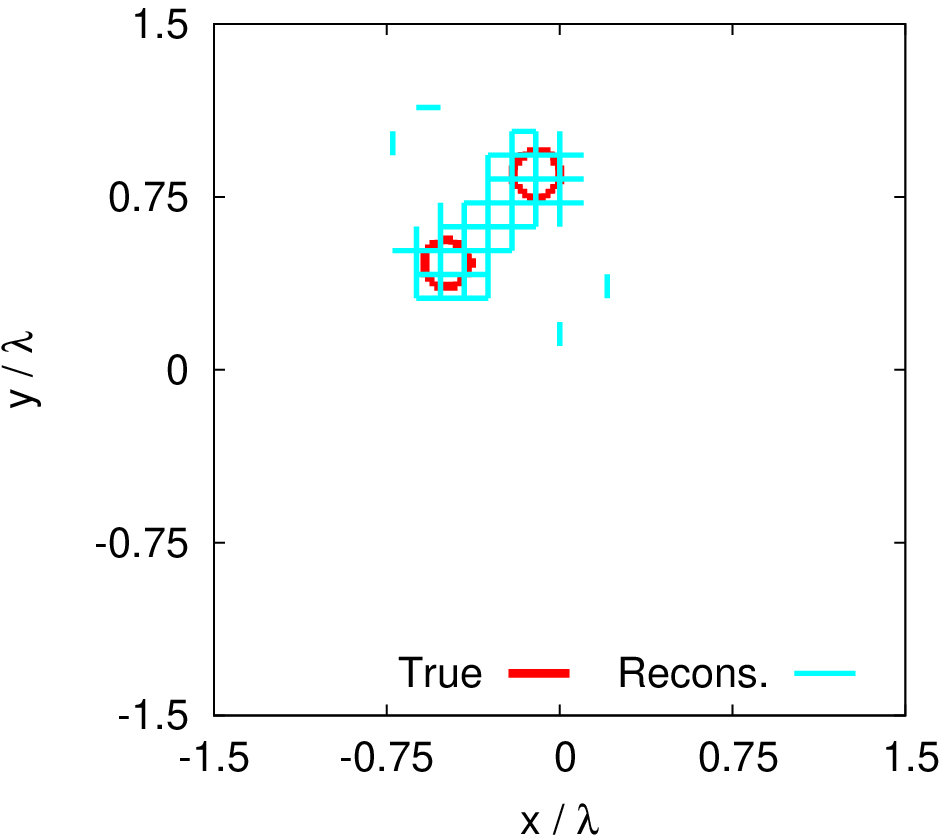}\tabularnewline
&
\emph{~~~~~~~~~~}(\emph{c})&
\emph{~~~~~~~~~~}(\emph{d})\tabularnewline
\begin{sideways}
\emph{~~~~~~~~~~~~~~~~~}~~~$D=0.30\,\,\lambda$%
\end{sideways}&
\includegraphics[%
  width=0.40\columnwidth]{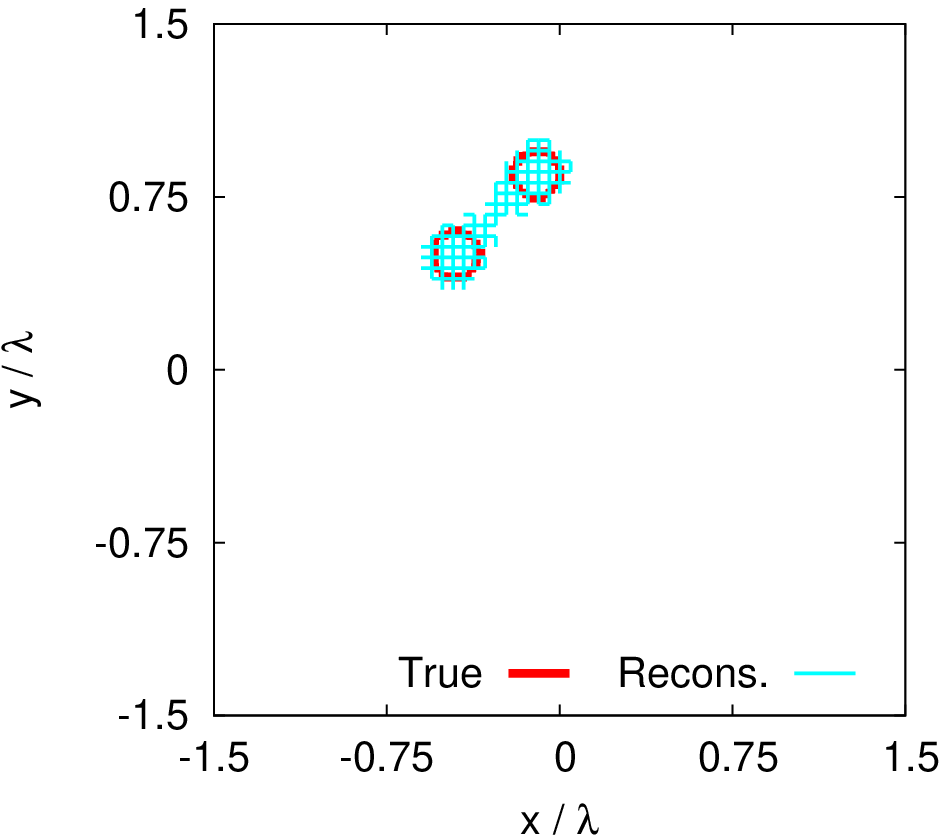}&
\includegraphics[%
  width=0.40\columnwidth]{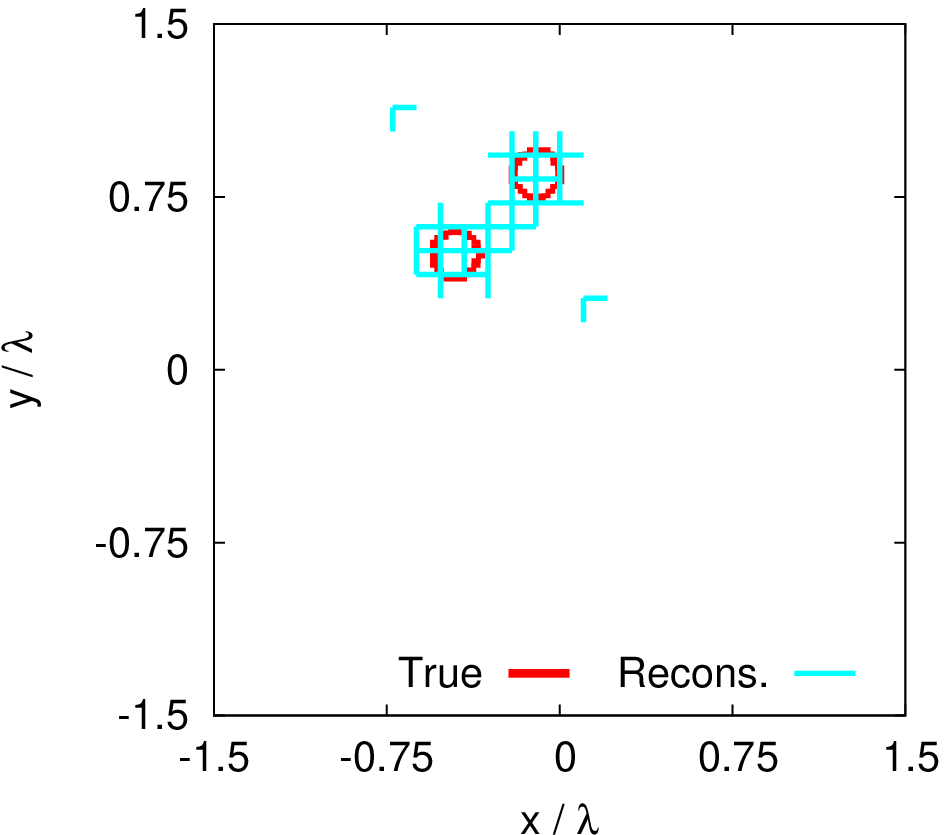}\tabularnewline
&
\emph{~~~~~~~~~~}(\emph{e})&
\emph{~~~~~~~~~~}(\emph{f})\tabularnewline
\end{tabular}\end{center}

\begin{center}~\vfill\end{center}

\begin{center}\textbf{Fig. 13 - Ye et} \textbf{\emph{al.}}\textbf{,}
\textbf{\emph{{}``}}Multi-Resolution Subspace-Based ...''\end{center}

\newpage
\begin{center}~\vfill\end{center}

\begin{center}\begin{tabular}{cc}
\emph{~~~~~~~~~~IMSA-SOM}&
\emph{~~~~~~~~~~SOM}\tabularnewline
\includegraphics[%
  width=0.40\columnwidth,
  keepaspectratio]{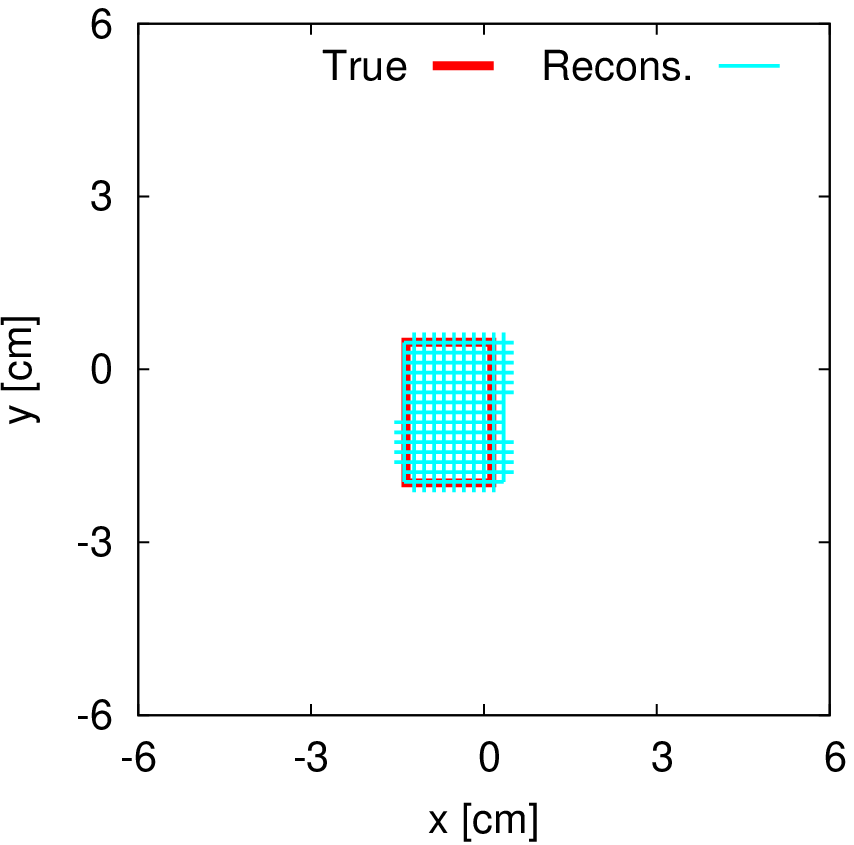}&
\includegraphics[%
  width=0.40\columnwidth,
  keepaspectratio]{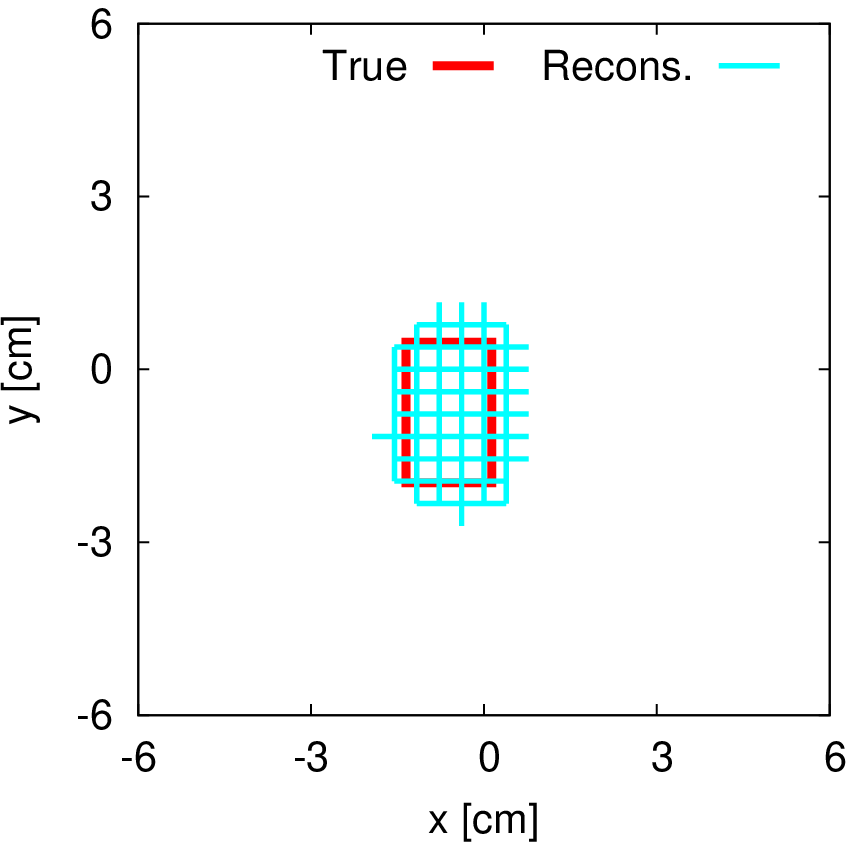}\tabularnewline
\emph{~~~~~~~~~~}(\emph{a})&
\emph{~~~~~~~~~~}(\emph{b})\tabularnewline
\end{tabular}\end{center}

\begin{center}~\vfill\end{center}

\begin{center}\textbf{Fig. 14 - Ye et} \textbf{\emph{al.}}\textbf{,}
\textbf{\emph{{}``}}Multi-Resolution Subspace-Based ...''\end{center}

\newpage
\begin{center}~\vfill\end{center}

\begin{center}\begin{tabular}{cc}
\emph{~~~~~~~~~~IMSA-SOM}&
\emph{~~~~~~~~~~SOM}\tabularnewline
\includegraphics[%
  width=0.40\columnwidth,
  keepaspectratio]{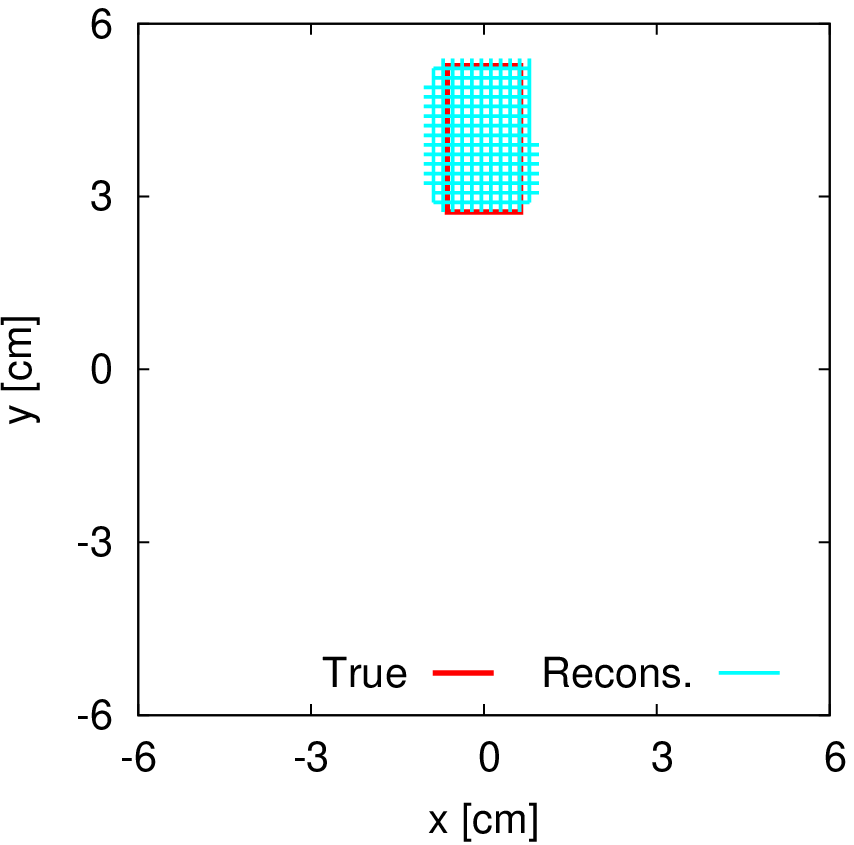}&
\includegraphics[%
  width=0.40\columnwidth,
  keepaspectratio]{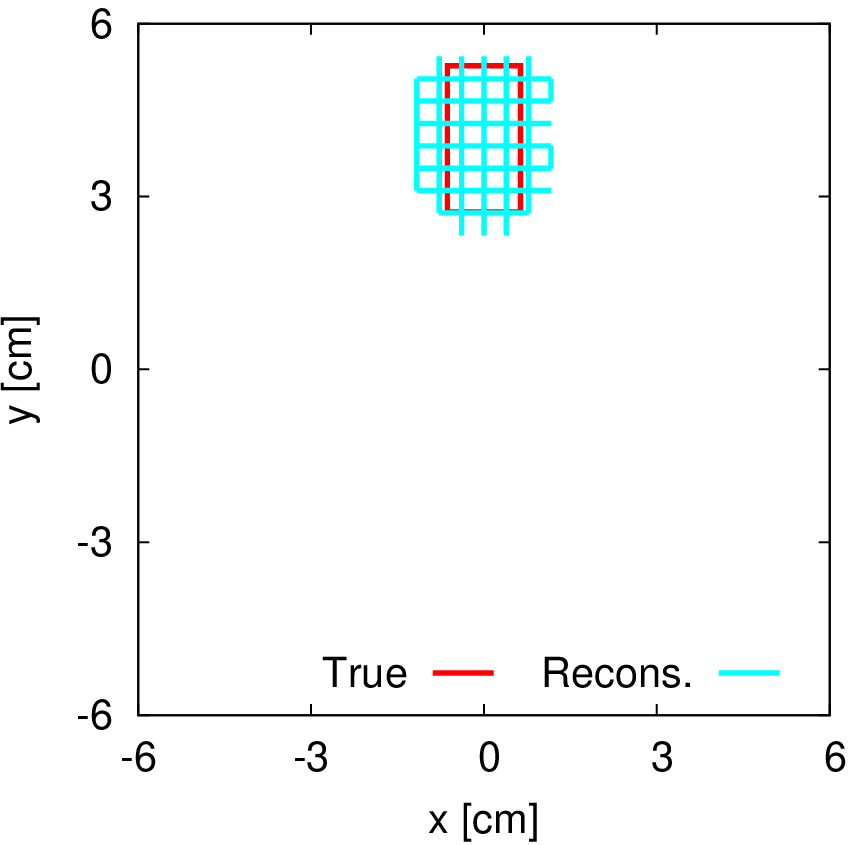}\tabularnewline
\emph{~~~~~~~~~~}(\emph{a})&
\emph{~~~~~~~~~~}(\emph{b})\tabularnewline
\end{tabular}\end{center}

\begin{center}~\vfill\end{center}

\begin{center}\textbf{Fig. 15 - Ye et} \textbf{\emph{al.}}\textbf{,}
\textbf{\emph{{}``}}Multi-Resolution Subspace-Based ...''\end{center}

\newpage
\begin{center}~\vfill\end{center}

\begin{center}\begin{tabular}{|c||c|c||c|c|}
\hline 
&
\multicolumn{2}{c||}{$\Xi_{tot}$~~~~{[}$\times10^{-2}${]} }&
\multicolumn{2}{c|}{ $\mathcal{T}$~~{[}min{]}}\tabularnewline
\cline{2-3} \cline{4-5} 
&
\emph{IMSA-SOM}&
\emph{BARE-SOM}&
\multicolumn{1}{c|}{\emph{IMSA-SOM}}&
\emph{BARE-SOM}\tabularnewline
\hline
\hline 
{}``rectTM\_cent''&
$1.67$&
$3.71$&
$124$&
$550$\tabularnewline
\hline 
{}``rectTM\_dece''&
$1.39$&
$2.80$&
$120$&
$430$\tabularnewline
\hline
\end{tabular} \end{center}

\begin{center}~\vfill\end{center}

\begin{center}\textbf{Tab. I - Ye et} \textbf{\emph{al.}}\textbf{,}
\textbf{\emph{{}``}}Multi-Resolution Subspace-Based ...''\end{center}

\begin{thebibliography}{10}
\bibitem{Liu 2019}T. Liu, Y. Zhao, Y. Wei, Y. Zhao, and S. Wei, {}``Concealed object
detection for activate millimeter wave image,'' \emph{IEEE Trans.
Ind. Electron.}, vol. 66, no. 12, pp. 9909-9917, Dec. 2019.
\bibitem{Briquech 2020}Z. Briqech, S. Gupta, A.-A. Beltay, A. Elboushi, A.-R. Sebak, and
T. A. Denidni, {}``57-64 GHz imaging/detection sensor--part II: experiments
on concealed weapons and threatening materials detection,'' \emph{IEEE
Sensors J}., vol. 20, no. 18, pp. 10833-10840, Sep. 2020.
\bibitem{Zhuralev 2021}A. Zhuravlev, V. Razevig, M. Chizh, G. Dong, and B. Hu, {}``A new
method for obtaining radar images of concealed objects in microwave
personnel screening systems,'' \emph{IEEE Trans. Microw. Theory Techn.},
vol. 69, no. 1, pp. 357-364, Jan. 2021.
\bibitem{Wang 2021}C. Wang, J. Shi, Z. Zhou, L. Li, Y. Zhou, and X. Yang, {}``Concealed
object detection for millimeter-wave images with normalized accumulation
map,'' \emph{IEEE Sensors J.}, vol. 21, no. 5, pp. 6468-6475, Mar.
2021.
\bibitem{Ok 2014}G. Ok, H. J. Kim, H. S. Chun, and S.-W. Choi, {}``Foreign-body detection
in dry food using continuous sub-terahertz wave imaging,'' \emph{Food
Control}, vol. 42, pp. 284-289, Aug. 2014.
\bibitem{Amineh 2008}R. K. Amineh, N. K. Nikolova, J. P. Reilly, and J. R. Hare, {}``Characterization
of surface-breaking cracks using one tangential component of magnetic
leakage field measurements,'' \emph{IEEE Trans. Magn.}, vol. 44,
no. 4, pp. 516-524, Apr. 2008.
\bibitem{Norouzi 2021}M. Norouzi, N. Masoumi, and H. Jahed, {}``Nondestructive phase variation-based
chipless sensing methodology for metal crack monitoring,'' \emph{IEEE
Trans. Instrum. Meas.}, vol. 70, pp. 1-11, Mar. 2021.
\bibitem{Chen 2021}Z. Chen, X. Q. Lin, Y. H. Yan, F. Xiao, M. T. Khan, and S. Zhang,
\char`\"{}Noncontact group-delay-based sensor for metal deformation
and crack detection,\char`\"{} \emph{IEEE Trans. Ind. Electron.},
vol. 68, no. 8, pp. 7613-7619, Aug. 2021.
\bibitem{Sakthivel 2014}M. Sakthivel, B. George, and M. Sivaprakasam, {}``A new inductive
proximity sensor based guiding tool to locate metal shrapnel during
surgery,'' \emph{IEEE Trans. Instrum. Meas.}, vol. 63, no. 12, pp.
2940-2949, Dec. 2014.
\bibitem{Bourgeois 1998}J. R. Bourgeois and G. S. Smith, {}``A complete electromagnetic simulation
of the separated-aperture sensor for detecting buried land mines,''
\emph{IEEE Trans. Antennas Propag.}, vol. 46, no. 10, pp. 1419-1426,
Oct. 1998.
\bibitem{Bansal 2015}R. Bansal, {}``Of mice and men {[}Microwave surfing{]},'' \emph{IEEE
Microw. Mag}., vol. 16, no. 11, pp. 18-20, Dec. 2015.
\bibitem{Shaw 2005}M. R. Shaw, S. G. Millard, T. C. K. Molyneaux, M. J. Taylor, and J.
H. Bungey, {}``Location of steel reinforcement in concrete using
ground penetrating radar and neural networks,'' \emph{NDT E Int}.,
vol. 38, no. 3, pp. 203-212, Apr. 2005.
\bibitem{Chen 2006}Y. Chen, Y. Chen, C. Chiu, and C. Chang, {}``Image reconstruction
of a buried perfectly conducting cylinder illuminated by transverse
electric waves,'' \emph{Int. J. Imag. Syst. Technol}., vol. 15, no.
6, pp. 261-265, Apr. 2006.
\bibitem{Zitouni 2006}A. Zitouni, L. Beheim, R. Huez, and F. Belloir, {}``Smart electromagnetic
sensor for buried conductive targets identification,'' \emph{IEEE
Sensors J.}, vol. 6, no. 6, pp. 1580-1591, Dec. 2006.
\bibitem{Sheiki 2015}F. Sheikhi, D. Spinello, and W. Gueaieb, {}``Renyi entropy filter
for anomaly detection with eddy current remote field sensors,'' \emph{IEEE
Sensors J.}, vol. 15, no. 11, pp. 6399-6408, Nov. 2015.
\bibitem{Chew 1992}W. C. Chew and G. P. Otto, {}``Microwave imaging of multiple conducting
cylinders using local shape functions,'' \emph{IEEE Microwave Guided
Wave Lett.}, Vol. 2, pp. 284-286, Jul. 1992.
\bibitem{Zhou 2002}Y. Zhou and H. Ling, {}``Electromagnetic inversion of ipswich objects
with the use of the genetic algorithm,'' \emph{Microw. Opt. Technol.
Lett.}, vol. 33, no. 6, pp. 457-459, Jun. 2002.
\bibitem{Takenaka 1997}T. Takenaka, Z. Q. Meng, T. Tanaka, and W. C. Chew, {}``Local shape
function combined with genetic algorithm applied to inverse scattering
for strips,'' \emph{Microw. Opt. Technol. Lett.}, vol. 16, no. 6,
pp. 337-341, Dec. 1997.
\bibitem{Weedon 1993}W. H. Weedon and W. C. Chew, {}``Time-domain inverse scattering using
the local shape function (LSF) method,'' \emph{Inv. Prob.}, vol.
9, pp. 551-564, Oct. 1993.
\bibitem{Otto 1994}G. P. Otto and W. C. Chew, {}``Microwave inverse scattering---Local
shape function imaging for improved resolution of strong scatterers,''
\emph{IEEE Trans. Microwave Theory Tech.}, vol. 42, no. 1, pp. 137-141,
Jan. 1994.
\bibitem{Poli 2013}L. Poli, G. Oliveri, and A. Massa, {}``Imaging sparse metallic cylinders
through a local shape function Bayesian compressive sensing approach,''
\emph{J. Opt. Soc. Amer. A.}, vol. 30, no. 6, pp. 1261, Jun. 2013.
\bibitem{Stevanovic 2016}M. N. Stevanovic, L. Crocco, A. R. Djordjevic, and A. Nehorai, {}``Higher
order sparse microwave imaging of PEC scatterers,'' \emph{IEEE Trans.
Antennas Propag.}, vol. 64, no. 3, pp. 988-997, Mar. 2016.
\bibitem{Ye 2013}X. Z. Ye, X. Chen, Y. Zhong, and R. C. Song, {}``Simultaneous reconstruction
of dielectric and perfectly conducting scatterers via T-matrix method,''
\emph{IEEE Trans. Antennas Propag.}, vol. 61, no. 7, pp. 3774-3781,
Jul. 2013.
\bibitem{Yu 2005}C. Yu, L. P. Song, and Q. H. Liu, {}``Inversion of multi-frequency
experimental data for imaging complex objects by a DTA-CSI method,''
\emph{Inverse Probl.}, vol. 21, no. 6, pp. S165-S178, Dec. 2005.
\bibitem{Azaro 2006}R. Azaro, M. Donelli, D. Franceschini, and A. Massa, {}``Multiscaling
reconstruction of metallic targets from TE and TM experimental data,''
\emph{Microw. Opt. Technol. Lett.}, vol. 48, no. 2, pp. 322-324, Feb.
2006.
\bibitem{Sun 2018}S. Sun, B. J. Kooij, and A. G. Yarovoy, {}``A linear model for microwave
imaging of highly conductive scatterers,'' \emph{IEEE Trans. Microw.
Theory Techn.}, vol. 66, no. 3, pp. 1149-1164, Mar. 2018.
\bibitem{Roger 1981}A. Roger, {}``Newton-Kantorovitch algorithm applied to an electromagnetic
inverse problem,'' \emph{IEEE Trans. Antennas Propag.}, vol. 29,
no. 2, pp. 232-238, Mar. 1981.
\bibitem{Chiu 1996}C. C. Chiu and P. T. Liu, {}``Image reconstruction of a perfectly
conducting cylinder by the genetic algorithm,'' \emph{IEE Proc. Microw}.
\emph{Antennas Propag}., vol. 143, no. 3, pp. 249, Jun. 1996.
\bibitem{Chien 2006 a}W. Chien, C. H. Huang, and C. C. Chiu, {}``Cubic-spline expansion
for a two-dimensional periodic conductor in free space,'' \emph{Int.
J. Appl. Electromagn. Mechanics}, vol. 24, no. 1-2, pp. 105-114, Dec.
2006.
\bibitem{Chien 2006 b}W. Chien, C. C. Chiu, and C. L. Li, {}``Cubic-spline expansion for
a conducting cylinder buried in a slab medium,'' \emph{Electromagnetics},
vol. 26, no. 5, pp. 329-343, 2006.
\bibitem{Qing 2003}A. Qing, {}``Electromagnetic inverse scattering of multiple two-dimensional
perfectly conducting objects by the differential evolution strategy,''
\emph{IEEE Trans. Antennas Propag.}, vol. 51, no. 6, pp. 1251-1262,
Jun. 2003.
\bibitem{Zhou 2003}Y. Zhou, J. Li, and H. Ling, {}``Shape inversion of metallic cavities
using hybrid genetic algorithm combined with tabu list,'' \emph{Electron.
Lett.}, vol. 39, no. 3, pp. 280, 2003.
\bibitem{Qing 2004}A. Qing, {}``Electromagnetic inverse scattering of multiple perfectly
conducting cylinders by differential evolution strategy with individuals
in groups (GDES),'' \emph{IEEE Trans. Antennas Propag}., vol. 52,
no. 5, pp. 1223-1229, May 2004.
\bibitem{Chien 2005}W. Chien and C.-C. Chiu, {}``Using NU-SSGA to reduce the searching
time in inverse problem of a buried metallic object,'' \emph{IEEE
Trans. Antennas Propag.}, vol. 53, no. 10, pp. 3128-3134, Oct. 2005.
\bibitem{Litman 1998}A. Litman, D. Lesselier, and F. Santosa, {}``Reconstruction of a
two-dimensional binary obstacle by controlled evolution of a level-set,''
\emph{Inverse Probl.}, vol. 14, no. 3, pp. 685-706, Jun. 1998.
\bibitem{Ye 2010}X. Ye, X. Chen, Y. Zhong, and K. Agarwal, {}``Subspace-based optimization
method for reconstructing perfectly electric conductors,'' \emph{Prog.
Electromag. Res}., vol. 100, pp. 119-128, 2010.
\bibitem{Ye 2011}X. Z. Ye, Y. Zhong, and X. Chen, {}``Reconstructing perfectly electric
conductors by the subspace-based optimization method with continuous
variables,'' \emph{Inverse Probl}., vol. 27, no. 5, pp. 055011, May
2011.
\bibitem{Chen 2018}X. Chen, \emph{Computational methods for electromagnetic inverse scattering}.
Singapore: Wiley, 2018.
\bibitem{Shen 2013}J. Shen, Y. Zhong, X. Chen, and L. Ran, {}``Inverse scattering problems
of reconstructing perfectly electric conductors with TE illumination,''
\emph{IEEE Trans. Antennas Propag.}, vol. 61, no. 9, pp. 4713-4721,
Sep. 2013.
\bibitem{Caorsi 2003}S. Caorsi, M. Donelli, D. Franceschini, and A. Massa, {}``A new methodology
based on an iterative multiscaling for microwave imaging,'' \emph{IEEE
Trans. Microw. Theory Techn.}, vol. 51, no. 4, pp. 1162-1173, Apr.
2003.
\bibitem{Rocca 2009}P. Rocca, M. Donelli, G. L. Gragnani, and A. Massa, {}``Iterative
multi-resolution retrieval of non-measurable equivalent currents for
the imaging of dielectric objects,'' \emph{Inverse Probl.}, vol.
25, no. 5, pp. 055004, May 2009.
\bibitem{Donelli 2006}M. Donelli, G. Franceschini, A. Martini, and A. Massa, {}``An integrated
multiscaling strategy based on a particle swarm algorithm for inverse
scattering problems,'' \emph{IEEE Trans. Geosci. Remote Sens.}, vol.
44, no. 2, pp. 298-312, Feb. 2006.
\bibitem{Oliveri 2011}G. Oliveri, Y. Zhong, X. Chen, and A. Massa, {}``Multiresolution
subspace-based optimization method for inverse scattering problems,''
\emph{J. Opt. Soc. Amer. A}, vol. 28, no. 10, pp. 2057-2069, Oct.
2011.
\bibitem{Ye 2015}X. Ye, L. Poli, G. Oliveri, Y. Zhong, K. Agarwal, A. Massa, and X.
Chen, {}``Multi-resolution subspace-based optimization method for
solving three-dimensional inverse scattering problems,'' \emph{J.
Opt. Soc. Amer. A}, vol. 32, no. 11, pp. 2218-2226, Nov. 2015.
\bibitem{Salucci 2017a}M. Salucci, G. Oliveri, N. Anselmi, F. Viani, A. Fedeli, M. Pastorino,
and A. Randazzo, {}``Three-dimensional electromagnetic imaging of
dielectric targets by means of the multiscaling inexact-newton method,''
\emph{J. Opt. Soc. Amer. A.}, vol. 34, no. 7, pp. 1119, Jul. 2017.
\bibitem{Salucci 2017b}M. Salucci, L. Poli, N. Anselmi and A. Massa, {}``Multifrequency
particle swarm optimization for enhanced multiresolution GPR microwave
imaging,'' \emph{IEEE Trans. Geosci. Remote Sens.}, vol. 55, no.
3, pp. 1305-1317, Mar. 2017.
\bibitem{Anselmi 2018}N. Anselmi, L. Poli, G. Oliveri, and A. Massa, {}``Iterative multiresolution
Bayesian CS for microwave imaging,'' \emph{IEEE Trans. Antennas Propag.},
vol. 66, no. 7, pp. 3665-3677, Jul. 2018.
\bibitem{Zhong 2020}Y. Zhong, M. Salucci, K. Xu, A. Polo, and A. Massa, {}``A multiresolution
contraction integral equation method for solving highly nonlinear
inverse scattering problems,'' \emph{IEEE Trans. Microw. Theory Techn.},
vol. 68, no. 4, pp. 1234-1247, Apr. 2020.
\bibitem{Salucci 2021}M. Salucci, C. Estatico, A. Fedeli, G. Oliveri, M. Pastorino, S. Povoli,
A. Randazzo, and P. Rocca, \char`\"{}2-D TM GPR imaging through a
multiscaling multifrequency approach in $L^{p}$ spaces,\char`\"{}
\emph{IEEE Trans. Geosci. Remote Sens.}, vol. 59, no. 12, pp. 10011-10021,
Dec. 2021.
\bibitem{Bucci 1997}O. M. Bucci and T. Isernia, {}``Electromagnetic inverse scattering:
retrievable information and measurement strategies,'' \emph{Radio
Sci.}, vol. 32, no. 6, pp. 2123-2137, 1997.
\bibitem{Chen 2009}X. Chen, {}``Application of signal-subspace and optimization methods
in reconstructing extended scatterers,'' \emph{J. Opt. Soc. Amer.
A}, vol. 26, no. 4, pp. 1022-1026, Apr. 2009.
\bibitem{Peterson 1998}A. F. Peterson, S. L. Ray, and R. Mittra, \emph{Computational methods
for electromagnetics}. New York, NY: IEEE Press, 1998.
\bibitem{Belkebir 2001}K. Belkebir and M. Saillard, \char`\"{}Special section: testing inversion
algorithms against experimental data,\char`\"{} \emph{Inverse Probl}.,
vol. 17, no. 6, pp. 1565-1571, Dec. 2001.
\end{thebibliography}
\end{document}